\documentclass[11pt,oneside]{article}

\usepackage[showpreprintnumbers]{aj}

\usepackage{footnote}
\makesavenoteenv{tabular}
\makesavenoteenv{table}

\title{First Order Galilean Superfluid Dynamics}

\author[a]{Nabamita Banerjee}{nabamita@iiserpune.ac.in}
\author[b]{Suvankar Dutta}{suvankar@iiserb.ac.in}
\author[c]{Akash Jain}{akash.jain@durham.ac.uk, ajainphysics@gmail.com}

\affiliation[a]{Dept. of Physics, Indian Institute of Science Education and Research (IISER) Pune --
  411008, Maharashtra, India}
\affiliation[b]{Dept. of Physics, Indian Institute of Science Education and Research (IISER) Bhopal
  -- 462030,
  Madhya Pradesh, India}
\affiliation[c]{Centre for Particle Theory \& Dept. of Mathematical Sciences, Durham University,
  Durham DH1 3LE, United Kingdom}

\abstract{ We study dynamics of (anomalous) Galilean superfluid up to first order in derivative
  expansion, both in parity-even and parity-odd sectors. We construct a relativistic system -- null
  superfluid, which is a null fluid (introduced in \cite{Banerjee:2015hra}) with a spontaneously
  broken global $\rmU(1)$ symmetry. A null superfluid is in one to one correspondence with Galilean
  superfluid in one lower dimension, i.e. they have same symmetries, thermodynamics, constitutive
  relations and are related to each other by a mere choice of basis. The correspondence is based on
  null reduction, which is known to reduce the Poincar\'{e} symmetry of a theory to Galilean
  symmetry in one lower dimension. To perform this analysis, we use offshell formalism of
  (super)fluid dynamics, adopting it appropriately to null (super)fluids. }

\preprint{DCPT-16/51} 

\def\Ndot{{%
    \setbox0\hbox{$\N$}%
    \rlap{\hbox to \wd0{\hss \raisebox{4pt}{\scriptsize {$\centerdot$}}\hss}}\box0
}}  

\begin{document}

\maketitle

\section{Introduction and Summary}

Hydrodynamics is an effective description of low energy fluctuations of a quantum system around
thermodynamic equilibrium. In this description, we assume the hydrodynamic system, known as a
\emph{fluid}, to be at a finite temperature, and study its fluctuations at length scales much larger
than the mean free path of the system. In this limit and far away from any second order phase
transition point, a fluid can be described by a small number of degrees of freedom known as
hydrodynamic modes: temperature, chemical potential(s) and normalized fluid velocity. Various
conserved currents of the system can then be written in terms of these hydrodynamic modes, arranged
as a perturbative expansion in derivatives, known as \emph{fluid constitutive relations}.  At any
particular order in derivative expansion, constitutive relations contain all the possible
independent tensor structures allowed by symmetry at that order, multiplied with unknown
coefficients known as transport coefficients.
If the underlying quantum theory has a continuous global symmetry which is spontaneously broken in
the ground state, then the low energy fluctuations can contain massless Goldstone modes
corresponding to the broken symmetry. Therefore for fluids with a spontaneously broken symmetry,
known as \emph{superfluids}, hydrodynamic modes also contain these Goldstone modes. This leads to a
considerable modification of the constitutive relations, adding new tensor structures containing the
derivatives of the Goldstone modes and hence new transport coefficients. In this paper, we work out
the most generic constitutive relations of a Galilean superfluid up to first order in the derivative
expansion.

Superfluidity was first observed in liquid helium by \cite{1938Natur.141...74K,1938Natur.142..643A}
in 1938, while studying its flow through a thin capillary. They observed that liquid helium flows
through the capillary without any dissipation, hence inspiring the name ``superfluid''.
Other than this dissipationless flow, superfluids have many more striking features, such as upon
rotation they develop vortices (quasi-one-dimensional strings whose number is proportional to the
externally imposed angular momentum). Furthermore, their specific heat shows a sudden change in
behavior at a certain critical temperature. Above the critical temperature system behaves like an
ordinary fluid, though as the temperature drops below the critical temperature, system undergoes a
phase transition from the ordinary fluid phase to the superfluid phase.

Study of superfluid dynamics has been a topic of interest for a long time. First theory of
superfluid dynamics was written down by London \cite{1938Natur.141..643L} in 1938, followed by a
two-fluid model of superfluids proposed by Landau and Tisza \cite{PhysRev.60.356,PhysRev.72.838} in
1940s. They studied ideal superfluids in a non-relativistic setting, which was later generalized to
describe a relativistic superfluid by \cite{KHALATNIKOV198270,1981PhLA...86...79I,Israel198277,
  Carter:1992gmy,CARTER1992243,Son:2000ht}. The subject was recently revisited by
\cite{Bhattacharya:2011eea, Bhattacharya:2011tra, Bhattacharyya:2012xi}, who re-derived the
relativistic superfluid constitutive relations using the second law of thermodynamics and
equilibrium partition functions. Among other interesting results, they found that up to first order
in derivative expansion, a relativistic superfluid is characterized by pressure (at ideal
order), 23 parity-even and 7 parity-odd first order transport coefficients and 2 undetermined
constants including the anomaly constant (after imposing Onsager relations and CPT invariance these
numbers drop down to 16 parity-even and 6 parity odd transport coefficients and one anomaly
constant).  See \cref{summ} for a summary and \cref{sec2} for more details.

In this paper, we perform a similar exercise for Galilean superfluids. We derive the constitutive
relations for a Galilean superfluid consistent with the second law of thermodynamics, up to first
order in derivative expansion, both in parity even and odd sectors. Study of Galilean superfluids is
important because it provides a laboratory to probe many-body physics in extreme quantum regime with
high-precision \cite{Giorgini:2008zz}.  Relativistic effects are important in high-energy
superfluids, where mass of the constituents is small compared to their kinetic energy, e.g. quark
superfluidity in compact stars.  In contrast, for low-energy systems such as liquid helium and
ultra-cold atomic gases, a Galilean framework is more ideal.
 
Recently in \cite{Banerjee:2015uta,Banerjee:2015hra}, we established that one can derive the most
generic constitutive relations for an ordinary Galilean fluid starting from a relativistic system,
namely a \emph{null fluid} in one higher dimension, followed by a \emph{null
  reduction}\footnote{Null reduction of an ordinary relativistic fluid gives us a constrained
  Galilean fluid as found in \cite{Banerjee:2014mka}.} \cite{Rangamani:2008gi}. Loosely speaking,
null fluid is a fluid coupled to a background with fields: a metric $g_{\sM\sN}$, a $\rmU(1)$ gauge
field $A_\sM$ and a covariantly constant null isometry $\scrV = \{V^\sM,\L_V\}$ with
$V^\sM A_\sM + \L_V =$ constant. We call this background a \emph{null background}\footnote{Here,
  definition of null backgrounds has been adapted to a torsionless spacetime. For backgrounds with
  torsion, look at \cite{Jain:2015jla}.}. Theories on a null background, which we call \emph{null
  theories}, are demanded to be invariant under $\scrV$ preserving diffeomorphisms and gauge
transformations. Upon performing null reduction, i.e. choosing a basis
$\{x^\sM\} = \{ x^-, t, x^i \}$ such that $\scrV = \{V = \dow_-,\L_V = 0\}$, these restricted
transformations reduce to the well known Galilean transformations on the background spanned by
coordinates $\{t,x^i\}$. It suggests that null theories are entirely equivalent to Galilean
theories, and are related by merely this choice of basis. It follows that a fluid on null background
-- null fluid is entirely equivalent to a Galilean fluid. Their constitutive relations, conservation
laws, thermodynamics etc. match exactly to all orders in derivative expansion.  Due to presence of
an additional vector field $V^\sM$, constitutive relations of a null fluid are vastly different from
those of a relativistic fluid and contain many more transport coefficients. This accounts for the
additional transport coefficients in a Galilean fluid as compared to a relativistic fluid, while at
the same time establishing that the most generic Galilean fluid cannot be gained by null reduction
of an ordinary relativistic fluid.

In this paper, we take the construction of null fluids one step further to include null superfluids,
i.e. we construct a null fluid with a spontaneously broken $\rmU(1)$ symmetry. The corresponding
Goldstone mode is a new field in the theory and modifies the constitutive relations of an ordinary
null fluid. Once we have the constitutive relations for a null superfluid, corresponding Galilean
superfluid constitutive relations follow trivially via null reduction. We find that up to first
order in derivatives, a Galilean superfluid is described by pressure $P$ (at ideal order), a total
of 51 first order transport coefficients and two unknown constants including the anomaly
constant. Out of these 51 coefficients, 38 lie in parity-even sector while 13 are in parity-odd
sector. Furthermore, only 22 parity-even and 3 parity-odd coefficients are dissipative. Out of the
non-dissipative coefficients, 3 parity-even and 3 parity-odd coefficients describe equilibrium
physics, while the remaining 13 parity-even and 7 parity-odd coefficients describe non-dissipative
effects away from equilibrium. Finally, following the intuition from relativistic superfluids and
known Galilean results in \cite{landau1959fluid}, there are hints that the 7 parity-even
non-dissipative non-hydrostatic coefficients and 3 parity-odd dissipative coefficients are switched
off using Onsager relations (imposing microscopic reversibility of field theories). This would imply
that the parity-odd sector is purely non-dissipative. However, a detailed microscopic calculation is
required to establish confidence in these Galilean Onsager relations, which we do not perform in
this paper. In \cref{summ}, we have summarized the counting of transport coefficients for the most
generic Galilean superfluid, along with a comparison with relativistic superfluids reviewed in
\cref{sec2}, and known results for ordinary Galilean and relativistic fluids.

\begin{table}[t]
  \centering         
  \begin{tabular}[t]{|l|c|c|c|c|}
    \hline
    & Relativistic & Relativistic & Galilean & Galilean \\
    & Fluid & Superfluid & Fluid & Superfluid \\
    \hline
    Hydrostatic & $0 + \tilde 0$ & $2 + \tilde 2$ & $0 + \tilde 0$ & $3 + \tilde 3$ \\
    Non-hydrostatic non-diss. & $0^* + \tilde 0$ & $7^* + \tilde 4$ & $1^* + \tilde 0$ & $13^* + \tilde 7$ \\
    Dissipative & $2 + \tilde 0^*$ & $14 + \tilde 1^*$ & $5 + \tilde 0^*$ & $22 + \tilde 3^*$ \\
    \hline
    Total & $2 + \tilde 0 = 2$ & $23 + \tilde 7 = 30$ & $6 + \tilde 0 = 6$ & $38 + \tilde {13} = 51$
    \\
    \hline
    Total (with Onsager) & $2 + \tilde 0 = 2$ & $16 + \tilde 6 = 22$ & $5 + \tilde 0 = 5$ & $25 + \tilde {10} = 35$
    \\
    \hline\hline
    Hydrostatic Constants & $\tilde 3 + \tilde 1_{\text{anomaly}}$
                   & $\tilde 1 + \tilde 1_{\text{anomaly}}$ & $\tilde 4 + \tilde 1_{\text{anomaly}}$
                                             & $\tilde 1  + \tilde 1_{\text{anomaly}}$ \\
    \hline
  \end{tabular}
  \caption{\label{summ} Counting of the independent first order transport coefficients consistent
    with the second law of thermodynamics. The numbers with a ``tilde'' represent the parity-odd count
    (in $3$ spatial dimensions) while the ``un-tilde'' numbers are the parity-even count. The
    coefficients with an ``asterisk'' drop out on imposing Onsager relations (microscopic
    time-reversal invariance). Finally, in the last row we have given the number of undetermined
    constants including the anomaly constant. In both relativistic and Galilean cases, we have
    gotten rid of a hydrostatic coefficient by redefinition of the $\rmU(1)$ phase $\vf$.}
\end{table}

Another recent development in hydrodynamics is offshell formalism introduced by
\cite{Loganayagam:2011mu,Haehl:2014zda,Haehl:2015pja}, which streamlines the analysis of
constitutive relations in accordance with the second law of thermodynamics, up to arbitrarily high
orders in derivative expansion. We have reviewed this formalism in \cref{sec2}. In a nutshell,
for ordinary fluids the formalism requires us to consider a version of the second law of
thermodynamics which is valid for thermodynamically non-isolated fluids,
\begin{equation}\label{disc.offshellFreeE.ord}
  \N_\mu J_S^\mu + \frac{u_\mu}{T} \lb \N_\nu T^{\mu\nu} - F^{\nu\r} J_\r - \rmT_\rmH^{\mu\perp}\rb
  + \frac{\mu}{T} \lb\N_\mu J^\mu - \rmJ^\perp_\rmH \rb = \D \geq 0.
\end{equation}
Since the fluid is not thermodynamically isolated, it is allowed to interact with its surrounding
and hence conservation laws are no longer satisfied. Therefore the original second law
$\N_\mu J_S^\mu \geq 0$ has been modified with combinations of the conservation laws. We need to
find the most generic constitutive relations for $T^{\mu\nu}$, $J^\mu$ allowed by symmetries (modulo
terms related to each other by conservation laws) which satisfy \cref{disc.offshellFreeE.ord} for
some entropy current $J^\mu_S$ and $\D \geq 0$. When we move to superfluids, we have an additional
field $\vf$ (the Goldstone mode) which comes with its own equation of motion $K=0$, the Josephson
equation. Going offshell in $\vf$, conservation equations get modified by combinations of $K$, and
the second law of thermodynamics for thermodynamically non-isolated superfluids takes the form (see
\cite{Jain:2016rlz} for more details),
\begin{equation}\label{disc.offshellFreeE.null}
  \N_\mu J_S^\mu + \frac{u_\mu}{T} \lb \N_\nu T^{\mu\nu} - F^{\nu\r} J_\r - \rmT_\rmH^{\mu\perp} -
  \xi^\mu K\rb
  + \frac{\mu}{T} \lb\N_\mu J^\mu - \rmJ^\perp_\rmH + K \rb = \D \geq 0.
\end{equation}
Note that contrary to the philosophy of \cite{Loganayagam:2011mu,Haehl:2014zda,Haehl:2015pja},
though we have gone offshell in $\vf$ we have not modified the second law with a multiple of
$K$. Rather, we require the second law of thermodynamics to be satisfied even for offshell
configurations of $\vf$. Next, we find the most generic ``superfluid constitutive relations''
$T^{\mu\nu}, J^\mu$, $K$ allowed by symmetries (modulo terms related to each other by conservation
laws or the Josephson equation) which satisfy \cref{disc.offshellFreeE.null} for some entropy
current $J^\mu_S$ and $\D \geq 0$. In \cref{sec3}, we have extended this formalism to null
(super)fluids, and used it to work out the constitutive relations of a null/Galilean superfluid up
to first order in derivative expansion.

The paper is organized as follows: we start \cref{sec2} with a review of  offshell formalism for
relativistic hydrodynamics. Readers well familiar with this formalism can skip to \cref{ss2.2} where
we have reviewed offshell formalism for relativistic superfluids and used it to work out
respective constitutive relations up to first order in derivative expansion.  Next in
\cref{sec3}, we introduce offshell formalism for null superfluids and find respective
constitutive relations up to first order in derivative expansion. The null superfluid results
have been reduced to Galilean superfluids in \cref{sec4}. These are the main results of this
paper. Finally, we conclude with some discussion in \cref{sec5}. The paper contains two appendices:
in \cref{eqbPF} we present equilibrium partition function for null superfluids and in
\cref{calc.details} we give details of some computations glossed over in the main text.

\section{Revisiting Relativistic Superfluids}\label{sec2}

Before starting with null superfluids, it is instructive to revisit the relativistic superfluids
first. It will help us appreciate the similarities between the two systems, while at the same time
allowing for an isolation of the differences. Needless to say, all the results in this section have
already been worked out in the literature \cite{Bhattacharya:2011tra, Bhattacharyya:2012xi,
  Bhattacharya:2011eea}, however our approach will be slightly different. We will work in the
``offshell formalism of hydrodynamics'', which was introduced for ordinary (non-super) fluids in
\cite{Loganayagam:2011mu,Haehl:2015pja}, and later extended to superfluids in \cite{Jain:2016rlz}.

\subsection{Offshell Formalism for Relativistic Ordinary Fluids} 

Let us begin with ordinary relativistic fluids. Consider a $d$-dimensional manifold $\cM_d$ equipped
with the background fields: a metric $g_{\mu\nu}$ and a $\rmU(1)$ gauge field $A_\mu$. Physical
theories coupled to $\cM_d$ are required to be invariant under diffeomorphisms and $\rmU(1)$ gauge
transformations. These act on the said background fields as,
\begin{equation}\label{background.transformations_rel}
  \d_\scrX g_{\mu\nu}= \lie_\c  g_{\mu\nu} = \nabla_{\mu} \c_{\nu} + \N_\nu \c_\mu, \qquad 
  \d_\scrX A_\mu = \lie_\c A_\mu + \dow_\mu \L_\c = \partial_{\mu}\lb \L_\c + \c^\nu A_\nu\rb+ \c^\nu F_{\nu\mu},
\end{equation}
for some diffeomorphism and $\rmU(1)$ gauge parameters $\scrX = \{\c^\mu,\L_\c\}$ respectively. In
this work we will only be interested in a particular class of these theories -- \emph{fluids}, which
are are the universal near equilibrium limit of quantum field theories. Near equilibrium, the
spectrum of any quantum field theory on $\cM_d$ must contain an \emph{energy momentum tensor}
$T^{\mu\nu}$ and a \emph{charge current} $J^\mu$. These quantities satisfy a set of
\emph{conservation laws} (here $\N_\mu$ is the covariant derivative associated with $g_{\mu\nu}$,
$F_{\mu\nu} = \dow_\mu A_\nu - \dow_\nu A_\mu$ is the field strength associated with $A_{\mu}$ and
$\rmT_\rmH^{\mu\perp}, \rmJ^\perp_\rmH$ are Hall currents carrying the anomalous contribution to the
conservation equations),
\begin{equation}\label{conservation}
  \N_\mu T^{\mu\nu} - F^{\nu\r} J_\r - \rmT_\rmH^{\nu\perp} = 0, \qquad
  \N_\mu J^\mu - \rmJ^\perp_\rmH = 0,
\end{equation}
provided that the system is \emph{thermodynamically isolated}. In fact, \cref{conservation} can be
taken as a definition of thermodynamic isolation for near equilibrium quantum systems. The
conservation laws \cref{conservation} can also be thought of as a `near equilibrium version' of the
\emph{first law of thermodynamics}, which imposes the conservation of not just energy, but also
momentum and $\rmU(1)$ charge. Formally, we define an (ordinary) fluid as a near equilibrium system
characterized by the currents $T^{\mu\nu}$, $J^\mu$, with dynamics given by the conservation laws
\cref{conservation} imposed as the `equations of motion'. Since \cref{conservation} are $(d+1)$
equations in $d$ dimensions, they can provide dynamics for a fluid described by an arbitrary set of
$(d+1)$ variables. We choose these to be a normalized \emph{fluid velocity} $u^\mu$ (with
$u^\mu u_\mu = -1$), a \emph{temperature} $T$ and a \emph{chemical potential} $\mu$, collectively
known as the \emph{hydrodynamic fields (modes)}. A fluid hence is completely characterized by a
gauge-invariant expression of $T^{\mu\nu}$, $J^\mu$ in terms of $g_{\mu\nu}$, $A_\mu$, $u^\mu$, $T$,
$\mu$, known as the \emph{hydrodynamic constitutive relations}. The near equilibrium assumption
allows us to arrange these constitutive relations as a perturbative expansion in derivatives (known
as \emph{derivative} or \emph{gradiant expansion}), consistently truncated at a finite order in
derivatives.

Being a thermodynamic system, a fluid is also required to satisfy a version of the \emph{second law
  of thermodynamics}, stating that there must exist an \emph{entropy current} $J_S^\mu$ whose
divergence is positive semi-definite everywhere, i.e.,
\begin{equation}\label{onshell2ndlaw_rel}
  \N_\mu J_S^\mu = \D \geq 0,
\end{equation}
as long as the fluid is thermodynamically isolated (i.e. conservation laws \cref{conservation} or
equivalently the first law(s) of thermodynamics are satisfied). The job of hydrodynamics now is to
find the most general constitutive relations $T^{\mu\nu}$, $J^\mu$ and an associated $J_S^\mu$, $\D$
order by order in derivative expansion, such that \cref{onshell2ndlaw_rel} is satisfied for
thermodynamically isolated fluids. This task has been extensively undertaken in the literature
\cite{Son:2009tf,Banerjee:2012iz,Jensen:2012jh,Bhattacharyya:2014bha,Bhattacharyya:2013lha,Bhattacharyya:2012nq}.

The problem stated in this language however, turns out to be increasingly hard to solve as we go to
2nd or higher orders in derivative expansion \cite{Banerjee:2015vxa}. Fortunately, it was
realized in \cite{Loganayagam:2011mu} that most of the complication in the aforementioned
computation comes from the fact that we need to maintain the thermodynamic isolation (i.e. satisfy
the conservation equations) perturbatively at every order. A much easier problem to solve is to
allow for the fluid to interact with its surroundings, i.e. break the conservation laws
\cref{conservation} by introducing an arbitrary external momentum $P^{\mu}_{ext}$ and a charge
$Q_{ext}$ source,
\begin{equation}
\N_\mu T^{\mu\nu} - F^{\nu\r} J_\r - \rmT_\rmH^{\nu\perp} = P_{ext}^\nu, \qquad
\N_\mu J^\mu - \rmJ^\perp_\rmH = Q_{ext}.
\end{equation}
The LHS of the second law in \cref{onshell2ndlaw_rel} will also need to be augmented with an
arbitrary combination of $P_{ext}^\mu$, $Q_{ext}$ for the inequality to be satisfied,
\begin{align}\label{offshell_2ndlaw}
  \N_\mu J_S^\mu + \b_\nu P^\nu_{ext} + \lb \L_\b + A_\mu  \b^\mu\rb Q_{ext} &= \D \geq 0, \nn\\
  \implies \N_\mu J_S^\mu + \b_\nu \lb \N_\mu T^{\mu\nu} - F^{\nu\r} J_\r - \rmT_\rmH^{\mu\perp} \rb
  + \lb \L_\b + A_\mu  \b^\mu\rb \lb\N_\mu J^\mu - \rmJ^\perp_\rmH \rb &= \D \geq 0,
\end{align}
for some fields $\scrB = \{\b^\mu,\L_\b\}$. This version of the second law is known as the
\emph{offshell second law of thermodynamics}, because the conservation laws, which are imposed as
equations of motion on the fluid, are not required to be satisfied. \Cref{offshell_2ndlaw} can be
recasted into a yet another useful form by defining a \emph{free energy current} $G^\mu$ as,
\begin{equation}\label{nmudef}
  - \frac{G^\mu}{T} = N^\mu = J_S^\mu + \b_\nu T^{\mu\nu} + \lb \L_\b + A_\nu \b^\nu\rb J^\mu, \quad
  - \frac{\rmG^\perp_\rmH}{T} = \rmN^\perp_\rmH = \b_\mu\rmT_\rmH^{\mu\perp} + \lb \L_\b + A_\nu \b^\nu\rb \rmJ_\rmH^\perp.
\end{equation}
Having done that, \cref{offshell_2ndlaw} implies a \emph{free energy conservation},
\begin{equation}\label{freeEcons}
  \N_\mu N^\mu - \rmN^\perp_\rmH = \half T^{\mu\nu} \d_\scrB g_{\mu\nu} + J^\mu\d_\scrB A_\mu + \D,
  \qquad \D\geq 0,
\end{equation}
where similar to \cref{background.transformations_rel} we have defined, 
\begin{equation}
  \d_\scrB g_{\mu\nu}= \lie_\b  g_{\mu\nu}= \nabla_{\mu}\b_{\nu} + \nabla_{\nu}\b_{\mu}, \qquad 
  \d_\scrB A_\mu = \lie_\b A_\mu + \dow_\mu \L_\b = \partial_{\mu}\lb \L_\b + \b^\nu A_\nu\rb+ \b^\nu F_{\nu\mu}.
\end{equation}
Recall that the hydrodynamic fields $u^\mu$, $T$, $\mu$ introduced earlier were some arbitrary
$(d+1)$ fields chosen to describe the fluid. Like in any field theory, they are permitted to admit
an arbitrary redefinition among themselves without changing the physics. This huge amount of freedom
can be fixed by explicitly choosing,
\begin{equation}\label{fixingu}
  T = \frac{1}{\sqrt{-\b^\nu \b_\nu}}, \qquad
  u^\mu = \frac{\b^\mu}{\sqrt{-\b^\nu \b_\nu}}, \qquad
  \mu = \frac{\L_\b + A_\mu\b^\mu}{\sqrt{-\b^\nu \b_\nu}},
\end{equation}
or conversely,
\begin{equation}\label{fixingbeta}
  \b^\mu = \frac{1}{T} u^\mu, \qquad
  \L_\b = \frac{1}{T}\mu - A_\mu\b^\mu.
\end{equation}
As a consequence of this choice, $\scrB = \{\b^\mu,\L_\b\}$ is just a renaming of the
hydrodynamic fields. Finally, we can find the most general gauge-invariant expression of the
currents $T^{\mu\nu}$, $J^\mu$ in terms of $g_{\mu\nu}$, $A_\mu$, $\b^\mu$, $\L_\b$ arranged in a
derivative expansion, along with an associated $N^\mu$, $\D$ such that \cref{freeEcons} is
satisfied. There however is a caveat in this way of thinking: these $T^{\mu\nu}$, $J^\mu$ are not
just the constitutive relations of a fluid; they also contain information about the external sources
$P^\mu_{ext}$, $Q_{ext}$. One way to circumvent this problem is to pick a set of terms which might
potentially appear in $T^{\mu\nu}$, $J^\mu$ and can be eliminated using the conservation laws, and
only consider the solutions $T^{\mu\nu}$, $J^\mu$ of \cref{freeEcons} (for some $N^\mu$, $\D$) which
do not involve these terms or their derivatives. $T^{\mu\nu}$, $J^\mu$ thus obtained are guaranteed
to be the constitutive relations of a fluid, as they will be free from any $P^\mu_{ext}$, $Q_{ext}$
dependence.

Authors in \cite{Haehl:2014zda,Haehl:2015pja} illustrated a consistent mechanism to find the most
generic constitutive relations of a fluid up to arbitrarily high orders in derivative expansion,
which satisfies \cref{freeEcons}. They further classified these constitutive relations in eight
exhaustive classes, which we will not have scope to review here. Instead, in the following
subsection we will review the offshell analysis of relativistic superfluids which has been
introduced in \cite{Jain:2016rlz}, and later adapt it to Galilean superfluids.

\subsection{Offshell Formalism for Relativistic Superfluids}\label{ss2.2}
  
Let us now review some essential aspects of the offshell formalism for a relativistic superfluid
following the work of \cite{Jain:2016rlz}, and use it to re-derive the respective constitutive
relations up to first order in derivative expansion \cite{Bhattacharya:2011tra,
  Bhattacharyya:2012xi, Bhattacharya:2011eea}. As we have already mentioned in the introduction, a
superfluid is a phase of the fluid where the global $\rmU(1)$ symmetry of the microscopic theory
gets spontaneously broken in the ground state due to condensation of a charged scalar operator. The
$\rmU(1)$ phase $\vf$ of the scalar operator becomes a new field in the theory, along with $u^\mu$,
$T$, $\mu$ on which the respective constitutive relations can depend. Under an infinitesimal gauge
transformation and diffeomorphism, $\vf$ transforms as $\d_\scrX \vf = \c^\mu \dow_\mu \vf -
\L_\c$, with covariant derivative,
\begin{equation}
\xi_\mu= \partial_{\mu}\vf + A_{\mu},
\end{equation}
commonly known as the \emph{superfluid velocity}.  Just like the dynamics of $u^\mu$, $T$, $\mu$ is
given by the conservation equations \cref{conservation}, $\vf$ comes with its own equation of motion
\footnote{$K=0$ should be thought of as a placeholder for the Josephson junction condition
  $u^\mu \xi_\mu = \mu + \cO(\dow)$, which provides dynamics for the $\rmU(1)$ phase $\vf$ in the
  conventional treatment of superfluids. At the moment however, we will allow for an arbitrary $K$
  treating it as yet another `current' besides $T^{\mu\nu}$, $J^\mu$ in the theory, and will later
  establish that the second law of thermodynamics forces $K$ to take the Josephson form.},
\begin{equation}\label{phi.eom}
  K = 0,
\end{equation}
We will be particularly interested in the `offshell' configurations of the field $\vf$, which we
define as the superfluid configurations for which $K\neq 0$. As was suggested by
\cite{Jain:2016rlz}, conservation laws for these configurations modify to,
\begin{equation}\label{eom_spr}
  \N_\mu T^{\mu\nu} = F^{\nu\r} J_\r + \rmT_\rmH^{\nu\perp} + \xi^\nu K, \qquad
  \N_\mu J^\mu = \rmJ^\perp_\rmH - K,
\end{equation}
which trivially turn back to their original form in \cref{conservation} when $K=0$. The claim is
that \emph{even the $\vf$-offshell configurations of a superfluid satisfy the second law of
  thermodynamics}, i.e. there exists an entropy current $J_S^\mu$ whose divergence is positive
semi-definite, i.e.,
\begin{equation}\label{onshell2ndlaw_spr}
  \N_\mu J_S^\mu = \D \geq 0,
\end{equation}
as long as the superfluid is thermodynamically isolated (i.e. \cref{eom_spr} are satisfied),
irrespective of $K$ being zero. Rest of the analysis follows exactly like ordinary fluids; on
allowing the superfluid to interact with its surroundings, the second law modifies to,
\begin{equation}\label{offshell_2ndlaw_spr}
  \N_\mu J_S^\mu + \b_\nu \lb \N_\mu T^{\mu\nu} - F^{\nu\r} J_\r - \rmT_\rmH^{\mu\perp} - \xi^\nu K \rb
  + \lb \L_\b + A_\s  \b^\s\rb \lb\N_\mu J^\mu - \rmJ^\perp_\rmH + K \rb = \D \geq 0.
\end{equation}
In terms of free energy current however, we get,
\begin{equation}\label{adiabaticity_anomaly_spr}
  \N_\mu N^\mu - \rmN^\perp_\rmH = \half T^{\mu\nu} \d_\scrB g_{\mu\nu} + J^\mu\d_\scrB A_\mu + K
  \d_\scrB\vf + \D,  \qquad \D\geq 0,
\end{equation}
where,
\begin{equation}\label{d_Bvf}
  \d_\scrB \vf = \b^\mu \dow_\mu \vf - \L_\b = \frac{1}{T} \lb u^\mu \xi_\mu - \mu \rb.
\end{equation}
Similar to the ordinary fluid, we should now consider the most generic expressions for $T^{\mu\nu}$,
$J^\mu$, $K$ in terms of $g_{\mu\nu}$, $A_\mu$, $\b^\mu$, $\L_\b$, $\vf$ arranged in a derivative
expansion, along with an associated $N^\mu$, $\D$ such that \cref{adiabaticity_anomaly_spr} is
satisfied. However, these $T^{\mu\nu}$, $J^\mu$, $K$ will not be the constitutive relations of a
superfluid, as they will also have information about the surroundings. The true constitutive
relations will be gained by considering those solutions to \cref{adiabaticity_anomaly_spr} which do
not involve a chosen set of terms that can be eliminated using the conservation equations
\cref{eom_spr} and the $\vf$ equation of motion \cref{phi.eom}.

We will now embark on the quest of finding these constitutive relations up to first order in the
derivative expansion. \cite{Jain:2016rlz} provides a complete classification and construction of the
superfluid constitutive relations satisfying \cref{adiabaticity_anomaly_spr} up to arbitrarily high
orders in derivative expansion. In this work however, we are only concerned with the (Galilean)
superfluids up to first derivative order, which can be analyzed directly by brute force without
involving the technicalities of \cite{Jain:2016rlz}.

\subsubsection{Josephson Equation}

In the study of superfluids, the $\rmU(1)$ phase $\vf$ is generally taken to be order $-1$ in the
derivative expansion, while its covariant derivative $\xi_\mu$ is taken to be order 0. The reason
being that the true dynamical degrees of freedom are encoded in the fluctuations of $\vf$ along the
$\rmU(1)$ circle, and not in $\vf$ itself. It implies that the $K\d_\scrB\vf$ term in the free
energy conservation \cref{adiabaticity_anomaly_spr} is allowed to be order zero, if $K$ has an order
0 term. This gives us the unique solution to \cref{adiabaticity_anomaly_spr} at zero derivative
order,
\begin{equation}
  N^\mu, T^{\mu\nu}, J^\mu = \cO(\dow^0), \qquad
  K = -\a \d_\scrB\vf + \cO(\dow), \qquad
  \D = \a (\d_\scrB\vf)^2 + \cO(\dow),
\end{equation}
for some ``transport coefficient'' $\a\geq0$. Note that the $\vf$ equation of motion at this order
will read $K = -\a\d_\scrB\vf + \cO(\dow) = 0$, implying,
\begin{equation}
  \d_\scrB\vf = \frac{1}{T} \lb u^\mu \xi_\mu - \mu \rb = \cO(\dow) \quad \implies\quad
  u^\mu \xi_\mu = \mu + \cO(\dow).
\end{equation}
This is the well known Josephson equation. This condition also ensures that $\D$ is at least
$\cO(\dow)$, avoiding ``ideal superfluid dissipation''. From this point onward, it would be
beneficial to think of $\d_{\scrB}\vf$ as an order $1$ data in derivative expansion rather than
$0$.

\subsection{Ideal Relativistic Superfluids}

Let us now move on to the ideal superfluids, i.e. superfluid constitutive relations that satisfy the
free energy conservation \cref{adiabaticity_anomaly_spr} at first derivative order. At ideal order,
the most generic tensorial form of various quantities appearing in \cref{adiabaticity_anomaly_spr}
can be written as,
\begin{align}\label{ideal.consti}
  T^{\m\n} &= (E + P)u^\m u^\n + P g^{\m\n}
              + R_s\x^\m \x^\n + \l (u^\m\x^\n+u^\n\x^\m) + \cO(\dow), \nn\\
  J^\m &= Q u^\m + Q_s \x^\mu + \cO(\dow), \nn\\
  K &= -\a \d_\scrB\vf + K_{ideal} + \cO(\dow), \nn\\
  N^\mu &= N u^\mu + N_s \xi^\mu + \cO(\dow), \nn\\
  \D &= (\a \d_\scrB\vf)^2 + \D_{ideal} + \cO(\dow^2),
\end{align}
where $E$, $P$, $R_s$, $\l$, $Q$, $Q_s$, $K_{ideal}$, $N$, $N_s$ are functions of $T$, $\mu$ and
$\mu_s \equiv -\frac{1}{2}\xi^\mu \xi_\mu$. We have omitted the only other possible scalar
$\d_\scrB\vf$ in the functional dependence, because using the $\vf$ equation of motion we know that
it is no longer an independent quantity.
Plugging \cref{ideal.consti} in \cref{adiabaticity_anomaly_spr} we can find,
\begin{multline}\label{freeE.rel.ideal.expand}
  (Q_s + R_s) \x^\mu \lb \N_\mu \nu + \frac{1}{T} u^\nu F_{\nu\mu} \rb
  + \l \x^\m \lb \frac{1}{T^2}\N_\mu T + u^\nu \N_\nu \bfrac{u_\mu}{T} \rb \\
  + \N_\mu \lb \lb \frac{P}{T} - N \rb u^\mu \rb
  + \frac{1}{T} u^\mu \lb \N_\mu E - T \N_\mu S - \mu \N_\mu Q + R_s \N_\mu \mu_s \rb \\
  + \N_\mu\lb \d_\scrB\vf R_s\x^\m - N_s\xi^\mu\rb
  + \lb K_{ideal} - \N_\mu (R_s\x^\m ) \rb \d_\scrB\vf
  + \D_{ideal} = 0,
\end{multline}
where we have defined $S$ through the ``Euler equation'',
\begin{equation}
  E + P = ST + Q\mu.
\end{equation}
\Cref{freeE.rel.ideal.expand} will imply a set of relations among various coefficients,
\begin{equation}
  Q_s = -R_s, \quad
  \l = 0, \quad
  N = \frac{P}{T}, \quad
  N_s = \d_\scrB\vf R_s, \quad
  K_{ideal} =  \N_\mu (R_s\x^\m ), \quad
  \D_{ideal} = 0,
\end{equation}
and the ``first law of thermodynamics'',
\begin{equation}
  \df E = T\df S + \mu \df Q - R_s \df \mu_s,
\end{equation}
giving physical meaning to the quantities we have introduced in \cref{ideal.consti}. Finally, we
have the full set of superfluid constitutive relations up to ideal order satisfying the second law,
\begin{align}\label{ideal.consti_final}
  T^{\m\n} &= (E + P)u^\m u^\n + P g^{\m\n} + R_s\x^\m \x^\n + \cO(\dow), \nn\\
  J^\m &= Q u^\m - R_s \x^\mu + \cO(\dow), \nn\\
  K &= -\a \d_\scrB\vf + \N_\mu (R_s\x^\m ) + \cO(\dow), \nn\\
  N^\mu &= \frac{P}{T} u^\mu + \d_\scrB\vf R_s \xi^\mu + \cO(\dow), \nn\\
  J_S^\mu &= N^\mu - \frac{1}{T} \lb T^{\mu\nu} u_\nu + \mu J^\mu \rb = S u^\mu + \cO(\dow), \nn\\
  \D &= \cO(\dow^2).
\end{align}
These are the well known ideal superfluid constitutive relations. Note that we have included first
order terms in $K$, $N^\mu$ which can be ignored when talking about the ideal order, but are
required for internal consistency with \cref{adiabaticity_anomaly_spr}. The $\vf$ equation of motion
$K=0$ will imply,
\begin{equation}
  \a \d_\scrB\vf = \N_\mu (R_s\xi^\mu) + \cO(\dow) \quad\implies\quad
  u^\mu \xi_\mu = \mu + \frac{T}{\a} \N_\mu (R_s\x^\m ) + \cO(\dow),
\end{equation}
which is a first order correction to the Josephson equation. Note however that this equation can
admit further one derivative corrections due to the first order constitutive relations
discussed in the next subsection; the correction mentioned here is only how the ideal superfluid
transport affects the Josephson equation. The conservation laws on the other hand are complete up to
the first order in derivatives,
\begin{align}\label{EOM1st}
  \frac{1}{\sqrt{-g}} \d_\scrB \lb \sqrt{-g} (E+P)T^2 \b_\mu \rb + QT\d_\scrB A_\mu
    &= - \xi_\nu\a\d_\scrB\vf + \cO(\dow^2), \nn\\
  \frac{1}{\sqrt{-g}} \d_\scrB \lb \sqrt{-g} QT \rb
  &= \a\d_\scrB \vf  + \cO(\dow^2).
\end{align}
These equations provide a set of relations between $\d_\scrB\vf$, $\d_\scrB g_{\mu\nu}$ and
$\d_\scrB A_\mu$, which can be used to eliminate a vector $u^\mu \d_{\scrB} g_{\mu\nu}$ and a scalar
$u^\mu \d_\scrB A_\mu$ (see \cref{relativistic-data}) from the first order constitutive
relations. On the other hand, we choose to eliminate the scalar data $\N_\mu (R_s\xi^\mu)$ using the
$\vf$ equation of motion.

\subsection{First Derivative Corrections to Relativistic Superfluids}

\begin{table}[p]
  \centering
  \begin{tabular}[t]{|c|c|c|}
    \hline
    \multicolumn{3}{|c|}{Vanishing at Equilibrium -- Onshell Independent} \\
    \hline
    $S_1$ & $\frac{T}{2} \tilde P^{\mu\nu} \d_{\scrB} g_{\mu\nu}$ & $\tilde P^{\mu\nu}\N_\mu
                                                                    u_\nu$ \\
    $S_2$ & $\frac{T}{2}\z^\mu \z^\nu \d_\scrB g_{\mu\nu}$ & $\z^\mu \z^\nu \N_\mu u_\nu$ \\
    $S_3$ & $T \z^\mu \d_{\scrB} A_\mu$ & $\z^\mu \lb T \N_\mu \nu + u^\nu F_{\nu\mu}\rb$ \\
    $S_4$ & $T\d_\scrB\vf$ & $u^\mu \xi_\mu - \mu$ \\
    \hline
    $V^\mu_{1}$ & $T \tilde P^{\mu\nu} \z^\r \d_{\scrB} g_{\nu\r}$ & $2 \tilde P^{\mu\nu} \z^\r
                                                                     \N_{(\nu} u_{\r)}$ \\
    $V^\mu_{2}$ & $T \tilde P^{\mu\nu} \d_{\scrB} A_\mu$ & $\tilde P^{\mu\nu}\lb T\N_\nu \nu +
                                                           u^\s F_{\s\nu} \rb$ \\
    \hline
    $\s^{\mu\nu}$ & $\frac{T}{2} \tilde P^{\r\langle \mu}\tilde P^{\nu\rangle\s}\d_\scrB g_{\r\s}$
  & $\tilde P^{\mu\r}\tilde P^{\nu\s}\lb \N_{(\r} u_{\s)} - \frac{1}{d-2} \tilde P_{\r\s} S_1
    \rb$ \\
    \hline
    $\tilde V^\mu_{1}$ & \multicolumn{2}{c|}{$\e^{\mu\nu\r\s} u_\nu \z_\r V_{1,\s}$} \\
    $\tilde V^\mu_{2}$ & \multicolumn{2}{c|}{$\e^{\mu\nu\r\s} u_\nu \z_\r V_{2,\s}$} \\
    \hline
    $\tilde\s^{\mu\nu}$ & \multicolumn{2}{c|}{$\e^{(\mu|\r\s\t} u_\r \z_\s \s_{\t}^{\ \nu)}$} \\
    \hline\hline
    \multicolumn{3}{|c|}{Vanishing at Equilibrium -- Onshell Dependent} \\
    \hline
    $S_5$ & $\frac{T}{2}u^\mu u^\nu \d_{\scrB} g_{\mu\nu}$ & $\frac{1}{T} u^\mu \N_\mu T$ \\
    $S_6$ & $Tu^\mu \d_{\scrB} A_\mu$ & $T u^\mu \N_\mu \nu$ \\
    $S_7$ & $T \z^{\mu} u^{\nu} \d_{\scrB} g_{\mu\nu}$ & $\z^\nu \lb \frac{1}{T} \N_\nu T + u^\s
            \N_\s u_\nu \rb$ \\
    \hline
    $V_3^\mu$ & $T \tilde P^{\mu\nu} u^\r \d_{\scrB} g_{\nu\r}$
                                                                  & $\tilde P^{\mu\nu}\lb \frac{1}{T} \N_\nu T + u^\s \N_\s u_\nu \rb$ \\
    \hline
    $\tilde V^\mu_{3}$ & \multicolumn{2}{c|}{$\e^{\mu\nu\r\s} u_\nu \z_\r V_{3,\s}$} \\
    \hline\hline
    \multicolumn{3}{|c|}{Surviving at Equilibrium} \\
    \hline
    $S_{e,1}$ & \multicolumn{2}{c|}{$\frac{1}{T}\z^\mu \dow_\mu T$} \\
    $S_{e,2}$ & \multicolumn{2}{c|}{$T\z^\mu \dow_\mu \nu$} \\
    $S_{e,3}$ & \multicolumn{2}{c|}{$\z^\mu \dow_\mu \hat\mu_s$} \\
    $S_{e,4}$ & \multicolumn{2}{c|}{$\N_\mu \z^\mu$} \\
    $\vdots$ & \multicolumn{2}{c|}{$\vdots$} \\
    \hline
    $V^\mu_{e,1}$ & \multicolumn{2}{c|}{$\frac{1}{T}\tilde P^{\mu\nu} \dow_\nu T$} \\
    $V^\mu_{e,2}$ & \multicolumn{2}{c|}{$T \tilde P^{\mu\nu} \dow_\nu \nu$} \\
    $\vdots$ & \multicolumn{2}{c|}{$\vdots$} \\
    \hline
    $\tilde S_{e,1}$ & \multicolumn{2}{c|}{$T \e^{\mu\nu\r\s}\z_\mu u_\nu \dow_\r u_\s$} \\
    $\tilde S_{e,2}$ & \multicolumn{2}{c|}{$\half T \e^{\mu\nu\r\s}\z_\mu u_\nu F_{\r\s}$} \\
    $\vdots$ & \multicolumn{2}{c|}{$\vdots$} \\
    \hline
    $\tilde V^\mu_{e,1}$ & \multicolumn{2}{c|}{$T \tilde P^{\mu}_{\ \t} \e^{\t\nu\r\s} u_\nu
                           \dow_\r u_\s$} \\
    $\tilde V^\mu_{e,2}$ & \multicolumn{2}{c|}{$\half T \tilde P^{\mu}_{\ \t} \e^{\t\nu\r\s} u_\nu
                           F_{\r\s}$} \\
    $\tilde V^\mu_{e,3}$ & \multicolumn{2}{c|}{$T \tilde P^{\mu}_{\ \t} \e^{\t\nu\r\s} \xi_\nu
                           \dow_\r u_\s$} \\
    $\tilde V^\mu_{e,4}$ & \multicolumn{2}{c|}{$\half T \tilde P^{\mu}_{\ \t} \e^{\t\nu\r\s} \xi_\nu
                           F_{\r\s}$} \\
    $\vdots$ & \multicolumn{2}{c|}{$\vdots$} \\
    \hline
  \end{tabular}
  \caption{\label{relativistic-data} Independent first order data for relativistic superfluids. We
    have not enlisted, neither would we need, all the independent data surviving at equilibrium.}
\end{table}

Moving on to the one derivative superfluids, let us schematically represent various quantities
appearing in \cref{adiabaticity_anomaly_spr} up to the first order in derivatives as,
\begin{align}\label{rel.consti.1d}
  T^{\m\n} &= \Big[ (E + P)u^\m u^\n + P g^{\m\n} + R_s\x^\m \x^\n \big] + \cT^{\mu\nu} + \cO(\dow^2), \nn\\
  J^\m &= \big[ Q u^\m - R_s \x^\mu \big] + \cJ^\mu + \cO(\dow^2), \nn\\
  K &= \big[ -\a \d_\scrB\vf + \N_\mu (R_s\x^\m ) \big] + \cK + \cO(\dow^2), \nn\\
  N^\mu &= \lB\frac{P}{T} u^\mu + \d_\scrB\vf R_s \xi^\mu\rB + \cN^\mu + \cO(\dow^2), \nn\\
  \D &= \a (\d_\scrB \vf)^2 + \cD,
\end{align}
where the corrections $\cT^{\mu\nu}$, $\cJ^\mu$, $\cK$, $\cN^\mu$, $\cD$ have exactly one derivative in
every term. Plugging these in the \cref{adiabaticity_anomaly_spr} we can get an equation among the
corrections,
\begin{equation}
  \N_\mu \cN^\mu - \rmN_\rmH^\perp
  = \half \cT^{\mu\nu} \d_\scrB g_{\mu\nu} + \cJ^\mu\d_\scrB A_\mu + \cK
  \d_\scrB\vf + \cD + \cO(\dow^3). \label{adiabaticity_first}
\end{equation}
We will now attempt to find all the solutions to this equation, hence recovering the superfluid
constitutive relations up to the first order in derivatives.

\subsubsection{Parity Even} 

We can find the most general parity even solution of \cref{adiabaticity_first} in 2 steps (note that
$\rmN^\perp_\rmH$ is parity odd): (1) first we write down the most general allowed parity-even
$\cN^\mu$ and find a set of constitutive relations pertaining to that, and (2) then find the most
general parity-even constitutive relations which satisfy \cref{adiabaticity_first} with
$\cN^\mu = 0$.

\begin{enumerate}
\item One can check that the most general form of $\cN^\mu$ (whose divergence only contains product
  of derivatives and has at least one $\d_\scrB$ per term) can be written as,
  \begin{equation}\label{rel.even.N} 
    \cN^\mu
    = 2f_1 u^{[\mu} \xi^{\nu]} \frac{1}{T^2}\dow_\nu T
    + 2f_2 u^{[\mu} \xi^{\nu]} \dow_\nu \bfrac{\mu}{T}
    + 2f_3 u^{[\mu} \xi^{\nu]} \dow_\nu R_s
    + \N_\nu \lb f_4 u^{[\mu} \xi^{\nu]} \rb,
  \end{equation}
  where $f$'s are functions of $T$, $\nu=\mu/T$ and $\hat\mu_s = -\half \z^\mu \z_\mu$ with
  $\z^\mu = P^{\mu\nu}\xi_\nu$ ($P^{\mu\nu} = g^{\mu\nu} + u^\mu u^\nu$ is the projection operator
  away from the fluid velocity). Note that,
  \begin{equation}
    \hat\mu_s = -\half \z^\mu \z_\mu
    = - \half \xi^\mu \xi_\mu - \half (\xi^\mu u_\mu)^2
    = \mu_s - \half (\mu + T\d_\scrB \vf)^2.
  \end{equation}
  Out of the four terms in \cref{rel.even.N}, the last one has trivially zero divergence and hence
  can be ignored. The third term on the other hand can be removed by elimination of
  $\N_\mu(R_s\xi^\mu)$ using the $\vf$ equation of motion. Computing the divergence of the remaining
  terms in $\cN^\mu$ and comparing them to \cref{adiabaticity_first}, we can directly read out the
  corresponding superfluid constitutive relations (the symbol `$\ni$' represents that they are not
  yet the complete solutions of \cref{adiabaticity_first}; we still have to add the terms with
  $\cN^\mu = 0$),
  \begin{align}\label{rel.even.N.consti}
    \cT^{\mu\nu}
    &\ni u^\mu u^{\nu} \lb
      \sum_{i=1}^2 \a_{E,i} S_{e,i}
      - \frac{1}{T}\N_\s (Tf_1 \z^{\s}) \rb
      + \lb\z^\mu \z^\nu - 2 (u^\r\xi_\r) u^{(\mu} \z^{\nu)}\rb \sum_{i=1}^2 \a_{R_s,i} S_{e,i} \nn\\
    &\qquad + \tilde P^{\mu\nu} \sum_{i=1}^2 f_i S_{e,i}
      - 2 \xi^{(\mu} \sum_{i=1}^2 f_i V_{e,1}^{\nu)}
      + 2 u^{(\mu} \z^{\nu)} \sum_{i=1}^2 f_i S_{4+i},
      \nn\\[0.1cm]
    \cJ^\mu
    &\ni u^{\mu} \lb \sum_{i=1}^2 \a_{Q,i} S_{e,i} - \frac{1}{T} \N_\nu (Tf_2 \z^{\nu}) \rb
      - \z^\mu \sum_{i=1}^2 \a_{R_s,i} S_{e,i}
      + \sum_{i=1}^2 f_i V^\mu_{e,i},
      \nn\\[0.1cm]
    \cK
    &\ni \N_\mu \lb \z^\mu \sum_{i=1}^2 \a_{R_s,i} S_{e,i}
    - \sum_{i=1}^2 f_i V^\mu_{e,i} \rb,
  \end{align}
  where $\tilde P^{\mu\nu} = g^{\mu\nu} + u^\mu u^\nu - \frac{1}{\z^\s \z_\s} \z^\mu \z^\nu$, and we
  have defined,
  \begin{equation}
    \df f_i = \frac{\a_{E,i}}{T} \df T + T \a_{Q,i} \df \nu + \lb \a_{R_s,i} - \frac{f_i}{2\hat\mu_s}
    \rb \df \hat\mu_s.
  \end{equation}
  The actual computation is not neat and we have presented the details in \cref{calc.details} for
  the aid of the readers interested in reproducing our results. Note that these constitutive
  relations are presented in terms of `data' which are natural for this sector; readers can modify
  these to their favorite basis and get results which might look considerably messier. Moreover,
  these results are written in a particular `hydrodynamic frame' chosen by aligning $u^\mu$, $T$,
  $\mu$ along $\b^\mu$, $\L_\b$, which again can be modified according to reader's preference.

\item Let us now look at the parity-even solutions to \cref{adiabaticity_first} with $\cN^\mu = 0$,
  \begin{equation}\label{LHS0_rel}
    0= \half \cT^{\mu\nu} \d_\scrB g_{\mu\nu} + \cJ^\mu\d_\scrB A_\mu + \cK \d_\scrB\vf + \cD.
  \end{equation}
  Every term in $\cT^{\mu\nu}$, $\cJ^\mu$, $\cK$ must either cancel or contribute to $\D$ which has
  to be a quadratic form. It follows that the terms in $\cT^{\mu\nu}$, $\cJ^\mu$, $\cK$ must be
  proportional to $\d_\scrB g_{\mu\nu}$, $\d_\scrB A_\mu$, $\d_\scrB \vf$. Recall however that we
  have chosen to eliminate $u^\mu \d_\scrB g_{\mu\nu}$, $u^\mu\d_\scrB A_\mu$ using the equations of
  motion. For $\D$ to be a quadratic form, it therefore implies that $\cT^{\mu\nu}$, $\cJ^\mu$
  cannot have a term like $\#^{(\mu} u^{\nu)}$, $\# u^\mu$ respectively for some vector $\#^\mu$ and
  scalar $\#$. With this input let us write down the most generic allowed form of the currents in
  terms of 20 new transport coefficients $[\b_{ij}]_{4\times 4}$ (with $\b_{44} = \a/T$),
  $[\k_{ij}]_{2\times 2}$ and $\eta$,
    \begin{align}
    \cT^{\mu\nu}
    &\ni - T\bigg[
      \lbr \b_{11} \tilde P^{\r\s} + \b_{12} \z^{\r} \z^{\s} \rbr \tilde P^{\mu\nu} 
      + \lbr \b_{21} \tilde P^{\mu\nu} + \b_{22} \z^{\r} \z^{\s} \rbr \z^{\mu} \z^{\nu}
      + 4\k_{11} \z^{(\mu} \tilde P^{\nu)(\r}\z^{\s)}  \nn\\
    &\qquad
      + \eta \tilde P^{\mu\langle\r} \tilde P^{\s\rangle\nu}
      \bigg] \half \d_\scrB g_{\r\s} 
      - T\bigg[
      \b_{13} \z^\r \tilde P^{\mu\nu}
      + \b_{23} \z^\r \z^{\mu}\z^{\nu}
      + 2\k_{12} \z^{(\mu} \tilde P^{\nu)\r}
      \bigg] \d_\scrB A_\r \nn\\
    &\qquad - T \bigg[ \b_{14} \tilde P^{\mu\nu} + \b_{24} \z^\mu\z^\nu \bigg] \d_\scrB\vf, \nn\\[0.1cm]
    &= - \tilde P^{\mu\nu} \sum_{i=1}^4\b_{1i} S_i
      - \z^{\mu} \z^{\nu} \sum_{i=1}^4\b_{2i} S_i
      - 2\z^{(\mu} \sum_{i=1}^2 \k_{1i} V_i^{\nu)}
      - \eta \s^{\mu\nu}.
  \end{align}
  \begin{align}
    \cJ^\mu
    &\ni -T\bigg[
      \lbr \b_{31} \tilde P^{\r\s} + \b_{32} \z^\r\z^\s \rbr \z^\mu
      + 2\k_{21} \tilde P^{\mu(\r} \z^{\s)} \bigg] \half \d_\scrB g_{\r\s} \nn\\
    &\quad - T\bigg[\b_{33} \z^\r \z^\mu + \k_{22} \tilde P^{\mu\r} \bigg] \d_\scrB A_\r
      -T\bigg[ \b_{34} \z^\mu \bigg] \d_\scrB\vf, \nn\\[0.1cm]
    &= 
      - \z^{\mu} \sum_{i=1}^4\b_{3i} S_i
      - \sum_{i=1}^2 \k_{2i} V_i^{\mu}.
  \end{align}
  \begin{equation}
    \cK \ni - T\bigg[\b_{41} \tilde P^{\r\s} + \b_{42} \z^{\r} \z^{\s} \bigg] \d_\scrB g_{\r\s}
    - T \bigg[\b_{43} \z^\r \bigg] \d_\scrB A_\r
    = -\sum_{i=1}^3 \b_{4i} S_i.
  \end{equation}
  Note that we did not include a term proportional to $\d_\scrB\vf$ in $\cK$, because such a term is
  already present in $K = -\a\d_\scrB\vf + \N_\mu(R_s\xi^\mu) + \cK + \cO(\dow^2)$. Defining
  $\b_{44}=\a/T$, we can read out the parity-even quadratic form
  $\D|_{even} = \a(\d_\scrB\vf)^2 + \cD|_{even}$,
  \begin{align}
    T\D|_{even} &=\sum_{i,j=1}^4 S_i\b_{ij}S_j + \sum_{i,j=1}^2 V_i^\mu\k_{ij} V_{i,\mu} + \eta
             \s^{\mu\nu}\s_{\mu\nu}, \nn\\
           &= \sum_{i,j=1}^4 S_i\b_{(ij)} S_j + \sum_{i,j=1}^2 V_i^\mu\k_{(ij)} V_{i,\mu} + \eta
             \s^{\mu\nu}\s_{\mu\nu}.
  \end{align}
  In the second step we have realized that only the symmetric parts of the matrices $\b_{ij}$ and
  $\k_{ij}$ will survive in this expression, and will contribute towards dissipation. Thus only 14
  out of 21 transport coefficients (including $\a$) are dissipative; the remaining 7 are
  non-dissipative.
\end{enumerate}

\subsubsection{Parity Odd (4 Dimensions)}

We can find the most general parity-odd solution of \cref{adiabaticity_first} in 3 steps: (1) first
we consider a particular set of solutions which takes care of the anomaly $\rmN^\perp_\rmH$ and
proceed towards the non-anomalous constitutive relations, (2) then we write down the most general
allowed parity-odd $\cN^\mu$ and find a set of constitutive relations pertaining to that, and (2)
finally find the most general parity-odd constitutive relations with zero $\cN^\mu$.

\begin{enumerate}
\item In $4$ dimensions at the first order in the derivatives $\rmT^{\mu\perp}_\rmH = 0$ and
  $\rmJ_\rmH^\perp = - \frac{3}{4}C^{(4)} \e^{\mu\nu\r\s} F_{\mu\nu}F_{\r\s}$, which implies,
  \begin{equation}
    \rmN_\rmH^\perp = -\frac{3}{4}\nu C^{(4)} \e^{\mu\nu\r\s} F_{\mu\nu}F_{\r\s}.
  \end{equation}
  A particular solution pertaining to \cref{adiabaticity_first} with this $\rmN_\rmH^\perp$ is given
  as (see e.g. \cite{Haehl:2015pja}),
  \begin{align}
    \cT^{\mu\nu} &\ni 2\mu^2 C^{(4)} u^{(\mu} \lb 3 B^{\nu)} + 2\mu \o^{\nu)} \rb, \nn\\
    \cJ^{\mu} &\ni \mu C^{(4)} \lb 6 B^\mu + 3 \mu \o^\mu \rb, \nn\\
    \cK &\ni 0, \nn\\
    \cN^{\mu} &\ni \frac{\mu^2}{T} C^{(4)} \lb 3 B^\mu + \mu \o^\mu \rb.
  \end{align}
  Here we have defined the magnetic field and fluid vorticity as,
  \begin{equation}
    B^\mu = \half \e^{\mu\nu\r\s} u_\nu F_{\r\s}, \qquad
    \o^\mu = \e^{\mu\nu\r\s} u_\nu \dow_{\r} u_{\s}.
  \end{equation}
  
\item One can check that the most general form of $\cN^\mu$ (whose divergence only contains product
  of derivatives and has at least one $\d_\scrB$ per term) can be written as,
  \begin{equation}\label{rel.odd.N}
    \cN^\mu
    =
    g_1 \lb \b^{\mu} \tilde S_{e,1} + \tilde V^\mu_{3} \rb
    + g_2 \lb \b^{\mu} \tilde S_{e,2} + \tilde V^\mu_{2} \rb
    + C_1 T^2 \o^\mu,
  \end{equation}  
  where $g$'s are functions of $T$, $\nu$, $\hat\mu_s$, and $C_1$ is a constant. From here we can
  directly read out the corresponding constitutive relations,
  \begin{align}\label{rel.odd.N.consti}
    \cT^{\mu\nu}
    &\ni u^\mu u^{\nu} \sum_{i=1}^2 \tilde\a_{E,i} \tilde S_{e,i}
    + \lb \z^\mu \z^\nu - 2 (u^\r\xi_\r) u^{(\mu} \z^{\nu)} \rb
      \sum_{i=1}^2 \tilde\a_{R_s,i} \tilde S_{e,i}
      - \z^\mu \z^\nu \sum_{i=1}^2 \frac{1}{2\hat\mu_s} g_{i} \tilde S_{e,i} \nn\\
    &\qquad
      - 2u^{(\mu} \sum_{i=1}^2 g_{i} \tilde V^{\nu)}_{e,2+i}
      - u^{(\mu} \lb  2P^{\nu)}_{\ \a} - u^{\nu)} u_\a \rb  \e^{\a\r\s\t} \N_\s \lb Tg_1
      u_\t\z_\r\rb 
      + 4 C_1 T^3 \o^{(\mu} u^{\nu)} \nn\\
    \cJ^\mu
    &\ni u^{\mu} \sum_{i=1}^2 \tilde\a_{Q,i} \tilde S_{e,i} 
      - \z^\mu \sum_{i=1}^2 \tilde\a_{R_s,i} \tilde S_{e,i} + \sum_{i=1}^2 g_{i} \tilde V^\mu_{e,i}
      + \e^{\mu\nu\r\s} \N_\nu (Tg_2\z_\r u_\s), \nn\\
    \cK
    &\ni \N_\mu \lb
      \z^\mu \sum_{i=1}^2 \tilde\a_{R_s,i} \tilde S_{e,i}
    - \sum_{i=1}^2 g_{i} \tilde V^\mu_{e,i} \rb,
  \end{align}
  where we have defined,
  \begin{equation}
    \df g_i = \frac{\tilde\a_{E,i}}{T} \df T + T \tilde\a_{Q,i} \df \nu + \lb \tilde\a_{R_s,i} -
    \frac{f_i}{2\hat\mu_s} \rb \df \hat\mu_s.
  \end{equation}
  The actual computation is not neat and we have presented the details in \cref{calc.details} for
  interested readers.

\item We should finally consider the parity-odd constitutive relations that satisfy
  \cref{adiabaticity_first} with zero LHS. Following our discussion in the parity-even sector, the
  allowed form of the constitutive relations can be written down in terms of 5 coefficients
  $[\tilde\k_{ij}]_{2\times 2}$ and $\tilde \eta$,
  \begin{align}
    \cT^{\mu\nu}
    &\ni
      - T u_\t \z_\k \bigg[ 4\tilde\k_{11} \z^{(\mu} \e^{\nu)\t\k(\r} \z^{\s)}
      + \tilde\eta \tilde P^{\l(\mu} \e^{\nu)\t\k(\r} \tilde P^{\s)}_{\ \ \l}
      \bigg] \half \d_\scrB g_{\r\s}
      - T u_\t \z_\k \bigg[ 2\tilde\k_{12} \z^{(\mu} \e^{\nu)\t\k\r} \bigg] \d_\scrB A_\r, \nn\\
    &= - 2\z^{(\mu} \sum_{i=1}^2 \tilde\k_{1i} \tilde V^{\nu)}_{i}
      - \tilde\eta \tilde\s^{\mu\nu}, \nn\\
    \cJ^\mu
    &\ni
      - T u_\t \z_\k \bigg[ 2 \tilde\k_{21} \e^{\mu\t\k(\r} \z^{\s)} \bigg] \half \d_\scrB g_{\r\s}
      - T u_\t \z_\k \bigg[ \tilde\k_{22} \e^{\mu\t\k\r} \bigg]  \d_\scrB A_\r, \nn\\
    &\ni - \sum_{i=1}^2 \tilde\k_{2i} \tilde V_{i}^{\mu},
 \nn\\
    \cK &\ni 0.
  \end{align}
  One can check that these constitutive relations trivially satisfy \cref{adiabaticity_first} with
  zero LHS and the quadratic form $\D\vert_{odd} = \cD\vert_{odd}$ is given as,
  \begin{align}
    T\D\vert_{odd}
    &= \e^{\mu\nu\t\k} u_\t \z_\k \lB \sum_{i=1}^2 V_{i,\mu} \tilde\k_{ij} V_{j,\nu} + \tilde\eta
      \s_{\mu\r} \s_{\nu}^{\ \r} \rB \nn\\
    &= \e^{\mu\nu\t\k} u_\t \z_\k \sum_{i=1}^2 V_{i,\mu} \tilde\k_{[ij]} V_{j,\nu}
    = 2\e^{\mu\nu\t\k} u_\t \z_\k V_{1,\mu} \tilde\k_{[12]} V_{2,\nu}.
  \end{align}
  It follows that out of the 5 transport coefficients, only 1 contribute to dissipation and the
  other 4 are non-dissipative.
\end{enumerate}

\subsubsection{Positivity Constraints} \label{pos.constr.rel}

The dissipative
transport coefficients are required to satisfy a set of inequalities to satisfy $\D =
\a(\d_{\scrB}\vf)^2 + \cD|_{even} + \cD|_{odd} \geq 0$,
\begin{equation}
  T\D
  = \sum_{i,j=1}^4 S_i\b_{(ij)} S_j
    + \lb \sum_{i,j=1}^2 V_i^\mu\k_{(ij)} V_{i,\mu} 
    + \sum_{i=1}^2 V_{i}^\mu \tilde\k_{[ij]} \tilde V_{j,\mu} \rb
    + \eta \s^{\mu\nu}\s_{\mu\nu}.
\end{equation}
We want this expression to be a quadratic form, which it nearly is except the parity-odd term in the
brackets. However this term can be made into a quadratic form by noticing that the square of a
parity odd term is parity-even, due to the identity,
\begin{equation}\label{epsilon.contraction_rel}
  (\e^{\mu\nu\r\s} u_\r \z_\s)(\e_{\t\nu\a\b} u^\a \z^\b) = \tilde P^\mu_{\ \t} \z^\nu \z_\nu = -2 \hat\mu_s
  \tilde P^\mu_{\ \t}.
\end{equation}
We define,
\begin{equation}\label{kappa'definition_rel}
  \begin{pmatrix}
    V'^\mu_1 \\ V'^\mu_2 
  \end{pmatrix}
  = \begin{pmatrix}
    V^\mu_1 \\ V^\mu_2 
  \end{pmatrix}
  + \begin{pmatrix}
    0 & a_{12} \\
    0 & 0
  \end{pmatrix}
  \begin{pmatrix}
    \tilde V^\mu_1 \\ \tilde V^\mu_2 
  \end{pmatrix}, \qquad
  \k'_{ij} = \k_{ij} + k_{ij}, \qquad k_{[ij]} = 0,
\end{equation}
such that,
\begin{equation}
  \sum_{i,j=1}^2 V'^\mu_i\k'_{(ij)} V'_{i,\mu} 
  = \sum_{i,j=1}^2 V_i^\mu\k_{(ij)} V_{i,\mu} 
  + \sum_{i=1}^2 V_{i}^\mu \tilde\k_{[ij]} \tilde V_{j,\mu}.
\end{equation}
Using the identity \cref{epsilon.contraction_rel}, the above equation can be easily solved to give,
\begin{equation}
  a_{12} = \frac{\tilde\k^{(a)}_{12}}{\k_{11}}, \qquad
  k_{1i} = k_{i1} = 0, \qquad
  k_{22} = 2 \hat\mu_s \frac{\tilde\k_{[12]}}{\k_{11}}.
\end{equation}
Consequently $\D$ will take the form,
\begin{equation}
  T\D
  = \sum_{i,j=1}^4 S_i\b_{(ij)} S_j
    + \sum_{i,j=1}^2 V'^\mu_i\k'_{(ij)} V'_{i,\mu} 
    + \eta \s^{\mu\nu}\s_{\mu\nu}.
\end{equation}
Given $T\geq0$, the condition $\D\geq 0$ implies that $\eta\geq 0$ and the matrices
$[\b_{(ij)}]_{4\times4}$, $[\k'_{(ij)}]_{2\times2}$ have all non-negative eigenvalues. This gives
$7$ inequalities among 15 dissipative transport coefficients, and 8 are completely arbitrary.

\subsection{Summary}

We have completed the analysis of a superfluid up to the first order in derivatives. Here we
summarize the results. We found that the entire superfluid transport up to the first order in
derivatives is characterized by an ideal order pressure $P$, 30 first order transport coefficients
which are functions of $T$, $\mu/T$, $\hat\mu_s$, and two constants $C_1$, $C^{(4)}$. $P$, $C_1$ and
$C^{(4)}$ along with 4 transport coefficients,
\begin{equation}
  \text{Parity Even (2):}\qquad f_1,\quad f_2, \qquad\qquad
  \text{Parity Odd (2):}\qquad g_1,\quad g_2,
\end{equation}
totally determine the \emph{hydrostatic} transport (part of the constitutive relations that survive
at equilibrium). \emph{Non-hydrostatic non-dissipative} transport (part that does not survive at
equilibrium but doesn't contribute to $\D\geq 0$ either) is given by 11 transport coefficients,
\begin{equation}\nn
  \text{Parity Even (7):}\qquad [\b_{[ij]}]_{4\times 4} \ \ \text{(antisymmetric)}, \quad [\k_{[ij]}]_{2\times2} \ \
  \text{(antisymmetric)},
\end{equation}
\begin{equation}
  \text{Parity Odd (4):}\qquad [\tilde\k_{(ij)}]_{2\times 2} \ \ \text{(symmetric)}, \quad \tilde\eta.
\end{equation}
Finally the entire \emph{dissipative} transport is given by $15$ transport coefficients,
\begin{equation}\nn
  \text{Parity Even (14):}\qquad [\b_{(ij)}]_{4\times 4} \ \ \text{(symmetric)}, \quad
  [\k_{(ij)}]_{2\times2} \ \  \text{(symmetric)}, \quad \eta,
\end{equation}
\begin{equation}
  \text{Parity Odd (1):}\qquad [\tilde\k_{[ij]}]_{2\times 2} \ \ \text{(antisymmetric)}.
\end{equation}
These dissipative transport coefficients follow a set of inequalities ($\k'$ is defined in \cref{kappa'definition_rel}),
\begin{equation}
  [\b_{(ij)}]_{4\times 4}, \quad [\k'_{(ij)}]_{2\times2}, \quad \eta \quad \geq\quad 0,
\end{equation}
where a `non-negative matrix' implies all its eigenvalues are non-negative. Using $P$, $f_i$, $g_i$
we define some new functions,
\begin{equation}\nn
  \df P = S \df T + Q\df \mu + R_s \df \mu_s, \qquad
  E + P = ST + Q\mu,
\end{equation}
\begin{equation}\nn
  \df f_i = \frac{\a_{E,i}}{T} \df T + T \a_{Q,i} \df \nu + \lb \a_{R_s,i} - \frac{f_i}{2\hat\mu_s}
  \rb \df \hat\mu_s, \qquad
  \a_{E,i} + f_i = \a_{S,i}T + \a_{Q,i}\mu,
\end{equation}
\begin{equation}
  \df g_i = \frac{\tilde\a_{E,i}}{T} \df T + T \tilde\a_{Q,i} \df \nu + \lb \tilde\a_{R_s,i} -
  \frac{f_i}{2\hat\mu_s} \rb \df \hat\mu_s, \qquad
\tilde\a_{E,i} + g_i = \tilde\a_{S,i}T + \tilde\a_{Q,i}\mu.
\end{equation}
In terms of these transport coefficients, corrections to the Josephson equation ($K=0$) coming from
the first order superfluid transport are given as  (here $\b_{44} = \a/T$),
\begin{multline}
  u^\mu \xi_\mu - \mu
  = \frac{1}{\b_{44}} \N_\mu \lb R_s\x^\m \rb
  - \sum_{i=1}^3 \frac{\b_{4i}}{\b_{44}} S_i \\
  + \frac{1}{\b_{44}} \N_\mu \bigg(
  \z^\mu \sum_{i=1}^2 \a_{R_s,i} S_{e,i}
  + \z^\mu \sum_{i=1}^2 \tilde\a_{R_s,i} \tilde S_{e,i}
  - \sum_{i=1}^2 f_i V^\mu_{e,i}
  - \sum_{i=1}^2 g_{i} \tilde V^\mu_{e,i} \bigg) + \cO(\dow^2),
\end{multline}
which can be seen as determining $u^\mu \xi_\mu$ in terms of the other superfluid variables. Note
that though this equation contains second order terms, it is only correct up to the first order in
derivatives, and will admit further corrections coming from higher order superfluid transport. The
energy-momentum tensor, charge current and entropy current up to first order in derivatives are however given as,
\begin{align}\label{summary.currents}
  T^{\mu\nu}
  &= (E + P)u^\m u^\n + P g^{\m\n} + R_s\x^\m \x^\n + \cT^{\m\n} + \cO(\dow^2), \nn\\
  J^\mu
  &= Q u^\m - R_s \x^\mu + \cJ^\mu + \cO(\dow^2), \nn\\
  J_S^\mu
  &= S u^\m + \cS^\mu + \cO(\dow^2),
\end{align}
where the higher derivative corrections are,
\begin{align}
  \cT^{\mu\nu}
  &= u^\mu u^{\nu} \Bigg[
    \sum_{i=1}^2 \a_{E,i} S_{e,i}
    + \sum_{i=1}^2 \tilde\a_{E,i} \tilde S_{e,i}
    - \frac{1}{T}\N_\s (Tf_1 \z^{\s})
    + \e^{\a\r\s\t} u_\a \N_\r \lb Tg_1 u_\s\z_\t\rb \Bigg] \nn\\
  &\quad
    + 2 u^{(\mu} \z^{\nu)} \Bigg[
    \sum_{i=1}^2 f_i S_{4+i}
    - (u^\r\xi_\r) \lb \sum_{i=1}^2 \a_{R_s,i} S_{e,i} + \sum_{i=1}^2 \tilde\a_{R_s,i} \tilde S_{e,i} \rb
    + \frac{1}{2\hat\mu_s} \e^{\a\r\s\t} \z_\a \N_\r \lb Tg_1 u_\s\z_\t\rb
    \Bigg] \nn\\
  &\quad + \z^\mu \z^\nu \Bigg[ \sum_{i=1}^2 \a_{R_s,i} S_{e,i}
    + \sum_{i=1}^2 \lb \tilde\a_{R_s,i} - \frac{g_i}{2\hat\mu_s} \rb \tilde S_{e,i}
    - \sum_{i=1}^4 \b_{2i} S_i \Bigg] \nn\\
  &\quad
    + 2u^{(\mu} \Bigg[
    (\xi^\s u_\s) \sum_{i=1}^2 f_i V_{e,i}^{\nu)}
    - \sum_{i=1}^2 g_{i} \tilde V^{\nu)}_{e,2+i}
    - \tilde P^{\nu)}_{\ \a}  \e^{\a\r\s\t} \N_\r \lb Tg_1 u_\s\z_\t\rb
    + 2 C_1 T^3 \o^{\nu)} \nn\\
  &\qquad 
    + C^{(4)}\mu^2 \lb 3 B^{\nu)} + 2\mu \o^{\nu)}\rb
    \Bigg]
    - 2\z^{(\mu} \Bigg[
    \sum_{i=1}^2 f_i V_{e,i}^{\nu)}
    + \sum_{i=1}^2 \k_{1i} V_i^{\nu)}
    + \sum_{i=1}^2 \tilde\k_{1i} \tilde V_{i}^{\nu)} \Bigg] \nn\\
  &\quad
    + \tilde P^{\mu\nu} \Bigg[ \sum_{i=1}^2 f_i S_{e,i} - \sum_{i=1}^4 \b_{1i} S_i \Bigg]
    - \eta \s^{\mu\nu}
    - \tilde\eta \tilde\s^{\mu\nu},
\end{align}
\begin{align}
  \cJ^\mu
  &= u^{\mu} \Bigg[
    \sum_{i=1}^2 \a_{Q,i} S_{e,i}
    + \sum_{i=1}^2 \tilde\a_{Q,i} \tilde S_{e,i}
    - \frac{1}{T} \N_\nu (Tf_2 \z^{\nu})
    + \e^{\a\nu\r\s} u_\a \N_\nu (Tg_2u_\r \z_\s)
    \Bigg] \nn\\
  &\quad - \z^\mu \Bigg[ \sum_{i=1}^2 \a_{R_s,i} S_{e,i}
    + \sum_{i=1}^2 \tilde\a_{R_s,i} \tilde S_{e,i}
    + \sum_{i=1}^4 \b_{3i} S_i
    - \frac{1}{2\hat\mu_s}\e^{\a\nu\r\s} \z_{\a}\N_\nu (Tg_2u_\r \z_\s)
    \Bigg]  \nn\\
  &\quad + \sum_{i=1}^2 f_i V^\mu_{e,i}
    + \sum_{i=1}^2 g_{i} \tilde V^\mu_{e,i}
    - \sum_{i=1}^2 \k_{2i} V_i^{\mu}
    - \sum_{i=1}^2 \tilde\k_{2i} \tilde V^\mu_{i}  \nn\\
  &\quad
    - \tilde P^\mu_{\ \a}\e^{\a\nu\r\s} \N_\nu (Tg_2u_\r \z_\s)
    + 3\mu C^{(4)} \lb 2 B^\mu + \mu \o^\mu \rb,
\end{align}
\begin{align}
  \cS^\mu
  &= g_1 \frac{1}{T}\e^{\mu\nu\r\s} u_\nu \z_\r \dow_\s T
    + g_2 T \e^{\mu\nu\r\s} u_\nu \xi_\r \dow_\s \nu
     + 3 C_1 T^2 \o^\mu  \nn\\
  &\quad + u^\mu \Bigg[
    \sum_{i=1}^2 \a_{S,i} S_{e,i}
    + \sum_{i=1}^2 \tilde\a_{S,i} \tilde S_{e,i}
    - \frac{1}{T^2}\N_\s (Tf_1 \z^{\s})
    + \frac{\mu}{T^2} \N_\nu (Tf_2 \z^{\nu}) \nn\\
  &\qquad
    + \frac{1}{T} \e^{\a\nu\r\s} u_\a \N_\nu \lb Tg_1 u_\r\z_\s\rb
    - \frac{\mu}{T} \e^{\a\r\s\t} u_\a \N_\nu \lb Tg_2 u_\r\z_\s\rb
    \Bigg] \nn\\
  &\quad
    + \frac{1}{T}\z^{\mu} \Bigg[
    \sum_{i=1}^4 \mu\b_{3i} S_i
    + \frac{1}{2\hat\mu_s} \e^{\a\r\s\t} \z_\a \N_\r \lb Tg_1 u_\s\z_\t\rb
    - \frac{\mu}{2\hat\mu_s} \z_{\a}\e^{\a\nu\r\s} \N_\nu (Tg_2u_\r \z_\s)
    \Bigg] \nn\\
  &\quad
    + \frac{\mu}{T} \sum_{i=1}^2 \k_{2i} V_i^{\mu}
    + \frac{\mu}{T} \sum_{i=1}^2 \tilde\k_{2i} \tilde V^\mu_{i}
    - \frac{1}{T} \tilde P^{\mu}_{\ \a}  \e^{\a\nu\r\s} \N_\nu \lb Tg_1 u_\r\z_\s\rb
    + \frac{\mu}{T} \tilde P^\mu_{\ \a}\e^{\a\nu\r\s} \N_\nu (Tg_2u_\r \z_\s).
\end{align}
The scalar $S_{4} = T\d_\scrB\vf = u^\mu\xi_\mu -\mu$ appearing here can be eliminated in favor of
$\N_\mu (R_s \xi^\mu)$ using the Josephson equation. We will like to reiterate that these results
are presented in a particular hydrodynamic frame (gained by aligning $u^\mu$, $T$, $\mu$ along
$\b^\mu$, $\L_\b$) and in a ``natural'' choice of basis for the independent data. They can be
transformed to any other preferred hydrodynamic frame or basis by a straight forward substitution.

In deriving these constitutive relations, we have only used the second law of thermodynamics. To
compare these results with the existing literature \cite{Bhattacharya:2011tra, Bhattacharyya:2012xi,
  Bhattacharya:2011eea}, one might need to further filter these results with requirements like
microscopic reversibility (Onsager relations), time reversal invariance and CPT invariance. For
example, Onsager relations are known to turn off 7 parity-even non-dissipative coefficients
$[\b_{[ij]}]_{4\times 4}$, $[\k_{[ij]}]_{2\times 2}$ and the only parity-odd dissipative coefficient
$[\tilde\k_{[ij]}]_{2\times 2}$ \cite{Bhattacharya:2011eea}. To avoid confusion, also note that
there is a coefficient $f_3$ appearing in \cref{rel.even.N} which we removed by using the $\vf$
equation of motion (or equivalently, by redefining $\vf$). This coefficient has been included in the
counting of independent transport coefficients in \cite{Bhattacharya:2011tra}.

\section{Null Superfluids}\label{sec3}

In \cite{Banerjee:2015hra} we proposed ``null fluids'' as a new viewpoint of Galilean fluids. In
this section, we will further extend this formalism to include Galilean superfluids. The main
benefit of working with ``null (super)fluids'' is that it is a ``relativistic embedding'' of
Galilean (super)fluids into one higher dimension and enables us to directly use the existing
relativistic machinery to read out the respective Galilean results. In this sense, our in-depth
review of relativistic superfluids in the previous section will be vital for our discussion of
null/Galilean superfluids. To make the transition from relativistic $\ra$ null $\ra$ Galilean
superfluids manifest, we will step by step imitate our relativistic discussion of the previous
section with appropriate accommodations for null superfluids. Later in \cref{sec4}, we will
translate our null superfluid results to the better known Newton-Cartan and conventional
non-covariant notations.

\subsection{Null Backgrounds and Null Superfluids}

Let us quickly recap null backgrounds \cite{Banerjee:2015uta,Banerjee:2015hra}, which are a natural
`embedding' of Galilean (Newton-Cartan) backgrounds into a relativistic spacetime of one higher
dimension. Consider a $(d+1)$-dimensional manifold $\cM_{(d+1)}$ equipped with a metric $g_{\sM\sN}$
and a $\rmU(1)$ gauge field $A_\sM$. Infinitesimal diffeomorphisms and gauge transformation with
parameters $\scrX = \{\c^\sM, \L_\c\}$ respectively, act on these background fields as,
\begin{equation}
  \d_\scrX g_{\sM\sN} = \N_\sM \c_\sN + \N_\sN \c_\sM, \qquad
  \d_\scrX A_\sM = \dow_\sM (\L_{\c} + \c^\sN A_\sN ) + \c^\sN F_{\sN\sM}.
\end{equation}
The characteristic feature of a null background is the existence of a compatible null isometry
$\scrV = \{V^\sM, \L_V\}$ which satisfies: $V^\sM V_\sM = 0$, $\N_\sM V^\sN = 0$,
$V^\sM A_\sM + \L_V = -1$ \footnote{This condition can be thought of as fixing a component of the
  $(d+1)$-dimensional gauge field $A_\sM$, leaving it with only $d$ independent components mapping
  bijectively to the $d$-dimensional Galilean gauge field. As opposed to the null backgrounds
  defined in \cite{Banerjee:2015hra} where we set $V^\sM A_\sM + \L_V = 0$, for superfluids we
  realize that it is more suitable to fix $V^\sM A_\sM + \L_V = -1$ instead.} and,
\begin{equation}
  \d_\scrV g_{\sM\sN} = \N_\sM V_\sN + \N_\sN V_\sM = 0, \qquad
  \d_\scrV A_\sM = \dow_\sM (\L_{V} + V^\sN A_\sN ) + V^\sN F_{\sN\sM} = V^\sN F_{\sN\sM} = 0.
\end{equation}
Since we will be interested in studying superfluids on this
background, we introduce a preferred $\rmU(1)$ phase $\vf$ which transforms under diffeomorphisms
and infinitesimal gauge transformations as $\d_\scrX \vf = \c^\sM \dow_\sM \vf - \L_\c$. The
covariant derivative of $\vf$ is known as the \emph{superfluid velocity},
\begin{equation}
  \xi_\sM = \dow_\sM \vf + A_\sM.
\end{equation}
We require $\vf$ to respect the null isometry $\scrV$, i.e.
$\d_\scrV \vf = V^\sM \dow_\sM \vf - \L_V = V^\sM \xi_\sM - 1 = 0$, which implies
$V^\sM \xi_\sM = -1$. The remainder of the story is exactly same as the relativistic case: any
theory coupled to a null background has an energy-momentum tensor $T^{\sM\sN}$ and a charge current
$J^\sM$ in its spectrum. The respective conservation laws are given as,
\begin{equation}\label{eom_gsf}
  \N_\sM T^{\sM\sN} = F^{\sN\sR} J_\sR + \rmT_\rmH^{\sN\perp} + \xi^\sM K, \qquad
  \N_\sM J^\sM = \rmJ_\rmH^\perp - K,
\end{equation}
where,
\begin{equation}\label{vfEOM_null}
  K = 0,
\end{equation}
is the $\vf$ equation of motion. Since \cref{eom_gsf,vfEOM_null} are $(d+3)$ equations in $(d+1)$
dimensions, they can provide dynamics for a superfluid described by an arbitrary set of $(d+2)$
variables in addition to the phase $\vf$. We choose these to be a normalized null \emph{fluid
  velocity} $u^\sM$ (with $u^\sM V_\sM = -1$, $u^\sM u_\sM = 0$), a \emph{temperature} $T$, a
\emph{mass chemical potential} $\mu_{n}$, and a \emph{chemical potential} $\mu$, known as the
\emph{hydrodynamic fields}. A null superfluid hence is completely characterized by gauge-invariant
expressions of $T^{\sM\sN}$, $J^\sM$, $K$ in terms of $g_{\sM\sN}$, $A_\sM$, $u^\sM$, $T$, $\mu_n$,
$\mu$ and $\xi_\sM$, known as the \emph{null superfluid constitutive relations}. The near
equilibrium assumption allows us to arrange these constitutive relations as a perturbative expansion
in derivatives (known as the \emph{derivative} or \emph{gradiant expansion}).

Same as the relativistic case, null superfluid is also required to satisfy a version of the
\emph{second law of thermodynamics}. It states that there must exist an \emph{entropy current}
$J_S^\sM$ whose divergence is positive semi-definite everywhere, i.e.,
\begin{equation}\label{onshell2ndlaw}
  \N_\sM J_S^\sM \geq 0,
\end{equation}
as long as the superfluid is thermodynamically isolated (i.e. conservation laws \cref{eom_gsf} are
satisfied), irrespective of $K$ being zero. The job of null superfluid dynamics now is to find the
most general constitutive relations $T^{\sM\sN}$, $J^\sM$, $K$ and an associated $J_S^\sM$, $\D$
order by order in derivative expansion, such that \cref{onshell2ndlaw} is satisfied for
thermodynamically isolated fluids. Owing to our previous experiences with the second law however, we
switch to the offshell formalism in the next subsection for simplicity.

\subsection{Offshell Formalism for Null (Super)fluids}

We couple the fluid to an external momentum $P^\sM_{ext}$ and charge $Q_{ext}$ source, so that the
conservation laws are no longer satisfied. Having done that, the second law \cref{onshell2ndlaw}
will be modified with an arbitrary combination of the conservation laws to get,
\begin{multline}\label{offshell_2ndlaw_null}
  \N_\sM J_S^\sM + \b_\sN \lb \N_\sM T^{\sM\sN} - F^{\sN\sR} J_\sR - \rmT_\rmH^{\sN\perp} - \xi^\sM K
  \rb \\
  + \lb \L_{\b} + A_\sM  \b^\sM\rb \lb\N_\sM J^\sM - \rmJ^\perp_\rmH + K \rb = \D \geq 0,
\end{multline}
where $\scrB = \{\b^\sM, \L_\b\}$ are some arbitrary fields. Recall that the hydrodynamic fields
$u^\sM$, $T$, $\mu_n$, $\mu$ were some arbitrary $(d+2)$ fields chosen to describe the fluid. Like
in any field theory, they are permitted to admit an arbitrary redefinition among themselves without
changing the physics. This huge amount of freedom can be fixed by explicitly choosing,
\begin{equation}
  u^\sM = -\frac{\b^\sM}{V_\sM \b^\sM} + \frac{\b^\sR \b_\sR V^\sM}{2(V_\sN \b^\sN)^2}, \quad
  T = -\frac{1}{V_\sM \b^\sM}, \quad
  \mu_n = \frac{\b^\sM U_\sM}{2(V_\sN \b^\sN)^2}, \quad
  \mu = -\frac{\L_{\b} + A_\sM \b^\sM}{V_\sN \b^\sN}.
\end{equation}
or conversely,
\begin{equation}
  \b^\sM = \frac{1}{T} \lb u^\sM - \mu_n V^\sM \rb, \qquad
  \L_\b = \frac{\mu}{T} - A_\sM u^\sM.
\end{equation}
We define a free energy current,
\begin{equation}
  -\frac{G^\sM}{T}= N^\sM = S^\sM + T^{\sM\sN} \b_\sN + \lb \L_{\b} + \b^\sN A_\sN\rb J^\sM, \quad
  - \frac{\rmG^\sM_\rmH}{T} = \rmN^\perp_\rmH = \b_\sM \rmT_\rmH^{\sM\perp} + \lb \L_{\b} + \b^\sM A_\sM \rb \rmJ_\rmH^\perp, 
\end{equation}
which turns the offshell second law in \cref{offshell_2ndlaw_null} to a free energy conservation equation,
\begin{equation}\label{adiabaticity_anomaly}
  \N_\sM N^\sM - \rmN^\perp_\rmH = \half T^{\sM\sN} \d_\scrB g_{\sM\sN} + J^\sM\d_\scrB A_\sM + K
  \d_\scrB \vf + \D, \qquad \D \geq 0.
\end{equation}
Now similar to our analysis of relativistic superfluids, we will try to find the most generic
$T^{\sM\sN}$, $J^\sM$, $K$ in terms of $g_{\sM\sN}$, $A_\sM$, $\b^\sM$, $\L_\b$, $\vf$ which solves
this equation for some $N^\sM$, $\D$. Again however, these expressions will be shy of being the null
superfluid constitutive relations because of their dependence on the external sources $P^\sM_{ext}$,
$Q_{ext}$. To fix this, we will only consider the expressions for $T^{\sM\sN}$, $J^\sM$, $K$ which
are independent of certain data that can be eliminated using the conservation laws. For the most
part, the following analysis along with the wordings would be exactly same as has been used in the
previous section for relativistic superfluids, except for certain modifications to accommodate the
compatible null isometry.

\subsubsection{Josephson Equation for Null (Galilean) Superfluids}

In the study of superfluids, the $\rmU(1)$ phase $\vf$ is generally taken to be order $-1$ in the
derivative expansion, while its covariant derivative $\xi_\sM$ is taken to be order 0. The reason
being that the true dynamical degrees of freedom are encoded in the fluctuations of $\vf$ along the
$\rmU(1)$ circle, and not in $\vf$ itself. It implies that the $K\d_\scrB\vf$ term in the free
energy conservation \cref{adiabaticity_anomaly} is allowed to be order zero, if $K$ has an order 0
term. This gives us the unique solution to \cref{adiabaticity_anomaly} at zero derivative order,
\begin{equation}
  N^\sM, T^{\sM\sN}, J^\sM = \cO(\dow^0), \quad
  K = -\a \d_\scrB\vf + \cO(\dow), \quad
  \D = \a (\d_\scrB\vf)^2 + \cO(\dow),
\end{equation}
for some ``transport coefficient'' $\a\geq0$. Note that the $\vf$ equation of motion at this order
will read $K = -\a\d_\scrB\vf + \cO(\dow) = 0$, implying,
\begin{equation}
  \d_\scrB\vf =  \frac{1}{T} \lb u^\sM \xi_\sM + \mu_n - \mu \rb = \cO(\dow)
  \quad\implies\quad u^\sM \xi_\sM = \mu - \mu_n + \cO(\dow).
\end{equation}
This is the Josephson equation for null superfluids. This condition also ensures that $\D$ is at
least $\cO(\dow)$, avoiding ``ideal superfluid dissipation''. From this point onward, it would be
beneficial to think of $\d_{\scrB}\vf$ as an order $1$ data in derivative expansion rather than
$0$.

\subsection{Ideal Null Superfluids}

Let us now move on to the ideal null superfluids, i.e. null superfluid constitutive relations that
satisfy the free energy conservation \cref{adiabaticity_anomaly} at first derivative order. At
ideal order, the most generic tensorial form of various quantities appearing in
\cref{adiabaticity_anomaly} can be written as,
\begin{align}\label{ideal.consti_null}
  T^{\sM\sN} &= R_n u^\sM u^\sN + 2 E u^{(\sM}V^{\sN)} + P P^{\sM\sN} + R_s
               \xi^\sM \xi^\sN + 2\l_1 \xi^{(\sM} V^{\sN)} + 2\l_2 \xi^{(\sM} u^{\sN)} + R_v V^\sM V^\sM
               + \cO(\dow), \nn\\ 
  J^\sM &= Q u^\sM + Q_s \xi^\sM + Q_v V^\sM + \cO(\dow), \nn\\
  K &= -\a \d_\scrB\vf + K_{ideal} + \cO(\dow), \nn\\
  N^\sM &= N u^\sM + N_s \xi^\sM + N_v V^\sM + \cO(\dow), \nn\\
  \D &= (\a \d_\scrB\vf)^2 + \D_{ideal} + \cO(\dow^2),
\end{align}
where $R_n$, $E$, $P$, $R_s$, $\l_1$, $\l_2$, $Q$, $Q_s$, $K_{ideal}$, $N$, $N_s$ are functions of
$T$, $\mu$, $\mu_n$ and $\mu_s \equiv -\frac{1}{2}\xi^\sM \xi_\sM$. We have omitted the only other
possible scalar $\d_\scrB\vf$ in the functional dependence, because using the $\vf$ equation of
motion we know that it is no longer an independent quantity. The coefficients $R_v$, $Q_v$, $N_v$ do
not contain any physical information, because their contribution to the conservation laws trivially
vanish owing to $\scrV$ being an isometry. Plugging \cref{ideal.consti_null} in
\cref{adiabaticity_anomaly} we can find,
\begin{multline}\label{freeE.null.ideal.expand}
  (Q_s + R_s) \xi^\sM \lb \N_\sM \nu + \frac{1}{T} u^\sN F_{\sN\sM} \rb
  + \frac{\l_1}{T^2} \xi^\sM \N_{\sM} T
  + \l_2 \xi^{\sN} \lb \N_{\sN} \nu_n + u^{\sM} \N_{\sM} U_\sN \rb\\
  \N_{\sM} \lb\lb \frac{P}{T} - N\rb u^\sN \rb
  + \frac{1}{T} u^\mu \lb \N_\mu E - T \N_{\sM} S - \mu_n \N_{\sM} R_n - \mu\N_\sM Q + R_s \N_{\sN} \mu_s \rb\\
  + \N_{\sM} \lb \lb\d_\scrB\vf R_s - N_s \rb\xi^\sM \rb
  + \lb K_{ideal} - \N_\sM (R_s \xi^\sM) \rb \d_\scrB \vf + \D_{ideal} = 0,
\end{multline}
where we have defined $S$ through the ``Euler equation'',
\begin{equation}
  E + P = ST + Q\mu + R_n\mu_n.
\end{equation}
\Cref{freeE.null.ideal.expand} will imply a set of relations among various coefficients,
\begin{equation}
  Q_s = -R_s, \quad
  \l_1 = \l_2 = 0, \quad
  N = \frac{P}{T}, \quad
  N_s = \d_\scrB\vf R_s, \quad
  K_{ideal} = \N_\sM (R_s \xi^\sM), \quad
  \D_{ideal} = 0,
\end{equation}
and the ``first law of thermodynamics'',
\begin{equation}
  \df E = T\df S + \mu \df Q + \mu_n \df R_n - R_s \df \mu_s,
\end{equation}
giving physical meaning to the quantities we have introduced in \cref{ideal.consti_null}. Finally,
we have the full set of null superfluid constitutive relations up to ideal order satisfying the
second law,
\begin{align}
  T^{\sM\sN} &= R_n u^\sM u^\sN + 2 E u^{(\sM}V^{\sN)} + P P^{\sM\sN} + R_s
  \xi^\sM \xi^\sN + R_v V^\sM V^\sN + \cO(\dow), \nn\\
  J^\sM &= Q u^\sM - R_s \xi^\sM + Q_v V^\sM  + \cO(\dow), \nn\\
  K &= -\a \d_\scrB\vf + \N_\sM (R_s \xi^\sM)  + \cO(\dow), \nn\\
  N^\sM &= \frac{P}{T} u^\sM + \d_\scrB\vf R_s \xi^\sM + N_v V^\sM  + \cO(\dow), \nn\\
  J_S^\sM &=  N^\sM - \frac{1}{T} \lb T^{\sM\sN} u_\sN - \mu_n
                  T^{\sM\sN} V_\sN + \mu J^\sM \rb = S u^\sM + S_v V^\sM  + \cO(\dow).
\end{align}
Here $S_v = N_v + \frac{1}{T} \lb R_v - \mu_n E - \mu Q_v \rb$, which again doesn't contain any
physical information. These are the ideal null superfluid constitutive relations. Note that we have
included first order terms in $K$, $N^\sM$ which can be ignored when talking about the ideal order,
but are required for internal consistency with \cref{adiabaticity_anomaly}. The $\vf$ equation
of motion $K=0$ will imply,
\begin{equation}
  \a \d_\scrB\vf = \N_\sM (R_s \xi^\sM)  + \cO(\dow) \quad\implies\quad
  u^\sM \xi_\sM = \mu - \mu_n + \frac{T}{\a} \N_\sM (R_s \xi^\sM) + \cO(\dow),
\end{equation}
which is a first order correction to the Josephson equation. Note however that this equation can
admit further one derivative corrections due to the first order constitutive relations discussed in
the next subsection; the correction mentioned here is only how the ideal null superfluid transport
affects the Josephson equation. The conservation laws on the other hand are complete up to the first
order in derivatives,
\begin{align}
  \frac{1}{\sqrt{-g}} \d_\scrB \lb\sqrt{-g} \lb T(E+P) V_M + RT u_\sM \rb \rb + QT \d_\scrB A_\sM
  &= - \xi_\sM \a \d_\scrB\vf  + \cO(\dow^2), \nn\\
  \frac{1}{\sqrt{-g}} \d_\scrB \lb\sqrt{-g} QT\rb &= \a\d_\scrB\vf + \cO(\dow^2).
\end{align}
These equations provide a set of relations between $\d_\scrB\vf$, $\d_\scrB g_{\sM\sN}$ and
$\d_\scrB A_\sM$, which can be used to eliminate a vector $u^\sM \d_{\scrB} g_{\sM\sN}$ and a scalar
$u^\sM \d_\scrB A_\sM$ (see \cref{data.null}) from the first order null constitutive relations. On
the other hand, we choose to eliminate the scalar data $\N_\sM (R_s\xi^\sM)$ using the $\vf$
equation of motion.

\subsection{First Derivative Corrections to Null Superfluids} 

\begin{table}[p]
  \centering
  \begin{tabular}[t]{|c|c|c|}
        \hline
    \multicolumn{3}{|c|}{Vanishing at Equilibrium -- Onshell Independent} \\
    \hline
    $S_1$ & $\frac{T}{2} \tilde P^{\sM\sN} \d_\scrB g_{\sM\sN}$ & $\tilde P^{\sM\sN} \N_\sM u_\sN$ \\
    $S_2,S_{e,1}$ \footnote{Null and Newton-Cartan geometries behave more naturally in presence of a
    minimal temporal torsion $H_{\sM\sN} = 2\dow_{[\sM} V_{\sN]}$ (cref. TTNC geometries
    \cite{}). In presence of $H_{\sM\sN}$, the data $S_2 = \z^\sM \lb \frac{1}{T} \dow_\sM T + u^\sN
    H_{\sN\sM}\rb$ vanishes at equilibrium while $S_{e,1} = \frac{1}{T} \z^\sM\dow_\sM T$
    survives. However when $H_{\sM\sN} = 0$, $S_2 = S_{e,1}$.}
          & $TV^\sM \z^\sN \d_\scrB g_{\sM\sN}$ & $\frac{1}{T}\z^\sM
                                              \N_\sM T$ \\
    $S_3$ & $\frac{T}{2}\z^\sM \z^\sN \d_\scrB g_{\sM\sN}$ &
                                                         $\z^\sM\z^{\sN} \N_\sM u_\sN$ \\
    $S_4$ & $T \z^\sM \d_\scrB A_{\sM}$ & $\z^\sM \lb T \N_\sM \nu + u^\sN
                                      F_{\sN\sM} \rb$ \\
    $S_5$ & $T \d_\scrB \vf$ & $u^\sM \xi_\sM + \mu_n - \mu$ \\
    \hline
    $V^\sM_1,V^\sM_{e,1}$ & $T\tilde P^{\sM\sR} V^\sN \d_\scrB g_{\sR\sN}$ & $\frac{1}{T} \tilde P^{\sM\sN} \N_\sN
                                                             T$ \\
    $V^\sM_2$ & $T\tilde P^{\sM\sR} \z^\sN \d_\scrB g_{\sR\sN}$ & $2 \tilde P^{\sM\sR} \z^\sN \N_{(\sR}
                                                              u_{\sN)}$
                                                                                             \\
    $V^\sM_3$ & $T \tilde P^{\sM\sN} \d_\scrB A_{\sN}$ & $\tilde P^{\sM\sN} \lb T \N_\sN \nu + u^\sR
                                                     F_{\sR\sN} \rb$ \\
    \hline
    $\s^{\sM\sN}$ & $\frac{T}{2} \tilde P^{\sR\langle\sM}\tilde P^{\sN\rangle\sS} \d_\scrB
                          g_{\sR\sS}$ & $\tilde P^{\sM\sR}\tilde
                              P^{\sN\sS} \lb \N_{(\sR} u_{\sS)} -
                                        \frac{\tilde P_{\sR\sS}}{d-1}
                                        S_1 \rb$ \\
    \hline
    $\tilde V_{1}^\sM$ & \multicolumn{2}{c|}{$\e^{\sM\sN\sR\sS\sT} V_\sN u_\sR \z_\sS V_{1,\sT} $}
  \\
    $\tilde V_{2}^\sM$ & \multicolumn{2}{c|}{$\e^{\sM\sN\sR\sS\sT} V_\sN u_\sR \z_\sS V_{2,\sT} $}
  \\
    $\tilde V_{3}^\sM$ & \multicolumn{2}{c|}{$\e^{\sM\sN\sR\sS\sT} V_\sN u_\sR \z_\sS V_{3,\sT} $}
  \\
    \hline
    $\tilde\s^{\sM\sN}$ & \multicolumn{2}{c|}{$\e^{(\sM|\sR\sS\sT\sP} V_\sR u_\sS \z_\sT \s_{\sP}^{\
                          \sN)}$} \\
    \hline\hline
    \multicolumn{3}{|c|}{Vanishing at Equilibrium -- Onshell Dependent} \\
    \hline
    $S_6$ & $Tu^\sM V^\sN \d_\scrB g_{\sM\sN}$ & $\frac{1}{T} u^\sM \N_\sM T$ \\
    $S_7$ & $Tu^\sM \d_\scrB A_{\sM}$ & $T u^\sM \N_\sM \nu$ \\
    $S_8$ & $\frac{T}{2}u^\sM u^\sN \d_\scrB g_{\sM\sN}$ & $T u^\sM \N_\sM \nu_n$ \\
    $S_9$ & $Tu^\sM \z^\sN \d_\scrB g_{\sM\sN}$ & $\z^\sM \lb T \N_\sM \nu_n + u^\sN \N_\sN u_\sM \rb$ \\
    \hline
    $V_4^\sM$ & $T \tilde P^{\sM\sR}u^{\sN} \d_\scrB g_{\sR\sN}$ & $\tilde P^{\sM\sN} \lb T \N_\sN \nu_n + u^\sR \N_\sR
                                                               u_\sN \rb$ \\
    \hline
    $\tilde V_{4}^\sM$ & \multicolumn{2}{c|}{$\e^{\sM\sN\sR\sS\sT} V_\sN u_\sR \z_\sS V_{4,\sT} $} \\
    \hline\hline
    \multicolumn{3}{|c|}{Surviving at Equilibrium} \\
    \hline
    $S_{e,2}$ & \multicolumn{2}{c|}{$T\z^\sM \dow_\sM \nu$} \\
    $S_{e,3}$ & \multicolumn{2}{c|}{$T\z^\sM \dow_\sM \nu_n$} \\
    $\vdots$ & \multicolumn{2}{c|}{$\vdots$} \\
    \hline
    $V^\sM_{e,2}$ & \multicolumn{2}{c|}{$T \tilde P^{\sM\sN} \dow_\sN \nu$} \\
    $V^\sM_{e,3}$ & \multicolumn{2}{c|}{$T \tilde P^{\sM\sN} \dow_\sN \nu_n$} \\
    $\vdots$ & \multicolumn{2}{c|}{$\vdots$} \\
    \hline
    $\tilde S_{e,1}$ & \multicolumn{2}{c|}{$T \e^{\sM\sN\sR\sS\sT} \z_\sM V_\sN u_\sR \dow_{\sS}u_\sT$} \\
    $\tilde S_{e,2}$ & \multicolumn{2}{c|}{$\half T\e^{\sM\sN\sR\sS\sT} \z_\sM V_\sN u_\sR F_{\sS\sT}$} \\
    $\vdots$ & \multicolumn{2}{c|}{$\vdots$} \\
    \hline
    $\tilde V^\sM_{e,1}$ & \multicolumn{2}{c|}{$T \tilde P^{\sM}_{\ \sK} \e^{\sK\sN\sR\sS\sT} V_\sN u_\sR \dow_{\sS}u_\sT$} \\
    $\tilde V^\sM_{e,2}$ & \multicolumn{2}{c|}{$\half T \tilde P^{\sM}_{\ \sK} \e^{\sK\sN\sR\sS\sT} V_\sN u_\sR F_{\sS\sT}$} \\
    $\tilde V^\sM_{e,3}$ & \multicolumn{2}{c|}{$T \tilde P^{\sM}_{\ \sK} \e^{\sK\sN\sR\sS\sT} \xi_\sN u_\sR \dow_{\sS}u_\sT$} \\
    $\tilde V^\sM_{e,4}$ & \multicolumn{2}{c|}{$\half T \tilde P^{\sM}_{\ \sK} \e^{\sK\sN\sR\sS\sT} \xi_\sN u_\sR F_{\sS\sT}$} \\
    $\vdots$ & \multicolumn{2}{c|}{$\vdots$} \\
    \hline
  \end{tabular} 
  \caption{\label{data.null} Independent first order data for null superfluids. We
    have not enlisted, neither would we need, all the independent data surviving at equilibrium.}
\end{table}

Moving on to the one derivative null superfluids, let us schematically represent various quantities
appearing in \cref{adiabaticity_anomaly} up to the first order in derivatives as,
\begin{align}\label{first.consti_null}
  T^{\sM\sN} &= \Big[ R_n u^\sM u^\sN + 2 E u^{(\sM}V^{\sN)} + P P^{\sM\sN} + R_s
  \xi^\sM \xi^\sN + R_v V^\sM V^\sN \big] + \cT^{\sM\sN} + \cO(\dow^2), \nn\\
  J^\sM &= \big[ Q u^\sM - R_s \xi^\sM + Q_v V^\sM \big] + \cJ^\sM + \cO(\dow^2), \nn\\
  K &= \big[ -\a \d_\scrB\vf + \N_\sM (R_s \xi^\sM) \big] + \cK + \cO(\dow^2), \nn\\
  N^\sM &= \lB \frac{P}{T} u^\sM + \d_\scrB\vf R_s \xi^\sM + N_v V^\sM\rB + \cN^\sM + \cO(\dow^2), \nn\\
  \D &= \a (\d_\scrB\vf)^2  + \cD,
\end{align}
where the corrections $\cT^{\sM\sN}$, $\cJ^\sM$, $\cK$, $\cN^\sM$, $\cD$ have exactly one derivative in
every term. Plugging these in the \cref{adiabaticity_anomaly} we can get an equation among the
corrections,
\begin{align}
  \N_\sM \cN^\sM - \rmN_\rmH^\perp
  &= \half \cT^{\sM\sN} \d_\scrB g_{\sM\sN} + \cJ^\sM\d_\scrB A_\sM + \cK
    \d_\scrB\vf + \cD + \cO(\dow^3). \label{null_adiabaticity_first}
\end{align}
We will now attempt to find all the solutions to this equation, hence recovering the null superfluid
constitutive relations up to the first order in derivatives.

\subsubsection{Parity Even}

We can find the most general parity even solution of \cref{null_adiabaticity_first} in 2 steps (note
that $\rmN^\perp_\rmH$ is parity odd): (1) first we write down the most general allowed parity-even
$\cN^\sM$ and find a set of constitutive relations pertaining to that, and (2) then find the most
general parity-even constitutive relations which satisfy \cref{null_adiabaticity_first} with
$\cN^\sM = 0$.

\begin{enumerate}
\item One can check that the most general form of $\cN^\sM$ (whose divergence only contains product
  of derivatives and has at least one $\d_\scrB$ per term) can be written as (see \cref{eqbPF} for
  more details),
  \begin{multline}\label{null.even.N}
    \cN^\sM =
    2 f_1 u^{[\sM} \z^{\sN]} \frac{1}{T^2} \dow_\sN T
    + 2 f_2 u^{[\sM} \z^{\sN]} \dow_\sN \nu
    + 2 f_3 u^{[\sM} \z^{\sN]} \dow_\sN \nu_n \\
    + 2 f_4 u^{[\sM} \z^{\sN]} \dow_\sN R_s
    + \N_{\sN} \lb f_5 u^{[\sM} \z^{\sN]} \rb,
  \end{multline}
  where $f$'s are functions of $T$, $\nu=\mu/T$, $\nu_n=\mu_n/T$ and $\hat\mu_s = -\half \z^\sM \z_\sM$ with
  $\z^\sM = P^{\sM\sN} \xi_\sN = \xi^\sM - u^\sM +(u^\sN \xi_\sN) V^\sM$
  ($P^{\sM\sN} = g^{\sM\sN} + 2u^{(\sM} V^{\sN)}$ is the projection operator away from the null
  fluid velocity). Note that,
  \begin{equation}
    \hat\mu_s = -\half \z^\sM \z_\sM = -\half \xi^\sM \xi_\sM + \xi^\sM u_\sM
    = \mu_s + \xi^\sM u_\sM = \mu_s - \mu_n + \mu + T\d_\scrB \vf.
  \end{equation}
  Out of the five terms in \cref{null.even.N}, the last one has trivially zero divergence and hence
  can be ignored. The forth term on the other hand can be removed by elimination of
  $\N_\sM(R_s\xi^\sM)$ using the $\vf$ equation of motion. Computing the divergence of the remaining
  terms in $\cN^\sM$ and comparing them to \cref{null_adiabaticity_first}, we can directly read out
  the corresponding null superfluid constitutive relations (the symbol `$\ni$' represents that they
  are not yet the complete solutions of \cref{null_adiabaticity_first}; we still have to add the
  terms with $\cN^\sM = 0$),
  \begin{align}\label{null.even.N.consti}
    \cT^{\sM\sN}
    &\ni u^\sM u^\sN \lb \sum_{i=1}^3 \a_{R_n,i} S_{e,i} - \frac{1}{T}\N_\sR (Tf_3 \z^\sR) \rb
      + 2 V^{(\sM} u^{\sN)} \lb \sum_{i=1}^3 \a_{E,i} S_{e,i} - \frac{1}{T} \N_\sR (Tf_1 \z^\sR) \rb \nn\\
    &\quad + \lb \z^\sM \z^\sN + 2 \z^{(\sM} u^{\sN)} - 2 \z^{(\sM}V^{\sN)} (u^\sR \xi_\sR) \rb \sum_{i=1}^3 \a_{R_s,i} S_{e,i} \nn\\
    &\quad - 2 \xi^{(\sM} \sum_{i=1}^3 f_i V_{e,i}^{\sN)}
      + \tilde P^{\sM\sN} \sum_{i=1}^3 f_i S_{e,i}
      + 2 \z^{(\sM} V^{\sN)}\sum_{i=1}^3 f_i S_{5+i}, \nn\\[0.1cm]
    \cJ^\sM
    &\ni u^\sM \lb \sum_{i=1}^3 \a_{Q,i} S_{e,i} - \frac{1}{T}\N_\sR (Tf_2 \z^\sR) \rb
      - \z^\sM \sum_{i=1}^3 \a_{R_s,i} S_{e,i} + \sum_{i=1}^3 f_i V_{e,i}^\sM, \nn\\[0.1cm]
    \cK
    &\ni \N_\sM \lb \z^\sM \sum_{i=1}^3 \a_{R_s,i} S_{e,i} - \sum_{i=1}^3 f_i V_{e,i}^\sM \rb,
  \end{align}
  where
  $\tilde P^{\sM\sN} = g^{\mu\nu} + 2u^{(\sM} V^{\sN)} - \frac{1}{\z^\sR \z_\sR} \z^\sM \z^\sN$, and
  we have defined,
  \begin{equation}
    \df f_i = \frac{\a_{E,i}}{T} \df T + T \a_{R_n,i} \df \nu_n + T \a_{Q,i} \df \nu
    + \lb \a_{R_s,i} - \frac{f_i}{2\hat\mu_s} \rb \df \hat\mu_s.
  \end{equation}
  The actual computation is not neat and we have presented the details in \cref{calc.details} for
  the aid of the readers interested in reproducing our results. Note that these constitutive
  relations are presented in terms of `data' which are natural for this sector; readers can modify
  these to their favorite basis and get results which might look considerably messier. Moreover,
  these results are written in a particular `hydrodynamic frame' chosen by aligning $u^\sM$, $T$,
  $\mu$, $\mu_n$ along $\b^\sM$, $\L_\b$, which again can be modified according to reader's
  preference.

\item Let us now look at the parity-even solutions to \cref{null_adiabaticity_first} with
  $\cN^\sM = 0$,
  \begin{equation}\label{LHS0_null}
    0= \half \cT^{\sM\sN} \d_\scrB g_{\sM\sN} + \cJ^\sM\d_\scrB A_\sM + \cK \d_\scrB\vf + \cD.
  \end{equation}
  Every term in $\cT^{\sM\sN}$, $\cJ^\sM$, $\cK$ must either cancel or contribute to $\D$ which has
  to be a quadratic form. It follows that the terms in $\cT^{\sM\sN}$, $\cJ^\sM$, $\cK$ must be
  proportional to $\d_\scrB g_{\sM\sN}$, $\d_\scrB A_\sM$, $\d_\scrB \vf$. Recall however that we
  have chosen to eliminate $u^\sM \d_\scrB g_{\sM\sN}$, $u^\sM\d_\scrB A_\sM$ using the equations of
  motion. For $\D$ to be a quadratic form, it therefore implies that $\cT^{\sM\sN}$, $\cJ^\sM$
  cannot have a term like $\#^{(\sM} u^{\sN)}$, $\# u^\sM$ respectively for some vector $\#^\sM$ and
  scalar $\#$. With this input let us write down the most generic allowed form of the currents in
  terms of 34 new transport coefficients $[\b_{ij}]_{5\times 5}$ (with $\b_{55} = \a/T$),
  $[\k_{ij}]_{3\times 3}$ and $\eta$,
  \begin{align}
  \cT^{\sM\sN}
  &\ni - T\bigg[
    \lbr \b_{11}\tilde P^{\sR\sS} + 2\b_{12}\z^{(\sR}V^{\sS)} + \b_{13}\z^\sR\z^\sS\rbr\tilde
    P^{\sM\sN}
    + \lbr \b_{21}\tilde P^{\sR\sS} + 2\b_{22}\z^{(\sR}V^{\sS)} + \b_{23}\z^\sR\z^\sS\rbr 2
    \z^{(\sM}V^{\sN)} \nn\\
  &\qquad + \lbr \b_{31}\tilde P^{\sR\sS} + 2\b_{32}\z^{(\sR}V^{\sS)} + \b_{33}\z^\sR\z^\sS\rbr
    \z^{\sM}\z^{\sN} \nn\\
  &\qquad
    + 4 \lbr \k_{11} V^{(\sR} + \k_{12} \z^{(\sR} \rbr \tilde P^{\sS)(\sM} V^{\sN)}
    + 4 \lbr \k_{21} V^{(\sR} + \k_{22} \z^{(\sR} \rbr \tilde P^{\sS)(\sM} \z^{\sN)}
    + \eta \tilde P^{\sM\langle\sR} P^{\sS\rangle\sN} \bigg] \half\d_\scrB g_{\sR\sS}
    \nn\\
  &\quad - T\bigg[
    \b_{14} \z^{\sR} \tilde P^{\sM\sN}
    + 2\b_{24} \z^{\sR} \z^{(\sM}V^{\sN)}
    + \b_{34} \z^\sR \z^{\sM}\z^{\sN}
    + 2\k_{13} \tilde P^{\sR(\sM} V^{\sN)}
    + 2\k_{23} \tilde P^{\sR(\sM} \z^{\sN)} \bigg] \d_\scrB A_\sR, \nn\\
  &\quad -T \bigg[ \b_{15} \tilde P^{\sM\sN} + 2\b_{25}\z^{(\sM}V^{\sN)} + \b_{35}\z^{\sM}\z^{\sN}
    \bigg] \d_\scrB\vf \nn\\[0.1cm] 
  &= - \tilde P^{\sM\sN} \sum_{i=1}^5 \b_{1i} S_i
    - 2\z^{(\sM}V^{\sN)} \sum_{i=1}^5 \b_{2i} S_i
    - \z^{\sM}\z^{\sN} \sum_{i=1}^5 \b_{3i} S_i
    - 2 V^{(\sM} \sum_{i=1}^3 \k_{1i} V^{\sN)}_i \nn\\
  &\qquad - 2 \z^{(\sM} \sum_{i=1}^3 \k_{2i} V^{\sN)}_i - \eta \s^{\sM\sN},
\end{align}
\begin{align}
\cJ^\sM
  &\ni - T\bigg[
    \lbr \b_{41}\tilde P^{\sR\sS} + 2\b_{42}\z^{(\sR}V^{\sS)} + \b_{43}\z^\sR\z^\sS\rbr\z^\sM
    + 2 \lbr \k_{31} V^{(\sR} + \k_{32} \z^{(\sR} \rbr \tilde P^{\sS)\sM} 
     \bigg] \half \d_\scrB g_{\sR\sS}  \nn\\
  &\quad - T\bigg[ \b_{44} \z^{\sM}\z^{\sN} + \k_{33} P^{\sM\sN} \bigg] \d_\scrB A_\sR
    -T \bigg[\b_{45} \z^\sM \bigg] \d_\scrB\vf, \nn\\
  &= - \z^\sM \sum_{i=1}^5 \b_{4i} S_i - \sum_{i=1}^3 \k_{3i} V^{\sM}_i,
\end{align}
\begin{equation}
  \cK \ni -T \bigg[ \b_{51}\tilde P^{\sR\sS} + 2\b_{52}\z^{(\sR}V^{\sS)} + \b_{53}\z^\sR\z^\sS
  \bigg]\d_\scrB g_{\sR\sS}
  - T \bigg[ \b_{54} \z^\sM \bigg] \d_\scrB A_\sM = -\sum_{i=1}^4 \b_{5i} S_i.
\end{equation}
Note that we did not include a term proportional to $\d_\scrB\vf$ in $\cK$, because such a term is
already present in $K = -\a\d_\scrB\vf + \N_\sM(R_s\xi^\sM) + \cK + \cO(\dow^2)$. Plugging these
back into \cref{LHS0_null} and defining $\b_{55}=\a/T$ we can read out the parity-even quadratic form
$\D|_{even} = \a(\d_\scrB\vf)^2 + \cD|_{even}$,
\begin{align}
  T\D|_{even}
  &= \sum_{i,j=1}^5 S_i\b_{ij}S_j + \sum_{i,j=1}^3 V^\sM_i\k_{ij}V_{j,\sM} + \eta
    \s^{\sM\sN}\s_{\sM\sN}, \nn\\
  &= \sum_{i,j=1}^5 S_i\b_{ij}^{(s)}S_j + \sum_{i,j=1}^3 V^\sM_i\k_{ij}^{(s)}V_{j,\sM} + \eta
  \s^{\sM\sN}\s_{\sM\sN}.
\end{align}
In the second step we have realized that only the symmetric parts of the matrices $\b_{ij}$ and
$\k_{ij}$ will survive in this expression, and will contribute towards dissipation. Thus only 22
out of 35 transport coefficients (including $\a$) are dissipative; the remaining 13 are
non-dissipative.
  
\end{enumerate}

\subsubsection{Parity-Odd (5 Dimensions)}

We can find the most general parity-odd solution of \cref{null_adiabaticity_first} in 3 steps: (1)
first we consider a particular set of solutions which takes care of the anomaly $\rmN^\perp_\rmH$
and proceed towards the non-anomalous constitutive relations, (2) then we write down the most
general allowed parity-odd $\cN^\sM$ and find a set of constitutive relations pertaining to that,
and (2) finally find the most general parity-odd constitutive relations with zero $\cN^\sM$.

\begin{enumerate}
\item In $4$ dimensions at the first order in the derivatives $\rmT_\rmH^{\sM\perp} = 0$ and
  $\rmJ_\rmH^\perp = - \frac{3}{4}C^{(4)} \e^{\sM\sN\sR\sS\sT} u_\sM F_{\sN\sR} F_{\sS\sR}$
  \cite{Banerjee:2015hra,Jain:2015jla}, which implies, 
  \begin{equation}
    \rmN_\rmH^\perp = - \frac{3}{4}C^{(4)} \frac{\mu}{T}\e^{\sM\sN\sR\sS\sT} u_\sM F_{\sN\sR} F_{\sS\sR} .
  \end{equation}
  A particular solution pertaining to \cref{null_adiabaticity_first} with this $\rmN_\rmH^\perp$ is
  given as (see \cite{Banerjee:2015hra}),
  \begin{equation}
    \cT^{\sM\sN} \ni 6 C^{(4)} \mu^2 V^{(\sM} B^{\sN)}, \quad
    \cJ^\sM \ni 6 C^{(4)} \mu B^\sM, \quad
    \cK \ni 0, \quad
    \cN^\sM \ni 3 C^{(4)} \frac{\mu^2}{T} B^\sM.
  \end{equation}
  Here we have defined the magnetic field and fluid vorticity as,
  \begin{equation}
    B^\sM = \half\e^{\sM\sN\sR\sS\sT} V_\sN u_\sR F_{\sS\sT}, \qquad
    \o^\sM = \e^{\sM\sN\sR\sS\s} V_\sN u_\sR \dow_\sS u_\sT.
  \end{equation}
  
\item One can check that the most general form of $\cN^\sM$ (whose divergence only contains product
  of derivatives and has at least one $\d_\scrB$ per term) can be written as (see \cref{eqbPF} for
  more details),
  \begin{equation}\label{null.odd.N}
    \cN^\sM
    = g_1 \lb \b^{\sM} \tilde S_{e,1} + \tilde V^\sM_{4} \rb
    + g_2 \lb \b^{\sM} \tilde S_{e,2} + \tilde V^\sM_{3} \rb
    + g_3 \tilde V^\sM_1
    + C_1 T \o^\sM,
  \end{equation}
  where $g$'s are functions of $T$, $\nu$, $\hat\mu_s$, and $C_1$ is a constant\footnote{It might be noted that since $\N_\sM \o^\sM = 0$, $C_1$ a
    priory can be an arbitrary function rather than a constant. However, if we do the same
    computation in presence of torsion and later turn it off, which allows for
    $\dow_{[\sM} V_{\sN]} \neq 0$, we will be forced to set $C_1$ to be a constant (see appendix (A)
    of \cite{Banerjee:2015hra}). Another way to see that $C_1$ should be a constant is using the
    equilibrium partition function discussed in \cref{eqbPF}.}. From here we
  can directly read out the corresponding constitutive relations,
  \begin{align}\label{null.odd.N.consti}
    \cT^{\sM\sN}
    &\ni u^\sM u^\sN \sum_{i=1}^2 \tilde\a_{R_n,i} \tilde S_{e,i}
      + 2 V^{(\sM} u^{\sN)} \sum_{i=1}^2 \tilde\a_{E,i} \tilde S_{e,i} \nn\\
    &\quad + \lb \z^\sM \z^\sN + 2 \z^{(\sM} u^{\sN)} - 2 \z^{(\sM}V^{\sN)} (u^\sR \xi_\sR) \rb
      \sum_{i=1}^2 \tilde\a_{R_s,i} \tilde S_{e,i}
      - \z^\sM \z^\sN \sum_{i=1}^2 \frac{g_i}{2\hat\mu_s} \tilde S_{e,i} \nn\\
    &\quad
      - 2 V^{(\sM} \sum_{i=1}^2 g_i \tilde V^{\sN)}_{e,i+2}
      - 2 u^{(\sM} \sum_{i=1}^2 g_i \tilde V^{\sN)}_{e,i}
      + 2 C_1T^2 V^{(\sM} \o^{\sN)}
      \nn\\
    &\quad
      + 2 u^{(\sM} P^{\sN)}_{\ \ \sP} \e^{\sP\sK\sR\sS\sT} \N_\sK \lb Tg_1 V_\sR u_\sS \xi_\sT \rb
      + 2 V^{(\sM} P^{\sN)}_{\ \ \sP} \N_\sK \lb g_3T \e^{\sP\sK\sR\sS\sT} V_\sR u_\sS \z_\sT \rb
      , \nn\\[0.1cm]
    \cJ^\sM
    &\ni u^\sM \sum_{i=1}^2 \tilde\a_{Q,i} \tilde S_{e,i}
      - \z^\sM \sum_{i=1}^2 \tilde\a_{R_s,i} \tilde S_{e,i}
      + \sum_{i=1}^2 g_i \tilde V_{e,i}^\sM
      + P^\sM_{\ \sK}\e^{\sK\sN\sR\sS\sT} \N_\sN \lb Tg_2 V_\sR u_\sS \xi_\sT \rb, \nn\\
    \cK
    &\ni \N_\sM \lb \z^\sM \sum_{i=1}^2 \tilde\a_{R_s,i} \tilde S_{e,i}
      - \sum_{i=1}^2 g_i \tilde V_{e,i}^\sM \rb,
  \end{align}
  where we have defined,
  \begin{equation}
    \df g_i = \frac{1}{T} \tilde\a_{E,i} \df T + T\tilde\a_{Q,i} \df\nu +
    T\tilde\a_{R_n,i} \df \nu_n + \lb \tilde\a_{R_s,i} - \frac{g_i}{2\hat\mu_s} \rb \df \hat\mu_s.
  \end{equation}
  The actual computation is not neat and we have presented the details in \cref{calc.details} for
  interested readers.

\item We should finally consider the parity-odd constitutive relations that satisfy
  \cref{null_adiabaticity_first} with zero LHS. Following our discussion in the parity-even sector,
  the allowed form of the constitutive relations can be written down in terms of 10 coefficients
  $[\tilde\k_{ij}]_{3\times 3}$ and $\tilde \eta$,
  \begin{align}
    \cT^{\sM\sN}
    &\ni - T  V_\sT u_\sK \z_\sL \bigg[
      4V^{(\sM}\e^{\sN)\sT\sK\sL(\sR} \lbr \tilde\k_{11}  V^{\sS)} + \tilde\k_{12}  \z^{\sS)} \rbr
      + 4\z^{(\sM}\e^{\sN)\sT\sK\sL(\sR} \lbr \tilde\k_{21}  V^{\sS)} + \tilde\k_{22}  \z^{\sS)}
      \rbr  \nn\\
    &\qquad
      + \tilde\eta\tilde P^{\sP(\sM}\e^{\sN)\sT\sK\sL(\sR} \tilde P^{\sS)}_{\ \ \sP} \bigg] \half \d_\scrB g_{\sR\sS}
    - T  V_\sT u_\sK \z_\sL \bigg[ 2 \tilde\k_{13} V^{(\sM}\e^{\sN)\sT\sK\sL\sR}
      + 2 \tilde\k_{23} \z^{(\sM}\e^{\sN)\sT\sK\sL\sR} \bigg] \d_\scrB A_\sR, \nn\\
    &= - 2V^{(\sM}\sum_{i=1}^3 \tilde\k_{1i} \tilde V^{\sN)}_{i}
      - 2\z^{(\sM}\sum_{i=1}^3 \tilde\k_{2i} \tilde V^{\sN)}_{i}
      - \tilde\eta\tilde \s^{\sM\sN}, \nn\\
    \cJ^\mu
    &\ni - T  V_\sT u_\sK \z_\sL \bigg[
      2\e^{\sM\sT\sK\sL(\sR} \lbr \tilde\k_{31}  V^{\sS)} + \tilde\k_{32}  \z^{\sS)} \rbr
      \bigg] \half \d_\scrB g_{\sR\sS}
      - T  V_\sT u_\sK \z_\sL \bigg[ \tilde\k_{33} \e^{\sM\sT\sK\sL\sR}\bigg] \d_\scrB A_\sR, \nn\\
    &= - \sum_{i=1}^3 \tilde\k_{3i} \tilde V^\sM_{i},
 \nn\\
    \cK &\ni 0.
  \end{align}
  One can check that these constitutive relations trivially satisfy \cref{null_adiabaticity_first} with
  zero LHS and the quadratic form $\D\vert_{odd} = \cD\vert_{odd}$ is given as,
  \begin{align}
    T\D\vert_{odd} &\ni - \e^{\sM\sN\sR\sS\sT} V_\sR u_\sS \z_\sT \lB \sum_{i,j=1}^3 V_{i,\sM} \tilde\k_{ij}
           V_{j,\sN} + \tilde\eta \s_{\sM\sP} \s_{\sN}^{\ \sP} \rB, \nn\\
         &= - \e^{\sM\sN\sR\sS\sT} V_\sR u_\sS \z_\sT \sum_{i,j=1}^3 V_{i,\sM} \tilde\k^{(a)}_{ij} V_{j,\sN}.
  \end{align}
  It follows that out of the 10 transport coefficients, only 3 contribute to dissipation and the
  other 7 are non-dissipative.
\end{enumerate}
  
\subsubsection{Positivity Constraints}

The dissipative transport coefficients are required to satisfy a set of inequalities to satisfy
$\D = \a(\d_{\scrB}\vf)^2 + \cD|_{even} + \cD|_{odd} \geq 0$,
\begin{equation}
  T\D
  = \sum_{i,j=1}^5 S_i\b_{(ij)}S_j + \lb \sum_{i,j=1}^3 V^\sM_i\k_{(ij)}V_{j,\sM}
  + \sum_{i,j=1}^3 V_{i}^\sM \tilde\k_{[ij]} \tilde V_{j,\sM}
\rb 
    + \eta\s^{\sM\sN}\s_{\sM\sN}.
\end{equation}
We want this expression to be a quadratic form, which it nearly is except the parity-odd terms in the
brackets. However this term can be made into a quadratic form by noticing that the square of a
parity odd term is parity-even, due to the identity,
\begin{equation}\label{epsilon.contraction_null}
  \lb\e^{\sM\sN\sR\sS\sT} V_\sR u_\sS \z_\sT \rb\lb\e_{\sM\sK\sL\sO\sP} V^\sL u^\sO \z^\sP\rb =
  \tilde P^{\sN}_{\ \sK} \z^\sM \z_\sM = - 2 \hat\mu_s \tilde P^{\sN}_{\ \sK},
\end{equation}
We define,
\begin{equation}\label{kappa'definition}
  \begin{pmatrix}
    V'^\sM_1 \\  V'^\sM_2 \\  V'^\sM_3
  \end{pmatrix}
  =
  \begin{pmatrix}
    V^\sM_1 \\  V^\sM_2 \\  V^\sM_3
  \end{pmatrix}
  + \begin{pmatrix}
    0 & a_{12} & a_{13} \\
    0 & 0 & a_{23} \\
    0 & 0 & 0
  \end{pmatrix}
  \begin{pmatrix}
    \tilde V'^\sM_{1} \\  \tilde V'^\sM_{2} \\  \tilde V'^\sM_{3}
  \end{pmatrix},
  \qquad
  \k'_{ij} = \k_{ij} + k_{ij}, \qquad
  k_{[ij]} = 0,
\end{equation}
such that,
\begin{equation}
  \sum_{i,j=1}^3 V'^\sM_i\k'_{(ij)}V'_{j,\sM}
  = \sum_{i,j=1}^3 V^\sM_i\k_{(ij)}V_{j,\sM}
  + \sum_{i,j=1}^3 V^\sM_{i} \tilde\k_{[ij]} \tilde V_{j,\sM}.
\end{equation}
Using the identity \cref{epsilon.contraction_null}, the above equation can be easily solved to give,
\begin{equation}
  [a_{ij}] =
  \begin{pmatrix}
    0 & \frac{\tilde\k_{[12]}}{\k_{11}} & \frac{
    \k_{11} \lb\k_{22}\tilde\k_{[13]} - \k_{(12)} \tilde\k_{[23]} \rb
    - \tilde\k_{[12]} \lb \k_{(12)} \k_{(13)} + \z^\sM \z_\sM\tilde\k_{[12]}\tilde\k_{[13]} \rb}{\k_{11}
    \lb\k_{11} \k_{22} - \k_{(12)}^2 - \z^\sM \z_\sM\tilde\k_{[12]}^2 \rb} \\
  0 & 0 & \frac{\k_{11} \tilde\k_{[23]} - \k_{(12)} \tilde\k_{[13]} +\k_{(13)} \tilde\k_{[12]}}{\k_{11} \k_{22}
    - \k_{(12)}^2 - \z^\sM \z_\sM\tilde\k_{[12]}^2} \\
  0 & 0 & 0
  \end{pmatrix},
\end{equation}
\begin{equation}
  [k_{ij}] =
  \begin{pmatrix}
    0 & 0 & 0 \\
    0 & -\z^\sM \z_\sM \frac{(\tilde\k_{[12]})^2}{\k_{11}} & - \z^\sM \z_\sM \frac{\tilde\k_{[12]}\tilde\k_{[13]}}{\k_{11}} \\
    0 & -\z^\sM \z_\sM \frac{\tilde\k_{[12]}\tilde\k_{[13]}}{\k_{11}}
    & -\z^\sM \z_\sM \lb \frac{(\tilde\k_{[13]})^2}{\k_{11}}
    + \frac{\lb\k_{11} \tilde\k_{[23]} - \k_{(12)} \tilde\k_{[13]} +\k_{(13)} \tilde\k_{[12]}\rb^2}
    {\k_{11}\lb\k_{11} \k_{22} - \k_{(12)}^2 - \z^\sM \z_\sM\tilde\k_{[12]}^2\rb} \rb
  \end{pmatrix}.
\end{equation}
Consequently $\D$ will take the form,
\begin{equation}
  T\D
  = \sum_{i,j=1}^5 S_i\b_{(ij)}S_j + \sum_{i,j=1}^3 V'^\sM_i\k'_{(ij)}V'_{j,\sM}
  + \eta \s^{\sM\sN}\s_{\sM\sN}.
\end{equation}
Given $T\geq0$, the condition $\D\geq 0$ implies that $\eta\geq 0$ and the matrices
$[\b_{(ij)}]_{5\times5}$, $[\k'_{(ij)}]_{3\times3}$ have all non-negative eigenvalues. This gives
$9$ inequalities among 25 dissipative transport coefficients, and 16 are completely arbitrary.

\subsection{Summary} \label{null-summary} 

We have completed the analysis of a null superfluid up to the first order in derivatives. Here we
summarize the results. We found that the entire null superfluid transport up to the first order in
derivatives is characterized by an ideal order pressure $P$, 51 first order transport coefficients
which are functions of $T$, $\mu/T$, $\mu_n/T$, $\hat\mu_s$, and two constants $C_1$,
$C^{(4)}$. $P$, $C_1$ and $C^{(4)}$ along with 6 transport coefficients,
\begin{equation}
  \text{Parity Even (3):}\qquad f_1,\quad f_2, \quad f_3, \qquad\qquad
  \text{Parity Odd (3):}\qquad g_1,\quad g_2, \quad g_3,
\end{equation}
totally determine the \emph{hydrostatic} transport (part of the constitutive relations that
survive at equilibrium). \emph{Non-hydrostatic non-dissipative} transport (part that does not
survive at equilibrium but doesn't contribute to $\D\geq 0$ either) is given by 20 transport coefficients,
\begin{equation}\nn
  \text{Parity Even (13):}\qquad [\b_{[ij]}]_{5\times 5} \ \ \text{(antisymmetric)}, \quad [\k_{[ij]}]_{3\times3} \ \
  \text{(antisymmetric)},
\end{equation}
\begin{equation}
  \text{Parity Odd (7):}\qquad [\tilde\k_{(ij)}]_{3\times 3} \ \ \text{(symmetric)}, \quad \tilde\eta.
\end{equation}
Finally the entire \emph{dissipative} transport is given by $25$ transport coefficients,
\begin{equation}\nn
  \text{Parity Even (22):}\qquad [\b_{(ij)}]_{5\times 5} \ \ \text{(symmetric)}, \quad
  [\k_{(ij)}]_{3\times3} \ \  \text{(symmetric)}, \quad \eta,
\end{equation}
\begin{equation}
  \text{Parity Odd (3):}\qquad [\tilde\k_{[ij]}]_{3\times 3} \ \ \text{(antisymmetric)}.
\end{equation}
These dissipative transport coefficients follow a set of inequalities ($\k'_{ij}$ is defined in \cref{kappa'definition}),
\begin{equation}
  [\b_{(ij)}]_{5\times 5}, \quad [\k'_{(ij)}]_{3\times3}, \quad \eta \quad \geq\quad 0,
\end{equation}
where a `non-negative matrix' implies all its eigenvalues are non-negative. Using $P$, $f_i$, $g_i$
we define some new functions,
\begin{equation}\nn\\
  \df P = S\df T + Q\df \mu + R_n\df\mu_n + R_s \df \mu_s, \qquad
  E + P = ST + Q\mu + R_n\mu_n,
\end{equation}
\begin{equation}\nn
  \df f_i = \frac{\a_{E,i}}{T} \df T + T \a_{R_n,i} \df \nu_n + T \a_{Q,i} \df \nu
  + \lb \a_{R_s,i} - \frac{f_i}{2\hat\mu_s} \rb \df \hat\mu_s, \quad
  \a_{E,i} + f_i = \a_{S,i}T + \a_{Q,i} \mu + \a_{R_n,i} \mu_n,
\end{equation}
\begin{equation}
  \df g_i = \frac{\tilde\a_{E,i}}{T} \df T + T \tilde\a_{R_n,i} \df \nu_n + T \tilde\a_{Q,i} \df \nu
  + \lb \tilde\a_{R_s,i} - \frac{g_i}{2\hat\mu_s} \rb \df \hat\mu_s, \quad
  \tilde\a_{E,i} + g_i = \tilde\a_{S,i}T + \tilde\a_{Q,i} \mu + \tilde\a_{R_n,i} \mu_n.
\end{equation}
In terms of these transport coefficients, corrections to the Josephson equation ($K=0$) coming from
the first order null superfluid transport are given as (here $\b_{55} = \a/T$),
\begin{multline}
 u^\sM \xi_\sM +\mu_n - \mu  = \frac{1}{\b_{55}}\N_\sM (R_s \xi^\sM)
 -\sum_{i=1}^4 \frac{\b_{5i}}{\b_{55}} S_i \\
 + \frac{1}{\b_{55}}\N_\sM \lb \z^\sM \sum_{i=1}^3 \a_{R_s,i} S_{e,i}
  + \z^\sM \sum_{i=1}^2 \tilde\a_{R_s,i} \tilde S_{e,i}
  - \sum_{i=1}^3 f_i V_{e,i}^\sM
  - \sum_{i=1}^2 g_i \tilde V_{e,i}^\sM \rb + \cO(\dow^2),
\end{multline}
which can be seen as determining $u^\sM \xi_\sM$ in terms of the other null superfluid
variables. Note that though this equation contains second order terms, it is only correct up to the
first order in derivatives, and will admit further corrections coming from higher order null
superfluid transport. The energy-momentum tensor, charge current and entropy current up to first
order in derivatives are however given as,
\begin{align}
  T^{\sM\sN}
  &= R_n u^\sM u^\sN + 2 E u^{(\sM}V^{\sN)} + P P^{\sM\sN} + R_s \xi^\sM \xi^\sN + \cT^{\sM\sN} +
    \cO(\dow^2), \nn\\
  J^\sM &= Q u^\sM - R_s \xi^\sM + \cJ^\sM + \cO(\dow^2) \nn\\
  J_S^\sM &= S u^\sM + \cS^\sM + \cO(\dow^2),
\end{align}
where the higher derivative corrections are,
\begin{align}
  \cT^{\sM\sN}
  &= u^\sM u^\sN \Bigg[
    \sum_{i=1}^3 \a_{R_n,i} S_{e,i}
    + \sum_{i=1}^2 \tilde\a_{R_n,i} \tilde S_{e,i}
    - \frac{1}{T}\N_\sR (Tf_3 \z^\sR)
    \Bigg] \nn\\
  &\quad+ 2 V^{(\sM} u^{\sN)} \Bigg[
    \sum_{i=1}^3 \a_{E,i} S_{e,i}
    + \sum_{i=1}^2 \tilde\a_{E,i} \tilde S_{e,i}
    - \frac{1}{T} \N_\sR (Tf_1 \z^\sR)
    \Bigg] \nn\\
  &\quad + 2 \z^{(\sM} u^{\sN)}
    \Bigg[
    \sum_{i=1}^3 \a_{R_s,i} S_{e,i}
    + \sum_{i=1}^2 \tilde\a_{R_s,i} \tilde S_{e,i}
    \Bigg] \nn\\
  &\quad + 2 \z^{(\sM}V^{\sN)}
    \Bigg[
    \sum_{i=1}^3 f_i S_{5+i}
    - (u^\sR \xi_\sR) \lb \sum_{i=1}^3 \a_{R_s,i} S_{e,i}
    + \sum_{i=1}^2 \tilde\a_{R_s,i} \tilde S_{e,i} \rb
    - \sum_{i=1}^5 \b_{2i} S_i
    \Bigg] \nn\\
  &\quad + \z^\sM \z^\sN
    \Bigg[
    \sum_{i=1}^3 \a_{R_s,i} S_{e,i}
    + \sum_{i=1}^2 \tilde\a_{R_s,i} \tilde S_{e,i}
    - \sum_{i=1}^2 \frac{g_i}{2\hat\mu_s} \tilde S_{e,i}
    - \sum_{i=1}^5 \b_{3i} S_i
    \Bigg] \nn\\
  &\quad
    + 2 u^{(\sM} \Bigg[
    - \sum_{i=1}^3 f_i V_{e,i}^{\sN)}
    - \sum_{i=1}^2 g_i \tilde V^{\sN)}_{e,i}
    + P^{\sN)}_{\ \ \sP} \e^{\sP\sK\sR\sS\sT} \N_\sK \lb Tg_1 V_\sR u_\sS \xi_\sT \rb
    \Bigg] \nn\\
  &\quad
    + 2 V^{(\sM} \Bigg[
    (u^\sR \xi_\sR) \sum_{i=1}^3 f_i V_{e,i}^{\sN)}
    - \sum_{i=1}^2 g_i \tilde V^{\sN)}_{e,i+2}
    - \sum_{i=1}^3 \k_{1i} V^{\sN)}_i
    - \sum_{i=1}^3 \tilde\k_{1i} \tilde V^{\sN)}_{i}
    + 3 C^{(4)} \mu^2 B^{\sN)} \nn\\
  &\qquad + P^{\sN)}_{\ \ \sP}  \e^{\sP\sK\sR\sS\sT} \N_\sK \lb Tg_3 V_\sR u_\sS \z_\sT \rb
    + C_1T^2 \o^{\sN)}
    \Bigg]
    + \tilde P^{\sM\sN} \Bigg[
    \sum_{i=1}^3 f_i S_{e,i}
    - \sum_{i=1}^5 \b_{1i} S_i
    \Bigg] 
    \nn\\
  &\quad
    - 2 \z^{(\sM} \Bigg[
    \sum_{i=1}^3 f_i V_{e,i}^{\sN)} + \sum_{i=1}^3 \k_{2i} V^{\sN)}_i
    + \sum_{i=1}^3 \tilde\k_{2i} \tilde V^{\sN)}_{i}
    \Bigg] 
    - \eta \s^{\sM\sN}
    - \tilde\eta \tilde\s^{\sM\sN},
\end{align}
\begin{align}
  \cJ^\sM
  &= u^\sM \Bigg[
    \sum_{i=1}^3 \a_{Q,i} S_{e,i}
    + \sum_{i=1}^2 \tilde\a_{Q,i} \tilde S_{q,i}
    - \frac{1}{T}\N_\sR (Tf_2 \z^\sR)
    \Bigg]
    + P^\sM_{\ \sK}\e^{\sK\sN\sR\sS\sT} \N_\sN \lb Tg_2 V_\sR u_\sS \xi_\sT \rb \nn\\
  &\quad - \z^\sM \Bigg[
    \sum_{i=1}^3 \a_{R_s,i} S_{e,i}
    + \sum_{i=1}^2 \tilde\a_{R_s,i} \tilde S_{e,i}
    + \sum_{i=1}^5 \b_{4i} S_i
    \Bigg] \nn\\
  &\quad
    + \sum_{i=1}^3 f_i V_{e,i}^\sM 
    + \sum_{i=1}^2 g_i \tilde V_{e,i}^\sM
    - \sum_{i=1}^3 \k_{3i} V^{\sM}_i
    - \sum_{i=1}^3 \tilde\k_{3i} \tilde V_{i}^\sM
    + 6 C^{(4)}\mu B^\sM,
\end{align}
\begin{align}
  \cS^\sM
  &= 
  u^\sM \Bigg[
    \sum_{i=1}^3 \a_{S,i} S_{e,i}
    + \sum_{i=1}^2 \tilde\a_{S,i} \tilde S_{e,i}
    - \frac{1}{T^2} \N_\sR (Tf_1 \z^\sR)
    + \frac{\mu_n}{T^2}\N_\sR (Tf_3 \z^\sR)
    + \frac{\mu}{T^2}\N_\sR (Tf_2 \z^\sR)
    \Bigg] \nn\\
  &\quad + \z^\sM
    \sum_{i=1}^5 \frac{\mu \b_{4i} - \b_{2i}}{T} S_i
    + \sum_{i=1}^3 \frac{\mu \k_{3i} - \k_{1i}}{T} V^{\sM}_i
    + \sum_{i=1}^3 \frac{\mu \tilde\k_{3i} - \tilde\k_{1i}}{T} \tilde V_{i}^\sM  \nn\\
  &\quad
    + T g_1 \e^{\sM\sN\sR\sS\sT} V_\sN u_\sR \z_\sS \dow_\sT \nu_n
    + T g_2 \e^{\sM\sN\sR\sS\sT} V_\sN u_\sR \z_\sS \dow_\sT \nu
    + 2 C_1 T \o^\sM \nn\\
  &\quad
    - P^{\sM}_{\ \ \sK} \e^{\sK\sN\sR\sS\sT} \lB
    \frac{\mu_n}{T} \N_\sN \lb Tg_1 V_\sR u_\sS \xi_\sT \rb
    + \frac{\mu}{T} \N_\sN \lb Tg_2 V_\sR u_\sS \xi_\sT \rb
    - \frac{1}{T} \N_\sN \lb Tg_3 V_\sR u_\sS \xi_\sT \rb \rB.
\end{align}
The scalar $S_{5} = T\d_\scrB\vf = u^\sM\xi_\sM + \mu_n -\mu$ appearing here can be eliminated in
favor of $\N_\sM (R_s \xi^\sM)$ using the Josephson equation. We will like to reiterate that these
results are presented in a particular hydrodynamic frame (gained by aligning $u^\sM$, $T$, $\mu_n$,
$\mu$ along $\b^\mu$, $\L_\b$) and in a ``natural'' choice of basis for the independent data. They
can be transformed to any other preferred hydrodynamic frame or basis by a straight forward
substitution.

\section{Null Reduction to Galilean Superfluids}\label{sec4}

We now reduce our null superfluid results to Galilean superfluids. The results are presented in the
covariant Newton-Cartan notation and the conventional non-covariant notation (for superfluids
coupled to flat space-time). For more details on the reduction, please refer \cite{Banerjee:2015hra}.

\subsection{Newton-Cartan Notation}

We start with a quick review of null reduction of null backgrounds to Newton-Cartan backgrounds; for
details see \cite{Banerjee:2015hra}. For an excellent review of Newton-Cartan geometries, please
refer the appendix of \cite{Jensen:2014aia}.

\paragraph*{Background and Hydrodynamic Fields:}

On our null background $\cM_{(d+1)}$, we choose a basis $\{x^\sM\} = \{x^-,x^\mu\}$ such that the
null isometry $\scrV = \{ V = \dow_-, \L_V = 0\}$. The fact that $\scrV$ is an isometry implies that
all the fields in the theory are independent of the $x^-$ coordinate. To perform the reduction, we
require an arbitrary null field $v^\sM$ normalized as $v^\sM v_\sM = 0$, $v^\sM V_\sM = -1$, which
can be interpreted as providing a ``Galilean frame of reference''. In the case of a null
(super)fluid, the null fluid velocity $v^\sM = u^\sM$ defines a special Galilean frame which we
refer to as the ``fluid frame of reference''. In an arbitrary Galilean frame, we decompose the
fields $V^\sM$, $v^\sM$, $g_{\sM\sN}$, $A_\sM$ in the chosen basis as,
\begin{equation}
  V^\sM = \begin{pmatrix} 1 \\ 0 \end{pmatrix}, \quad
  v^\sM = \begin{pmatrix} v^\mu B^{(v)}_\mu \\ v^\mu \end{pmatrix}, \quad
  g_{\sM\sN} = \begin{pmatrix}
    0 & - n_\nu \\
    - n_\mu & h_{\mu\nu} + 2 n_{(\mu} B^{(v)}_{\nu)}
  \end{pmatrix}, \quad
  A_{\sM} = \begin{pmatrix} -1 \\ A_\mu \end{pmatrix},
\end{equation}
along with,
\begin{equation}
  V_\sM = \begin{pmatrix} 0 \\ - n_\mu \end{pmatrix}, \quad
  v_{\sM} = \begin{pmatrix} -1 \\ B^{(v)}_\mu \end{pmatrix}, \quad
  g^{\sM\sN} = \begin{pmatrix}
    h^{\nu\r} B^{(v)}_\nu B^{(v)}_\r - 2 v^\mu B^{(v)}_\mu & h^{\nu\r}B^{(v)}_\r - v^\nu \\
    h^{\mu\nu} B^{(v)}_\nu - v^\mu & h^{\mu\nu}
  \end{pmatrix},
\end{equation}
such that,
\begin{equation}
  n_\mu v^\mu = 1, \qquad 
  v^\mu h_{\mu\nu} = 0, \qquad
  n_\mu h^{\mu\nu} = 0, \qquad
  h_{\mu\r}h^{\r\nu} + n_\mu v^\nu = \d_{\mu}^{\ \nu}.
\end{equation}
The collection of fields $\{n_\mu,v^\mu,h^{\mu\nu},h_{\mu\nu},B^{(v)}_\mu\}$ defines a Newton-Cartan
structure. The condition $\N_\sM V^\sN = 0$ implies that the ``time-metric''
$n = n_\mu \df x^\mu$ is a closed one-form, i.e. $\df n = 0$; this is known to be true for
torsionless Newton-Cartan structures. Note that after choosing the said basis, the residual
diffeomorphisms are $x^\mu \ra x^\mu + \c^\mu(x^\nu)$ and $x^- \ra \x^- + \c^- (x^\mu)$. The former
of these are just the Newton-Cartan diffeomorphisms, while the latter are known as ``mass gauge
transformations''. Only fields that transform under these mass gauge transformations are,
\begin{equation}
  \d_{\c^-} B^{(v)}_\mu = - \dow_\mu \c^-, \qquad
  \d_{\c^-} A_\mu = - \dow_\mu \c^-.
\end{equation}
$B^{(v)}_\mu$ is therefore known as the mass gauge field. On the other hand mass gauge
transformation of $A_\mu$ can be absorbed into its $\rmU(1)$ gauge transformation. We define the
volume element on a Newton-Cartan background as,
\begin{equation}
  \ve^{\mu\nu\r\s}
  = v_\sM\e^{\sM\mu\nu\r\s}
  = - \e^{-\mu\nu\r\s}.
\end{equation}
Note that the volume element is independent of the Galilean frame employed to define it. The
Levi-Civita connection $\G^\sR_{\ \sM\sN}$ decomposes in this basis as,
\begin{align}
  \label{NC_Connection}
  \G^{\l}_{\ \mu\nu}
  &= 
    v^\l \dow_{(\mu} n_{\nu)}
    + \half h^{\l\r} \lb \dow_\mu h_{\r\nu} + \dow_\nu h_{\r\mu} - \dow_\r h_{\mu\nu} \rb
    - \O^{(v)}_{\s(\mu} n_{\nu)} h^{\s\l}, \nn\\
  \G^{-}_{\ \mu\nu}
  &= 
    h_{\l(\mu} \Ndot_{\nu)}v^\l
    - \Ndot_{(\mu} B^{(v)}_{\nu)},
\end{align}
and all the remaining components zero. Here we have identified $\G^{\l}_{\ \mu\nu}$ as the (torsionless)
Newton-Cartan connection and denoted the respective covariant derivative by $\Ndot_\mu$. We have
also defined the (dual) frame vorticity and electromagnetic field strength as,
\begin{equation}
  \O^{(v)}_{\mu\nu} = 2h_{\s[\nu}\Ndot_{\mu]}v^\s = \dow_\mu B^{(v)}_\nu - \dow_\nu B^{(v)}_\mu,
  \qquad
  F_{\mu\nu} = \dow_\mu A_\nu - \dow_\nu A_\mu.
\end{equation}
The covariant derivative $\Ndot$ acts on the Newton-Cartan structure appropriately,
\begin{equation}
  \Ndot_\mu n_\nu = 0, \qquad
  \Ndot_\mu h^{\r\s} = 0, \qquad
  \Ndot_\mu h_{\nu\r} = - 2 n_{(\nu} h_{\r)\s} \Ndot_\mu v^\s.
\end{equation}
Note that $v^\sM$ was an arbitrary field chosen to perform the reduction, and one is allowed to
arbitrarily redefine it without changing the physics. This leads to the invariance of the system
under ``Milne transformations'' of the Newton-Cartan structure,
\begin{equation}\label{milne.trans}
  v^\mu \ra v^\mu + \p^\mu, \quad
  h_{\mu\nu} \ra h_{\mu\nu} - 2 n_{(\mu} \p_{\nu)} + n_\mu n_\nu \p^\r \p_\r, \quad
  B^{(v)}_\mu \ra B^{(v)}_\mu + \p_\mu - \half n_\mu \p^\r \p_\r,
\end{equation}
where $\p^\mu n_\mu = 0$, $\p_\mu = h_{\mu\nu}\p^\nu$. The fields $n_\mu$, $h^{\mu\nu}$,
$\G^\r_{\ \mu\nu}$ and $\ve^{\mu\nu\r\s}$ are Milne invariant. We can now decompose the
fluid velocity $u^\sM$ and the associated projector $P^{\sM\sN}$ as,
\begin{equation}
  u^\sM = \begin{pmatrix} u^\mu B_\mu \\ u^\mu \end{pmatrix}, \quad
  u_{\sM} = \begin{pmatrix} -1 \\ B_\mu \end{pmatrix}, \quad
  P_{\sM\sN} = \begin{pmatrix}
    0 & 0 \\
    0 & p_{\mu\nu}
  \end{pmatrix}, \quad
  P^{\sM\sN} = \begin{pmatrix}
    p^{\nu\r}B_\nu B_\r & p^{\mu\nu}B_\nu \\
    p^{\mu\nu} B_\nu & p^{\mu\nu}
  \end{pmatrix}.
\end{equation}
The fields $\{n_\mu, u^\mu, p^{\mu\nu},p_{\mu\nu},B_\mu\}$ define the Newton-Cartan structure
in the fluid frame of reference, satisfying,
\begin{equation}
  n_\mu u^\mu = 1, \qquad 
  u^\mu p_{\mu\nu} = 0, \qquad
  n_\mu p^{\mu\nu} = 0, \qquad
  p_{\mu\r}p^{\r\nu} + n_\mu u^\nu = \d_{\mu}^{\ \nu}.
\end{equation}
They can be re-expressed in terms of $\{n_\mu, v^\mu, h^{\mu\nu},h_{\mu\nu},B^{(v)}_\mu\}$ using
\cref{milne.trans} with $\p^\mu = \bar u^\mu = h^\mu{}_\nu u^\nu = u^\mu - v^\mu$,
\begin{equation}
  p^{\mu\nu} = h^{\mu\nu}, \quad
  p_{\mu\nu} = h_{\mu\nu} - 2 n_{(\mu} \bar u_{\nu)} + n_\mu n_\nu \bar u^\r \bar u_\r, \quad
  B_\mu = B^{(v)}_\mu + \bar u_\mu - \half n_\mu \bar u^\r \bar u_\r.
\end{equation}
The (dual) fluid vorticity is defined similar to the (dual) frame vorticity as,
\begin{equation}
  \O_{\mu\nu} = 2p_{\s[\nu}\Ndot_{\mu]}u^\s = \dow_\mu B_\nu - \dow_\nu B_\mu.
\end{equation}
For later use, we define the magnetic field and fluid vorticity,
\begin{equation}
  B^\mu = \half\ve^{\nu\r\s\mu} n_\nu F_{\r\s}, \qquad
  \o^\mu = \half\ve^{\nu\r\s\mu} n_\nu \O_{\r\s}.
\end{equation}
Finally the superfluid velocity can be decomposed as,
\begin{equation}
  \z^{\sM} = \begin{pmatrix} B_\mu \z^\mu \\ \z^\mu \end{pmatrix}, \qquad
  \xi^{\sM} = \begin{pmatrix} \mu_s + \half p_{\mu\nu} \z^\mu\z^\nu + B_\mu \xi^\mu \\ \xi^\mu =
    \z^\mu + u^\mu \end{pmatrix},
\end{equation}
where $\xi^\mu n_\mu = 1$, $\z^\mu n_\mu = 0$. We have treated the superfluid potential $\mu_s$ as
an independent component of $\xi^\sM$. The hatted superfluid potential is however given as
$\hat\mu_s = -\half \z^\mu\z_\mu$.  Decomposition of the projector $\tilde P^{\sM\sN}$ on the other
hand is,
\begin{equation}
  \tilde P_{\sM\sN} = \begin{pmatrix}
    0 & 0 \\
    0 & \tilde p_{\mu\nu} = p_{\mu\nu} - \frac{\z_\mu \z_\nu}{p^{\r\s}\z_\r\z_\s}
  \end{pmatrix}, \quad
  \tilde P^{\sM\sN} = \begin{pmatrix}
    \tilde p^{\nu\r} B_\nu B_\r & \tilde p^{\mu\nu}B_\nu \\
    \tilde p^{\mu\nu} B_\nu & \tilde p^{\mu\nu} = p^{\mu\nu} - \frac{\z^\mu \z^\nu}{p^{\r\s}\z_\r\z_\s}
  \end{pmatrix}.
\end{equation}



\paragraph*{Currents and Conservation:} The mass current $\r^\mu$, energy current $\e^\mu$, stress
tensor $t^{\mu\nu}$, charge
current $j^\mu$ and entropy current $s^\mu$ on Newton-Cartan backgrounds can be respectively read
out in terms of $T^{\sM\sN}$, $J^\sM$, $J_S^\sM$ as \cite{Banerjee:2015hra},
\begin{equation}\label{current.redn}
  \r^\mu = -T^{\mu\sM}V_\sM, \qquad
  \e^\mu = -T^{\mu\sM}u_\sM, \qquad
  t^{\mu\nu} = P^{\mu}_{\ \sM} P^{\nu}_{\ \sN} T^{\sM\sN}, \qquad
  j^\mu = J^\mu, \qquad
  s^\mu = J_S^\mu,
\end{equation}
with  $t^{\mu\nu} = t^{\nu\mu}$ and $t^{\mu\nu}n_\nu = 0$. They satisfy the conservation laws and
the second law of thermodynamics,
\begin{align}
  \text{Mass Conservation:} \qquad\qquad\qquad \ \Ndot_\mu \r^{\mu}
  &= 0, \nn\\
  \text{Energy Conservation:}  \qquad\qquad\qquad \ \Ndot_\mu \e^\mu
  &= - u^\nu F_{\nu\r} j^\r
    - \lb u^\mu \r^{\s} + t^{\mu\s} \rb p_{\s\nu} \Ndot_\mu u^\nu
    - \rmT_\rmH{}^{\perp}_{\ \nu} u^\nu, \nn\\
  \text{Momentum Conservation:} \ \ \Ndot_\mu (u^\mu p^{\s}_{\ \nu} \r^{\nu} + t^{\mu\s})
  &= p^{\s\nu} F_{\nu\r} j^\r - \r^\mu \Ndot_{\mu}u^\s +
    \rmT_\rmH{}^{\perp}_{\ \nu} p^{\s\nu}, \nn\\
  \text{Charge Conservation:}  \qquad\qquad\qquad \ \Ndot_\mu j^\mu &= \rmJ^\perp_\rmH, \nn\\
  \text{Second Law of Thermo.:}  \qquad\qquad\qquad \ \Ndot_\mu s^\mu &\geq 0.
\end{align}
The energy current $\e^\mu$ and the stress tensor $t^{\mu\nu}$ in \cref{current.redn} are defined in
the fluid frame of reference; we can define the respective quantities in an arbitrary frame of
reference,
\begin{align}
  \e^\mu_{(v)}
  &= -T^{\mu\sM}v_\sM
    = \e^\mu
    + u^\mu \bar u^\nu p_{\nu\r} \r^{\r}
    + \half \r^\mu \bar u^\r \bar u_\r
    + t^{\mu\nu} \bar u_\nu, \nn\\
  t^{\mu\nu}_{(v)}
  &= (P_{(v)})^{\mu}_{\ \sM} (P_{(v)})^{\nu}_{\ \sN} T^{\sM\sN}
    = t^{\mu\nu} + 2 \bar u^{(\mu}h^{\nu)}_{\ \ \s} \r^{\s} - \bar u^\mu \bar u^\nu
    \r^\s n_\s,
\end{align}
where $P_{(v)}^{\sM\sN} = g^{\sM\sN} + 2v^{(\sM}V^{\sN)}$. They satisfy the conservation laws,
\begin{align}
  \Ndot_\mu \e_{(v)}^\mu
  &= - v^\nu F_{\nu\r} j^\r
    - \lb v^\mu \r^{\s} + t_{(v)}^{\mu\s} \rb h_{\s\nu} \Ndot_\mu v^\nu
    - \rmT_\rmH{}^{\perp}_{\ \nu} v^\nu \nn\\
  \Ndot_\mu (v^\mu h^{\s}_{\ \nu} \r^{\nu} + t^{\mu\s}_{(v)})
  &= h^{\s\nu} F_{\nu\r} j^\r - \r^\mu \Ndot_{\mu}v^\s +
    \rmT_\rmH{}^{\perp}_{\ \nu} h^{\s\nu}.
\end{align}


\paragraph*{Galilean Superfluid Constitutive Relations:}

\begin{table}[p]
  \centering
  \begin{tabular}[t]{|c|c||c|c|}
    \hline
    \multicolumn{2}{|c||}{Newton-Cartan Data} & \multicolumn{2}{c|}{Non-Covariant Data} \\
    \hline\hline

    
    \multicolumn{4}{|c|}{Vanishing at Equilibrium -- Onshell Independent} \\
    \hline
    $S_1$ & $\tilde p^{\mu}_{\ \nu}\Ndot_{\mu}u^{\nu}$
          & $S_1$ & $\tilde p^{ij} \dow_i u_j$\\
    $S_2,S_{e,1}$ & $\frac{1}{T}\z^\mu \dow_\mu T$
          & $S_2,S_{e,1}$ & $ \frac{1}{T}\z^i \dow_i T$\\
    $S_3$ & $\z^\mu\z_{\nu}\Ndot_{\mu}u^{\nu}$
          & $S_3$ &$\z^i \z_j \dow_i u^j$\\
    $S_4$ & $\z^\mu \lb T \dow_\mu \nu + u^\nu F_{\nu\mu} \rb$
          & $S_4$ & $ \z^i(T \dow_i\nu - e_i + u^j\beta_{ji})$\\
    $S_5$ & $- \half \z^\mu \z_\mu - \mu_s + \mu_n - \mu$ & $S_5$
          & $- \half \z^k \z_k - \mu_s +\mu_n-\mu$\\
    \hline

    
    $V^\mu_1,V^\mu_{e,1}$ &$ \frac{1}{T} \tilde p^{\mu\nu} \dow_\nu T$
          & $V^i_1,V^i_{e,1}$ & $\frac{1}{T}\tilde p^{ij}\dow_jT$ \\
    $V^\mu_2$ & $2 \tilde  p^{\mu\nu}\z^\s p_{\r(\s}\Ndot_{\nu)}u^{\r}$&
   $V^i_2$ & $\tilde p^{ij}\z^k \dow_{(j}u_{k)}$\\
    $V^\mu_3$ &$\tilde p^{\mu\nu}\lb T \dow_\nu \nu + u^\r F_{\r\nu} \rb$
          & $V^i_3$ &  $\tilde p^{ij} (T \dow_j \nu - e_j + u^k \beta_{kj})$\\
    \hline

    
    $\s^{\mu\nu}$ & $\tilde p^{\mu\r}\tilde p^{\nu\s}\lb p_{\t(\r}\Ndot_{\s)}u^{\t}
                          - \frac{\tilde p_{\r\s}}{d-1} S_1 \rb $
          & $\s^{ij}$ & $\tilde p^{ik}\tilde p^{jl}(\dow_{(k}u_{l)}
                               + \frac{\tilde p_{kl}}{d-1} S_1 )$\\
    \hline

    
    $\tilde V_{1}^\mu$ & $-\ve^{\mu\nu\rho\sigma}n_\nu\z_\rho V_{1,\sigma}$
          &$\tilde V_{1}^i$ & $\ve^{ijk} \z_j V_{1,k}$ \\
    $\tilde V_{2}^\mu$ & {$-\ve^{\mu\nu\rho\sigma}n_\nu\z_\rho V_{2,\sigma} $}
          &$\tilde V_{2}^i$ & $\ve^{ijk} \z_j V_{2,k}$\\
    $\tilde V_{3}^\mu$ & {$-\ve^{\mu\nu\rho\sigma}n_\nu\z_\rho V_{3,\sigma} $}
          &$\tilde V_{3}^i$ & $\ve^{ijk}\z_j V_{3,k}$\\
    
    \hline


    $\tilde\s^{\mu\nu}$ & $-\ve^{(\mu|\r\s\t}n_\r\z_\s \s_{\t}^{\ \nu)}$
          & $\tilde\s^{ij}$ & $\ve^{(i|kl} \z_k \s_{l}^{\ j)}$ \\
    \hline\hline

    
    \multicolumn{4}{|c|}{Vanishing at Equilibrium -- Onshell Dependent} \\
    \hline
    
    $S_6$ & $\frac{1}{T} u^\mu \dow_\mu T$
          &$S_6$ & $\frac{1}{T}(\dow_tT+ u^i \dow_i T)$\\
    $S_7$ & $T u^\mu \dow_\mu \nu$& $S_7$ & $T (\dow_t\nu+ u^i \dow_i \nu)$\\
    $S_8$ & $T u^\mu \dow_\mu \nu_n$& $S_8$ & $T (\dow_t\nu_n+ u^i \dow_i \nu_n)$\\
    $S_9$ & $\z^\mu \lb T \dow_\mu \nu_n + u^\nu p_{\r\mu}\Ndot_{\nu}u^{\r} \rb$
          & $S_9$ & $\z^i(T \dow_i \nu_n+ \dow_t u_i + u^j \dow_j u_i)$\\
    \hline

    
    $V_4^\mu$ & $\tilde P^{\mu\nu}\lb T \dow_\nu \nu_n + u^\s p_{\r\nu}\Ndot_{\s}u^{\r} \rb$
          & $V_4^i$ & $\tilde p^{ij}(T\dow_j \nu_n+\dow_t u_j + u^k \dow_k u_j)$\\
    \hline
    $\tilde V_{4}^\mu$ & {$-\ve^{\mu\nu\rho\sigma}n_\nu\z_\rho V_{4,\sigma} $}
          & $\tilde V_4^i$ & $\ve^{ijk} \z_j V_{4,k}$\\
    \hline\hline
    
    
    \multicolumn{4}{|c|}{Surviving at Equilibrium} \\
    \hline
    $S_{e,2}$ &$T\z^\mu \dow_\mu \nu$& $S_{e,2}$ & $T \z^i \dow_i \nu$\\
    $S_{e,3}$ &$T\z^\mu \dow_\mu \nu_n$& $S_{e,3}$ & $T \z^i \dow_i \nu_n$\\
    $\vdots$ & $\vdots$ & $\vdots$ & $\vdots$ \\
    \hline

    
    $V^\mu_{e,2}$ & $T \tilde p^{\mu\nu} \dow_\nu \nu$
          & $V^i_{e,2}$ & $T\tilde p^{ij}\dow_j\nu$\\
    $V^\mu_{e,3}$ &$T \tilde P^{\mu\nu} \dow_\nu \nu_n$
          & $V^i_{e,3}$ & $T\tilde p^{ij}\dow_j\nu_n$\\
    $\vdots$ & $\vdots$ & $\vdots$ & $\vdots$ \\
    \hline

    
    $\tilde S_{e,1}$ & $ T\ve^{\mu\nu\rho\s} n_\mu \z_\nu \dow_{\rho}B_\s$
          & $\tilde S_{e,1}$ & $T \ve^{ijk} \z_i \dow_j u_k$\\
    $\tilde S_{e,2}$ & $ \frac{T}{2}\ve^{\mu\nu\rho\sigma} n_\mu  \z_\nu  F_{\rho\sigma}$
          & $\tilde S_{e,2}$ & $\frac{T}{2}\ve^{ijk}\z_i\beta_{jk}$\\
    $\vdots$ & $\vdots$ & $\vdots$ & $\vdots$ \\
    \hline

    
    $\tilde V^\mu_{e,1}$ &$- T\tilde p^\mu_{\ \t}  \ve^{\t\nu\r\s} n_\nu \dow_\rho B_\s$
          & $\tilde V^i_{e,1}$ & $T \tilde p^i_{\ l} \ve^{ljk}\dow_j u_k$\\
    $\tilde V^\mu_{e,2}$ &$- \frac{T}{2} \tilde p^\mu_{\ \t}  \ve^{\t\nu\r\s} n_\nu F_{\r\s}$
          & $\tilde V^i_{e,2}$ & $ \frac{T}{2} \tilde p^i_{\ l} \ve^{ljk} \beta_{jk}$\\
    
    $\tilde V^\mu_{e,3}$ &$T\tilde p^\mu_{\ \t}\ve^{\t\nu\r\s} \z_\nu \dow_\r B_\s
                           + (\mu_s + \half \z^\mu\z_\mu) \tilde V^\mu_{e,1} $
          & $\tilde V^i_{e,3}$ &
                                 \begin{minipage}[t]{5cm}
                                   \centering
                                   $- T \lb u^i \ve^{jkl}\z_j\dow_ku_l-\ve^{ijk}\z_j\dow_tu_k\rb$
                                   $+ (\mu_s + \half \z^k\z_k) \tilde V^i_{e,1}$
                                 \end{minipage} \\
    
    $\tilde V^\mu_{e,4}$ & $\frac{T}{2}\tilde p^\mu_{\ \t}\ve^{\t\nu\r\s} \z_\nu F_{\r\s}
                           + (\mu_s + \half \z^\mu\z_\mu) \tilde V^\mu_{e,2} $
          &$\tilde V^i_{e,4}$ &
                                \begin{minipage}[t]{5cm}
                                  \centering
                                  $- T \lb u^i \half\ve^{jkl}\z_j\beta_{kl}+\ve^{ijk} \z_j e_k\rb$
                                  $+ (\mu_s + \half \z^k\z_k) \tilde V^i_{e,2}$
                                \end{minipage} \\
    $\vdots$ & $\vdots$ & $\vdots$ & $\vdots$ \\
    \hline
  \end{tabular} 
  \caption{\label{data.Galilean} Independent null superfluid data at the first order in
    derivatives. Note that we have not, neither do we need to, enlist all the independent data that
    survives in equilibrium; the ones listed here are the only ones we use in the null superfluid
    constitutive relations.}
\end{table}

Finally, by a direct computation we can find that the Galilean superfluid constitutive relations
in the fluid frame take a structural form.
\begin{align}
  \r^\mu &= \r u^\mu + R_s \xi^\mu + \vs_\r^\mu, \nn\\
  \e^\mu &= \e u^\mu + R_s \lb \half \z^\mu\z_\mu + \mu_s \rb \xi^\mu + \vs_\e^\mu, \nn\\
  t^{\mu\nu} &= P p^{\mu\nu} + R_s \z^\mu \z^\nu + \vs_t^{\mu\nu}, \nn\\
  j^\mu &= q u^\mu - R_s \xi^\mu + \vs_q^\mu, \nn\\
  s^\mu &= s u^\mu + \vs_s^\mu.
\end{align}
While in an arbitrary frame, energy current and stress tensor are given as,
\begin{align}
  \e^\mu_{(v)}
  &= 
    u^\mu \lb \e + \half \r \bar u^2 + \vs_\r^\s \bar u_\s \rb
    + R_s \xi^\mu \lb  \half \bar \xi^2 + \mu_s \rb
    + P \bar u^\mu
    + \lb \vs_\e^\mu + \half \vs_\r^\mu \bar u^2 + \vs_s^{\mu\r} \bar u_\r \rb, \nn\\
  t^{\mu\nu}_{(v)} &= \r \bar u^\mu \bar u^\nu + R_s \bar\xi^\mu \bar\xi^\nu + P h^{\mu\nu}  +
                       \lb \vs_s^{\mu\nu} + 2\vs_\r^{(\mu} \bar u^{\nu)} \rb,
\end{align}
where $\bar u^\mu = h^{\mu}_{\ \nu} u^\nu = u^\mu - v^\mu$ and
$\bar \xi^\mu = h^{\mu}_{\ \nu} \xi^\nu = \xi^\mu - v^\mu$. Various quantities appearing in the
constitutive relations can be found via reduction as: fluid densities,
\begin{align}
  \r
  &= R_n
    + \sum_{i=1}^3 \a_{R_n,i} S_{e,i}
    + \sum_{i=1}^2 \tilde\a_{R_n,i} \tilde S_{e,i}
    - \frac{1}{T}\Ndot_\r (Tf_3 \z^\r), \nn\\
  \e
  &= E
    + \sum_{i=1}^3 \a_{E,i} S_{e,i}
    + \sum_{i=1}^2 \tilde\a_{E,i} \tilde S_{e,i}
    - \frac{1}{T} \Ndot_\r (Tf_1 \z^\r) \nn\\
  q
  &= Q
    + \sum_{i=1}^3 \a_{Q,i} S_{e,i}
    + \sum_{i=1}^2 \tilde\a_{Q,i} \tilde S_{q,i}
    - \frac{1}{T}\Ndot_\r (Tf_2 \z^\r), \nn\\
  s
  &= S 
    + \sum_{i=1}^3 \a_{S,i} S_{e,i}
    + \sum_{i=1}^2 \tilde\a_{S,i} \tilde S_{e,i}
    - \frac{1}{T^2} \Ndot_\r (Tf_1 \z^\r)
    + \frac{\mu_n}{T^2}\Ndot_\r (Tf_3 \z^\r)
    + \frac{\mu}{T^2}\Ndot_\r (Tf_2 \z^\r).
\end{align}
and dissipative currents,
\begin{align}
  \vs_\r^\mu
  &= \z^{\mu}
    \Bigg[
    \sum_{i=1}^3 \a_{R_s,i} S_{e,i}
    + \sum_{i=1}^2 \tilde\a_{R_s,i} \tilde S_{e,i} \Bigg]
    - \sum_{i=1}^3 f_i V_{e,i}^{\mu}
    - \sum_{i=1}^2 g_i \tilde V^{\mu}_{e,i}
    + \ve^{\mu\nu\r\s} \dow_\nu \lb Tg_1 n_\r \z_\s \rb, \nn\\
  \vs_\e^\mu
  &= \z^{\mu}
    \Bigg[
    \sum_{i=1}^3 f_i S_{5+i}
    + (\mu_s + \half \z^\mu \z_\mu) \lb \sum_{i=1}^3 \a_{R_s,i} S_{e,i}
    + \sum_{i=1}^2 \tilde\a_{R_s,i} \tilde S_{e,i} \rb
    - \sum_{i=1}^5 \b_{2i} S_i
    \Bigg] \nn\\
  &\quad
    + (\mu_s + \half \z^\mu \z_\mu) \sum_{i=1}^3 f_i V_{e,i}^{\mu}
    - \sum_{i=1}^2 g_i \tilde V^{\mu}_{e,i+2}
    - \sum_{i=1}^3 \k_{1i} V^{\mu}_i
    - \sum_{i=1}^3 \tilde\k_{1i} \tilde V^{\mu}_{i}
    + 3 C^{(4)} \mu^2 B^{\mu} \nn\\
  &\quad
    + \ve^{\mu\nu\r\s} \dow_\nu \lb Tg_3 n_\r \z_\s \rb
    + C_1T^2 \o^{\mu}
    , \nn\\
  \vs_t^{\mu\nu}
  &= \z^\mu \z^\nu
    \Bigg[
    \sum_{i=1}^3 \a_{R_s,i} S_{e,i}
    + \sum_{i=1}^2 \tilde\a_{R_s,i} \tilde S_{e,i}
    - \sum_{i=1}^2 \frac{g_i}{2\hat\mu_s} \tilde S_{e,i}
    - \sum_{i=1}^5 \b_{3i} S_i
    \Bigg]
    - \eta \s^{\mu\nu}
    - \tilde \eta \tilde\s^{\mu\nu} \nn\\
  &\quad
    - 2 \z^{(\mu} \Bigg[
    \sum_{i=1}^3 f_i V_{e,i}^{\nu)} + \sum_{i=1}^3 \k_{2i} V^{\nu)}_i
    + \sum_{i=1}^3 \tilde\k_{2i} \tilde V^{\nu)}_{i}
    \Bigg] 
    + \tilde p^{\mu\nu} \Bigg[
    \sum_{i=1}^3 f_i S_{e,i}
    - \sum_{i=1}^5 \b_{1i} S_i
    \Bigg], \nn\\
  \vs_q^\mu
  &= - \z^\mu \Bigg[
    \sum_{i=1}^3 \a_{R_s,i} S_{e,i}
    + \sum_{i=1}^2 \tilde\a_{R_s,i} \tilde S_{e,i}
    + \sum_{i=1}^5 \b_{4i} S_i
    \Bigg]
    + \ve^{\mu\nu\r\s} \dow_\nu \lb Tg_2 n_\r \z_\s \rb \nn\\
  &\quad
    + \sum_{i=1}^3 f_i V_{e,i}^\mu
    + \sum_{i=1}^2 g_i \tilde V_{e,i}^\mu
    - \sum_{i=1}^3 \k_{3i} V^{\mu}_i
    - \sum_{i=1}^3 \tilde\k_{3i} \tilde V_{i}^\mu
    + 6 C^{(4)} \mu B^\mu, \nn\\
  \vs_s^\mu
  &= \z^\mu \sum_{i=1}^5 \frac{\mu \b_{4i} - \b_{2i}}{T} S_i
    - \ve^{\mu\nu\r\s}  \lB
    \frac{\mu_n}{T} \dow_\nu \lb Tg_1 n_\r \z_\s \rb
    + \frac{\mu}{T} \dow_\nu \lb Tg_2 n_\r \z_\s \rb
    - \frac{1}{T} \dow_\nu \lb Tg_3 n_\r \z_\s \rb
    \rB\nn\\
  &\quad
    + \sum_{i=1}^3 \frac{\mu \k_{3i} - \k_{1i}}{T} V^{\mu}_i
    + \sum_{i=1}^3 \frac{\mu \tilde\k_{3i} - \tilde\k_{1i}}{T} \tilde V_{i}^\mu
    - T g_1 \ve^{\mu\nu\r\s} n_\nu \z_\r \dow_\s \nu_n
    - T g_2 \ve^{\mu\nu\r\s} n_\nu \z_\r \dow_\s \nu \nn\\
  &\quad + 2 C_1 T \o^\mu.
\end{align}
In addition, we also have the Josephson equation,
\begin{multline}
  - \half \z^\mu\z_\mu - \mu_s + \mu_n - \mu = \frac{1}{\b_{55}}\Ndot_\mu (R_s \xi^\mu)
 -\sum_{i=1}^4 \frac{\b_{5i}}{\b_{55}} S_i \\
 + \frac{1}{\b_{55}}\Ndot_\mu \lb \z^\mu \sum_{i=1}^3 \a_{R_s,i} S_{e,i}
  + \z^\mu \sum_{i=1}^2 \tilde\a_{R_s,i} \tilde S_{e,i}
  - \sum_{i=1}^3 f_i V_{e,i}^\mu
  - \sum_{i=1}^2 g_i \tilde V_{e,i}^\mu \rb,
\end{multline}
which is the derivative correction of the ideal order version $\mu_s = - \half \z^\mu\z_\mu - \mu +
\mu_n$. This completes our discussion of the first order Galilean (Newton-Cartan) superfluids;
counting of various transport coefficients appearing in the constitutive relations is same as the
null superfluid given in \cref{null-summary}. 

\subsection{Non-Covariant Notation (for Flat Spacetime)}

If the superfluid is coupled to a flat Galilean spacetime, it is fitting to re-express the results
in the conventional non-covariant notation where we treat the time and space indices distinctly. It
might help the reader to better relate the Galilean superfluid constitutive relations to the
existing Galilean literature, e.g. in \cite{landau1959fluid}.

\paragraph*{Background and Hydrodynamic Fields:} On the Newton-Cartan background, we choose a basis
$\{x^\mu\} = \{t,x^i\}$ such that the Galilean frame velocity $(v^\mu) = \dow_t$. A flat Galilean
background is defined by a particular choice of the Newton-Cartan structure in this basis,
\begin{equation}
  n_\mu = \begin{pmatrix} 1 \\ 0 \end{pmatrix}, \quad
  v^\mu = \begin{pmatrix} 1 \\ 0 \end{pmatrix}, \quad
  p^{\mu\nu} = \begin{pmatrix}
    0 & 0 \\ 0 & \d^{ij}
  \end{pmatrix}, \quad
  p_{\mu\nu} = \begin{pmatrix}
    0 & 0 \\ 0 & \d_{ij}
  \end{pmatrix}, \quad
  B^{(v)}_\mu = 0,
\end{equation}
where $\d^{ij} = \d_{ij}$ is the Kronecker delta. It can be checked that the respective
Newton-Cartan connection $\G^{\l}_{\ \mu\nu} = 0$, justifying the spacetime to be flat. The
Newton-Cartan structure in the fluid frame can be worked out from here to be,
\begin{equation}
  u^\mu = \begin{pmatrix} 1 \\ u^i \end{pmatrix}, \quad
  B_\mu = \begin{pmatrix} -\half u^k u_k \\ u_i \end{pmatrix}, \quad
  p^{\mu\nu} = \begin{pmatrix}
    0 & 0 \\ 0 & \d^{ij}
  \end{pmatrix}, \quad
  p_{\mu\nu} = \begin{pmatrix}
    u^k u_k & - u_j \\ - u_i & \d_{ij}
  \end{pmatrix}.
\end{equation}
We define the spatial volume element,
\begin{equation}
  \ve^{ijk} = n_\mu\ve^{\mu ijk} = \ve^{tijk}.
\end{equation}
The $\rmU(1)$ gauge field $A_\mu$ can be decomposed as $A_\mu \df x^\mu = A_t \df t + A_i \df
x^i$. The fluid vorticity and electromagnetic field strength on the other hand can be decomposed as,
\begin{equation}
  \O_{\mu\nu} = \begin{pmatrix}
    0 & (\dow_t + u^k \dow_k ) u_i + \o_{ik} u^k \\
    - (\dow_t + u^k \dow_k ) u_i - \o_{ik} u^k & \o_{ij} = \dow_i u_j - \dow_j u_i
  \end{pmatrix},
\end{equation}
\begin{equation}
  F_{\mu\nu} = \begin{pmatrix}
    0 & - e_i = \dow_t A_i - \dow_i A_t \\
    e_i = - \dow_t A_i + \dow_i A_t & \b_{ij} = \dow_i A_j - \dow_j A_i
  \end{pmatrix},
\end{equation}
where $\o_{ij}$ is the (dual) spatial vorticity, $e_i$ is the electric field and $\b_{ij}$ is the
dual magnetic field. For later use, we define the magnetic field and fluid vorticity,
\begin{equation}
  B^i = \half\ve^{ijk} \b_{jk}, \qquad
  \o^i = \half\ve^{ijk} \o_{jk}.
\end{equation}
Finally the superfluid velocity can be decomposed as,
\begin{equation}
  \z^\mu = \begin{pmatrix} 0 \\ \z^i \end{pmatrix}, \qquad
  \xi^\mu = \begin{pmatrix} 1 \\ \xi^i = u^i + \z^i \end{pmatrix}, \qquad
  \mu_s = - \xi_t - \half \xi^i\xi_i, \qquad
  \hat\mu_s = - \half \z^i \z_i,
\end{equation}
with the projection operators,
\begin{equation}
  \tilde p_{\mu\nu} = \begin{pmatrix}
    u^k u_k & - u_j \\
    - u_i & \tilde p_{ij} = \d_{ij} - \frac{\z_i \z_j}{\z^k\z_k}
  \end{pmatrix}, \quad
  \tilde p^{\mu\nu} = \begin{pmatrix}
    0 & 0 \\
    0 & \tilde p^{ij} = \d^{ij} - \frac{\z^i \z^j}{\z^k\z_k}
  \end{pmatrix}.
\end{equation}

\paragraph*{Densities, Currents and Conservation Laws:}

In flat spacetime, the conservation laws and the second law of thermodynamics take the well known
form,
\begin{align}
  \text{Mass Conservation:} \qquad \dow_t \r^t + \dow_i \r^i
  &= 0 \nn\\
  \text{Energy Conservation:} \ \ \dow_t \e_{(v)}^t + \dow_i \e_{(v)}^i
  &= j^i e_i - \rmT_\rmH{}^{\perp}_{\ t} \nn\\
  \text{Momentum Conservation:} \quad \ \dow_t \r^{j}
  + \dow_i t^{ij}_{(v)}
  &= \lb e^j j^t + \b^{jk} j_k \rb +
    \rmT_\rmH{}^{\perp j}, \nn\\
  \text{Charge Conservation:} \qquad \dow_t j^t + \dow_i j^i &= \rmJ^\perp_\rmH, \nn\\
  \text{Second Law of Thermodynamics:} \qquad \dow_t s^t + \dow_i s^i &\geq 0,
\end{align}
where we have identified various Galilean quantities: mass density $\r^t$, mass current $\r^i$,
energy density $\e^t_{(v)}$, energy current $\e^i_{(v)}$, stress tensor $t^{ij}_{(v)}$, charge
density $j^t$, charge current $j^i$, entropy density $s^t$ and entropy current $s^i$.


\paragraph*{Superfluid Constitutive Relations:}

Finally, we can read out the structural form of the Galilean superfluid constitutive relations in
non-covariant notation using reduction,
\begin{align}
  \r^t
  &= \r + R_s, \qquad
  \r^i
  = \r u^i + R_s \xi^i + \vs_\r^i, \nn\\
  \e^t_{(v)}
  &= 
    \e + R_s \mu_s + \half \r \vec u^2
    + \half R_s \vec \xi^2  + \vs_\r^i u_i, \nn\\
  \e^i_{(v)}
  &= 
    u^i \lb \e + P + \half \r \bar u^2 + \vs_\r^j u_j \rb
    + R_s \xi^i \lb  \half \bar \xi^2 + \mu_s \rb
    + \lb \vs_\e^i + \half \vs_\r^i \bar u^2 + \vs_s^{ij} u_j \rb, \nn\\
  t^{ij}_{(v)} &= \r u^i u^j + R_s \xi^i \xi^j + P \d^{ij}  +
                       \lb \vs_s^{ij} + 2\vs_\r^{(i} u^{j)} \rb, \nn\\
  j^t &= q - R_s, \qquad
        j^i = q u^i - R_s \xi^i + \vs_q^i, \nn\\
  s^t &= s, \qquad
  s^i = s u^i + \vs_s^i.
\end{align}
Various quantities appearing here can also be worked out using reduction: fluid densities,
\begin{align}
  \r
  &= R_n
    + \sum_{i=1}^3 \a_{R_n,i} S_{e,i}
    + \sum_{i=1}^2 \tilde\a_{R_n,i} \tilde S_{e,i}
    - \frac{1}{T}\dow_i (Tf_3 \z^i), \nn\\
  \e
  &= E
    + \sum_{i=1}^3 \a_{E,i} S_{e,i}
    + \sum_{i=1}^2 \tilde\a_{E,i} \tilde S_{e,i}
    - \frac{1}{T} \dow_i (Tf_1 \z^i) \nn\\
  q
  &= Q
    + \sum_{i=1}^3 \a_{Q,i} S_{e,i}
    + \sum_{i=1}^2 \tilde\a_{Q,i} \tilde S_{q,i}
    - \frac{1}{T}\dow_i (Tf_2 \z^i), \nn\\
  s
  &= S 
    + \sum_{i=1}^3 \a_{S,i} S_{e,i}
    + \sum_{i=1}^2 \tilde\a_{S,i} \tilde S_{e,i}
    - \frac{1}{T^2} \dow_i (Tf_1 \z^i)
    + \frac{\mu_n}{T^2}\dow_i (Tf_3 \z^i)
    + \frac{\mu}{T^2}\dow_i (Tf_2 \z^i),
\end{align}
and dissipative currents,
\begin{align}
  \vs_\r^i
  &= \z^{i}
    \Bigg[
    \sum_{i=1}^3 \a_{R_s,i} S_{e,i}
    + \sum_{i=1}^2 \tilde\a_{R_s,i} \tilde S_{e,i} \Bigg]
    - \sum_{i=1}^3 f_i V_{e,i}^{i}
    - \sum_{i=1}^2 g_i \tilde V^{i}_{e,i}
    + \ve^{ijk} \dow_j \lb Tg_1 \z_k \rb, \nn\\
  \vs_\e^i
  &= \z^{i}
    \Bigg[
    \sum_{i=1}^3 f_i S_{5+i}
    + (\mu_s + \half \z^k \z_k) \lb \sum_{i=1}^3 \a_{R_s,i} S_{e,i}
    + \sum_{i=1}^2 \tilde\a_{R_s,i} \tilde S_{e,i} \rb
    - \sum_{i=1}^5 \b_{2i} S_i
    \Bigg] \nn\\
  &\quad
    + (\mu_s + \half \z^k \z_k) \sum_{i=1}^3 f_i V_{e,i}^{i}
    - \sum_{i=1}^2 g_i \tilde V^{i}_{e,i+2}
    - \sum_{i=1}^3 \k_{1i} V^{i}_i
    - \sum_{i=1}^3 \tilde\k_{1i} \tilde V^{i}_{i}
    + 3 C^{(4)} \mu^2 B^{i}, \nn\\
  &\quad
    + \ve^{ijk} \dow_j (Tg_3 \z_k) + C_1 T^2 \o^i \nn\\
  \vs_t^{ij}
  &= \z^i \z^j
    \Bigg[
    \sum_{i=1}^3 \a_{R_s,i} S_{e,i}
    + \sum_{i=1}^2 \tilde\a_{R_s,i} \tilde S_{e,i}
    - \sum_{i=1}^2 \frac{g_i}{2\hat\mu_s} \tilde S_{e,i}
    - \sum_{i=1}^5 \b_{3i} S_i
    \Bigg]
    - \eta \s^{ij} - \tilde \eta \tilde\s^{ij} \nn\\
  &\quad
    - 2 \z^{(i} \Bigg[
    \sum_{i=1}^3 f_i V_{e,i}^{j)} + \sum_{i=1}^3 \k_{2i} V^{j)}_i
    + \sum_{i=1}^3 \tilde\k_{2i} \tilde V^{j)}_{i}
    \Bigg] 
    + \tilde p^{ij} \Bigg[
    \sum_{i=1}^3 f_i S_{e,i}
    - \sum_{i=1}^5 \b_{1i} S_i
    \Bigg] 
    , \nn\\
  \vs_q^i
  &= - \z^i \Bigg[
    \sum_{i=1}^3 \a_{R_s,i} S_{e,i}
    + \sum_{i=1}^2 \tilde\a_{R_s,i} \tilde S_{e,i}
    + \sum_{i=1}^5 \b_{4i} S_i
    \Bigg]
    + \ve^{ijk} \dow_j \lb Tg_2 \z_k \rb \nn\\
  &\quad
    + \sum_{i=1}^3 f_i V_{e,i}^i
    + \sum_{i=1}^2 g_i \tilde V_{e,i}^i
    - \sum_{i=1}^3 \k_{3i} V^{i}_i
    - \sum_{i=1}^3 \tilde\k_{3i} \tilde V_{i}^i
    + 6 C^{(4)} \mu B^i
    , \nn\\
  \vs_s^i
  &= \z^i \sum_{i=1}^5 \frac{\mu \b_{4i} - \b_{2i}}{T} S_i
    - \ve^{ijk} \lB
    \frac{\mu_n}{T} \dow_j \lb Tg_1 \z_k \rb
    + \frac{\mu}{T} \dow_j \lb Tg_2 \z_k \rb
    - \frac{1}{T} \dow_j \lb Tg_3 \z_k \rb
    \rB \nn\\
  &\quad
    + \sum_{i=1}^3 \frac{\mu \k_{3i} - \k_{1i}}{T} V^{i}_i
    + \sum_{i=1}^3 \frac{\mu \tilde\k_{3i} - \tilde\k_{1i}}{T} \tilde V_{i}^i
    + T g_1 \ve^{ijk} \z_j \dow_k \nu_n
    + T g_2 \ve^{ijk} \z_j \dow_k \nu
    + 2 C_1 T \o^i.
\end{align}
In addition, we have the Josephson equation,
\begin{multline}
  - \half \z^i \z_i - \mu_s + \mu_n - \mu = \frac{1}{\b_{55}} \lb \dow_t R_s + \dow_i (R_s \xi^i) \rb
 -\sum_{i=1}^4 \frac{\b_{5i}}{\b_{55}} S_i \\
 + \frac{1}{\b_{55}}\dow_k \lb \z^k \sum_{i=1}^3 \a_{R_s,i} S_{e,i}
  + \z^k \sum_{i=1}^2 \tilde\a_{R_s,i} \tilde S_{e,i}
 - \sum_{i=1}^3 f_i V_{e,i}^k
 - \sum_{i=1}^2 g_i \tilde V_{e,i}^k \rb,
\end{multline}
which is the derivative correction of the ideal order version $\mu_s = - \half \z^i \z_i + \mu_n -
\mu$. These equation can be compared with \cite{landau1959fluid} for which the $\rmU(1)$ chemical
potential $\mu = 0$. This completes our discussion of Galilean superfluids coupled to flat Galilean
spacetime, expressed in non-covariant notation.

\section{Discussion}\label{sec5}

We worked out the most generic constitutive relations of an (anomalous) Galilean superfluid up to
first order in derivative expansion, both in parity even and odd sectors. We extended the idea of
null fluid introduced in \cite{Banerjee:2015uta,Banerjee:2015hra} to null superfluid, which is a
relativistic embedding of Galilean superfluids in one higher dimension, and used these to obtain the
mentioned results. We found the spectrum of transport coefficients to be extremely rich with 38
coefficients in parity-even and 13 coefficients in parity-odd sector at first order, in addition to
two undetermined constants in parity-odd sector including the $\rmU(1)$ anomaly constant (see
\cref{summ}). Out of these, 3 parity-odd and 3 parity-even coefficients survive in equilibrium and
determine the hydrostatic physics, while 13 parity-even and 7 parity-odd coefficients govern
non-dissipative phenomenon away from equilibrium. On the other hand, 22 parity-even and 3 parity-odd
coefficients are dissipative. Though we did not discuss it in the main text, there are hints that 13
parity-even non-dissipative non-hydrostatic coefficients and 3 parity-odd dissipative coefficients
vanish on imposing Onsager relations (microscopic reversibility). To avoid confusion with counting,
we would like to note that we have removed one parity-even hydrostatic coefficient by redefinition
of the $\rmU(1)$ phase $\vf$.

An important point to note is that in this work we have only been interested in a broader class of
Galilean systems, and not the non-relativistic ones specifically. A system is said to be Galilean if
it respects Galilean symmetry transformations (as opposed to the Poincar\'{e} transformations for
the relativistic case). On the other hand a non-relativistic system is obtained by taking
$c\ra \infty$ limit. Every non-relativistic system is Galilean as it respects Galilean symmetry
transformations, but the converse might not be true, i.e. not every Galilean system necessarily
follows from $c\ra \infty$ limit of a relativistic system.  Keeping this in mind, most of the
existing literature on non-relativistic physics (e.g. non-relativistic fluid dynamics in
\cite{landau1959fluid}) actually refers to Galilean physics, as it is more natural to formulate a
theory with Galilean symmetry, than it is to take a relativistic system and perform a
non-relativistic limit (in addition, most of this literature was written before special relativity
was well explored).  Following this philosophy, here we have focused on Galilean (super)fluids,
with the ambition to return to a rigorous analysis of non-relativistic (super)fluids in near future.
At this point, we can only conclude that a non-relativistic superfluid obtained as a low energy
limit of some relativistic superfluid is at least a subsystem of the Galilean superfluid studied in
this paper.

Perhaps the most striking benefit of working in the offshell formalism is that it leads to a
complete classification of (super)fluid transport up to all orders in derivative expansion
\cite{Haehl:2014zda,Haehl:2015pja,Jain:2016rlz}, and provides a natural setting to attempt writing
down a Wilsonian effective action describing the entire (super)fluid dynamics
\cite{Grozdanov:2013dba, Kovtun:2014hpa, Harder:2015nxa, Haehl:2015pja, Haehl:2015uoc,
  Crossley:2015tka, deBoer:2015ija}. It will be interesting to undertake these ambitious problems in
context of null/Galilean (super)fluids, and we plan to return to these in future.

In this paper, we focused on breaking the internal $\rmU(1)$ symmetry of Galilean fluids and obtain
a null/Galilean superfluid. The same procedure can also be used to break spacetime symmetries, which
lead to the formation of boundaries/surfaces in (super)fluids \cite{Armas:2015ssd}. In an upcoming
paper \cite{superfluid.surfaces}, authors discuss the surface transport for relativistic and
Galilean superfluids. Finally, first order computations of this paper can also be easily extended to
higher orders; in an ongoing project \cite{ongoing} we are looking at some interesting second order
phenomenon in Galilean (super)fluids.

\subsection*{Acknowledgements}

We would like to thank Ashish Kakkar for initial collaboration during this project.  We would also
like to thank Jay Armas, Sayali Bhatkar, Jyotirmoy Bhattacharya and Felix Haehl for various helpful
discussions. The work of NB is supported by DST Ramanujan Fellowship. SD acknowledges hospitality at
ICTP and Simon's Fellowship. AJ would like to thank Durham Doctoral Scholarship for financial
support, and hospitality of ICTP where part of this project was done.  NB and SD would like to thank
the people of India for their generous support to basic science research.

\appendix



\section{Equilibrium Partition Function} \label{eqbPF}

It was realized by \cite{Banerjee:2012iz,Jensen:2012jh} that a huge part of the (super)fluid
constitutive relations can be fixed by requiring existence of an equilibrium partition function,
which generates the part of the constitutive relations that survive in equilibrium. In this
appendix, we will discuss the equilibrium partition function for Galilean superfluids. In
hydrodynamics, equilibrium is defined by a set of fields $\scrK = \{K^\sM,\L_K\}$ with $K^\sM K_\sM
< 0$, which act on the background fields $g_{\sM\sN}$, $A_\sM$ and the superfluid phase $\vf$ as an isometry,
\begin{equation}\nn
  \d_\scrK g_{\sM\sN} = \N_\sM K_\sN + \N_\sN K_\sM = 0, \qquad
  \d_\scrK A_\sM = \dow_\sM (\L_{K} + K^\sN A_\sN ) + K^\sN F_{\sN\sM} = 0,
\end{equation}
\begin{equation}
  \d_\scrK \vf = K^\sM \dow_\sM \vf - \L_K = K^\sM \xi_\sM - (\L_{K} + K^\sN A_\sN ) = 0.
\end{equation}
For simplicity, we choose a basis $\{x^\sM\} = \{ x^-, t, x^i \}$ such that the null isometry $\scrV
= \{V = \dow_-, \L_V = 0\}$ and the equilibrium isometry $\scrK = \{K = \dow_t, \L_K = 0\}$. The
fact that $\scrV$, $\scrK$ are isometries implies that all the fields are independent of $x^-$, $t$
coordinates. In this basis, we decompose the background fields as,
\begin{align}
  \df s^2 &= - 2 \E{-\F} (\df t + a_i \df x^i) (\df x^- - B_t \df t - B_i \df x^i) + g_{ij} \df x^i
            \df x^j, \nn\\
  A &= - \df x^- + A_t \df t + A_i \df x^i.
\end{align}
We will denote the covariant derivative associated with the spatial metric $g_{ij}$ by
$\acute\N_i$. After choosing the said basis, the residual diffeomorphisms are the spatial
diffeomorphisms $x^i \ra x^i + \c^i(x^j)$, mass gauge transformations $x^- \ra x^- + \c^- (x^i)$ and
Kaluza-Klein gauge transformations $t \ra t + \c^t (x^i)$. Under mass gauge transformations, only
fields that transform are,
\begin{equation}
  \d_{\c^-} B_i = - \dow_i \c^-, \qquad
  \d_{\c^-} A_i = - \dow_i \c^-,
\end{equation}
while under Kaluza-Klein gauge transformations,
\begin{equation}
  \d_{\c^+} a_i = \dow_i \c^+, \qquad
  \d_{\c^+} B_i = B_t \dow_i \c^+, \qquad
  \d_{\c^+} A_i = A_t \dow_i \c^+.
\end{equation}
We define the fields,
\begin{equation}
  \acute B_i = B_i - a_i B_t, \qquad
  \acute{A}_i = A_i - a_i A_t - \acute B_i.
\end{equation}
$\acute B_i$ is mass gauge field which is invariant under Kaluza-Klein guage
transformations. $\acute A_i$ on the other hand is invariant under both mass and Kaluza-Klein gauge
transformations, and only transforms under the $\rmU(1)$. $a_i$ is Kaluza-Klein gauge field.
Components of the superfluid velocity $\xi_\sM = \dow_\sM \vf + A_\sM$ can be found as,
\begin{equation}
  \xi_- = -1, \qquad
  \xi_t = A_t, \qquad
  \xi_i = \dow_i \vf + A_i.
\end{equation}
Out of these, $\xi_i$ is not mass or Kaluza-Klein gauge invariant due to presence of $A_i$. We can
write an invariant version as,
\begin{equation}
  \acute\xi_i = \dow_i \vf + \acute A_i.
\end{equation}
The superfluid potential can also be written in terms of these as,
\begin{equation}
  \mu_s = -\half \xi^\sM \xi_\sM = -\half \acute\xi^i \acute\xi_i - \E{\F} A_t + \E{\F} B_t,
\end{equation}
and we define $\acute\mu_s = -\half \acute\xi^i \acute\xi_i$. Finally, the fundamental variables at equilibrium
are,
\begin{equation}
  \F, \quad A_t, \quad B_t, \quad a_i, \quad \acute A_i, \quad \acute B_i, \quad
  g_{ij}, \quad \vf.
\end{equation}
The argument is that at equilibrium, constitutive relations should be derivable from an equilibrium
partition function written in terms of these fundamental fields. In covariant terms, variation of an
equilibrium partition function $W$ can be parametrized as,
\begin{equation}
  \d W = \int \{\df x^\sM\} \sqrt{-g} \ \lb \half T^{\sM\sN} \d g_{\sM\sN} + J^\sM \d A_\sM + K \d \vf \rb.
\end{equation}
In our chosen basis it decomposes as,
\begin{multline}
  \d W = \int \lbr \df x^i \rbr \sqrt{g_3} \bigg[ 
  \lb T_{t-} + T_{--} B_t \rb \d \F
  + \E{-\F} \lb T^i{}_{t} + J^i A_t \rb \d a_i
  + \half\E{-\F} T^{ij} \d g_{ij} \\
  + \lb T_{--} \d B_t - \E{-\F} (T^{i}{}_{-} - J^i) \d \acute B_i \rb
  - \lb J_- \d A_t - \E{-\F} J^i \d \acute A_i \rb
  + \E{-\F} K \d \vf
  \bigg],
\end{multline}
where $g_3 = \det g_{ij}$. Now, given the most generic partition function
$W[\F,A_t,B_t,a_i,\acute A_i, \acute B_i, g_{ij},\vf]$ as a gauge invariant scalar functional of the
fundamental fields, various components of the currents $T^{\sM\sN}$, $J^\sM$, $K$ can be read out in
terms of $W$ as,
\begin{equation}\nn
  T_{--} = \frac{1}{\sqrt{g_3}} \frac{\d W}{\d B_t}, \qquad
  T_{t-} = \frac{1}{\sqrt{g_3}} \lb \frac{\d W}{\d \F} - B_t \frac{\d W}{\d B_t} \rb,  
\end{equation}
\begin{equation}\nn
  T^i_{\ -} = - \frac{\E{\F}}{\sqrt{g_3}} \lb \frac{\d W}{\d \acute B_i} - \frac{\d W}{\d \acute
    A_i} \rb, \qquad
  T^i_{\ t} = \frac{\E{\F}}{\sqrt{g_3}} \lb \frac{\d W}{\d a_i} - A_t \frac{\d W}{\d \acute
    A_i} \rb, \qquad
  T^{ij} = \frac{2 \E{\F}}{\sqrt{g_3}} \frac{\d W}{\d g_{ij}},
\end{equation}
\begin{equation}
  J_- = -\frac{1}{\sqrt{g_3}} \frac{\d W}{\d A_t}, \qquad
  J^i = \frac{\E{\F}}{\sqrt{g_3}} \frac{\d W}{\d \acute A_i}.
\end{equation}
Since these expressions are already in a ``non-covariant notation'', we can easily perform null
reduction to read out the Galilean currents. We define a Galilean frame field to perform the reduction,
\begin{equation}
  v^\sM_{(K)} = -\frac{K^\sM}{V_\sM K^\sM} + \frac{K^\sR K_\sR V^\sM}{2(V_\sN K^\sN)^2}
  =
  \begin{pmatrix}
    \E{\F} B_t \\
    \E{\F} \\
    0
  \end{pmatrix}.
\end{equation}
In $v_{(K)}^\sM$ Galilean frame, the Galilean currents can be read out in terms of $W$ as,
\begin{equation}\nn
  \r = \frac{1}{\sqrt{g_3}} \frac{\d W}{\d B_t}, \qquad
  \r^i = \frac{\E{\F}}{\sqrt{g_3}} \lb \frac{\d W}{\d \acute B_i} - \frac{\d W}{\d \acute
    A_i} \rb, \qquad
  t^{ij}_{(v_K)} = \frac{2 \E{\F}}{\sqrt{g_3}} \frac{\d W}{\d g_{ij}},
\end{equation}
\begin{equation}\nn
  \e_{(v_K)} = \frac{\E{\F}}{\sqrt{g_3}} \frac{\d W}{\d \F}, \qquad
  \e^i_{(v_K)} = \frac{\E{2\F}}{\sqrt{g_3}} \lb - \frac{\d W}{\d a_i} + (A_t - B_t) \frac{\d W}{\d
    \acute A_i} + B_t \frac{\d W}{\d \acute B_i} \rb,
\end{equation}
\begin{equation}\label{eqbPF_varformula}
  j = \frac{1}{\sqrt{g_3}} \frac{\d W}{\d A_t}, \qquad
  j^i = \frac{\E{\F}}{\sqrt{g_3}} \frac{\d W}{\d \acute A_i}.
\end{equation}
Finally, we can write down the most general equilibrium partition function $W$ up to first order in
derivatives as,



\begin{multline}\label{eqbPF_full}
  W = \int \{\df x^i\}\sqrt{g_3} \bigg[
  \E{-\F} P
  + \E{-\F} f_1 \acute\xi^i\dow_i \F
  + f_2 \acute\xi^i\dow_i A_t
  + f_3 \acute\xi^i\dow_i B_t
  + f_4 \acute\N_i\lb \acute\xi^i \frac{\dow P}{\dow \acute\mu_s} \rb
  + \acute\N_i (f_5 \acute\xi^i) \\
  + (g_1 + g_2) \ve^{ijk} \acute\xi_i \dow_j \acute B_k
  + g_2 \ve^{ijk} \acute\xi_i \dow_j \acute A_k
  + (g_1 B_t + g_2 A_t - \E{-\F} g_3 ) \ve^{ijk} \acute\xi_i \dow_j a_k
  - C_1 \ve^{ijk} a_i \dow_j \acute B_k
  \bigg],
\end{multline}
where the coefficients $P$, $f_i$, $g_i$ are arbitrary functions of the scalars $\F$, $A_t$, $B_t$
and $\acute\mu_s$. $C_1$ on the other hand has to be a constant, so that integral of the term
coupling to it is gauge invariant. The term coupling to $f_4$ is multiplied with the first order
equation of motion of $\vf$ and hence can be neglected. On the other hand, term coupling to $f_5$ is
a total derivative. Acute reader might note that we have not included a term like to
$C_0 \ve^{ijk} \acute B_i \dow_j \acute B_k$. The reason is that this term does not have a
``covariant analogue'' and hence is switched off by the second law of thermodynamics
\cite{Banerjee:2015hra}. Finally, this equilibrium partition function does not account for
anomalies; for a discussion on anomalous partition function for null fluids see
\cite{Banerjee:2015hra,Jain:2015jla}.

Varying the partition function $W$ in \cref{eqbPF_full} and using \cref{eqbPF_varformula}, we can
read out the equilibrium constitutive relations. We will not perform the explicit variation here,
but one can check that the constitutive relations gained are the same as the ones derived in the
bulk of the paper, after identifying the equilibrium values of the hydrodynamic fields,
\begin{equation}
  u^\sM |_{eqb} = v^\sM_{(K)}, \qquad
  T  |_{eqb} = \E{\F}, \qquad
  \mu_n |_{eqb} = \E{\F} B_t, \qquad
  \mu |_{eqb} = \E{\F} A_t.
\end{equation}
These can also be summarized as $\scrB |_{eqb} = \{\b^\sM, \L_\b\}_{eqb} = \{K^\sM, \L_K\}  =
\scrK$. Having established that, the equilibrium value of the projected superfluid velocity is given as,
\begin{equation}
  \z_\sM  |_{eqb} = P_{\sM\sN}\xi^\sN  |_{eqb} =
  \begin{pmatrix}
    0 \\ 0 \\ \acute \xi_i
  \end{pmatrix},
\end{equation}
and hence $\hat\mu_s |_{eqb} = \acute\mu_s$. This finishes our discussion of equilibrium partition
function for null/Galilean superfluids.

\section{Calculational Details}\label{calc.details}

In this appendix, we will give details of the computation regarding divergence of the free energy
current, glossed over in the main text. We will find the following identities useful in the
following computation: let $S$ be a scalar and $\b^\mu$ be a vector, then,
\begin{equation}
  \N_\mu (\b^\mu S) = \frac{1}{\sqrt{-g}} \lie_\b \lb \sqrt{-g} S \rb = \half S g^{\mu\nu}
  \lie_\b g_{\mu\nu}  + \lie_\b S.
\end{equation}
There is a corresponding null background version of this identity,
\begin{equation}
  \N_\sM (\b^\sM S) = \frac{1}{\sqrt{-g}} \lie_\b \lb \sqrt{-g} S \rb = \half S g^{\sM\sN}
  \lie_\b g_{\sM\sN}  + \lie_\b S.
\end{equation}
Given a tensor $X^{\mu\nu}$, we have,
\begin{align}
  \N_{\mu}\N_{\nu} X^{[\mu\nu]}
  &= \half \lb \N_{\mu}\N_{\nu} - \N_{\nu}\N_{\mu} \rb X^{\mu\nu} \nn\\
  &= \half \lb R_{\mu\nu}{}^{\mu}{}_{\r} X^{\r\nu} + R_{\mu\nu}{}^{\nu}{}_{\r} X^{\mu\r} \rb
    = \half \lb R_{\nu\r} X^{\r\nu} - R_{\mu\r} X^{\mu\r} \rb = 0.
\end{align}
Similarly,
\begin{equation}
  \N_{\sM}\N_{\sN} X^{[\sM\sN]} = 0.
\end{equation}

\paragraph*{Relativistic Superfluid Free Energy Current:}

Let us start with relativistic superfluids. The $\d_\scrB$ variation of hydrodynamic and superfluid
fields can be computed to be,
\begin{equation}\nn
  \d_\scrB T = \frac{T}{2} u^\mu u^\nu \d_\scrB g_{\mu\nu}, \qquad
  \d_\scrB \bfrac{\mu}{T} = \frac{1}{T}u^\mu \d_\scrB A_\mu, \qquad
  \d_\scrB \mu_s = \frac{1}{2} \xi^\mu \xi^\nu \d_\scrB g_{\mu\nu} - \xi^\mu \d_\scrB
  A_\mu - \xi^\mu \N_\mu \d_\scrB \vf,
\end{equation}
\begin{equation}\nn
  \d_\scrB \hat\mu_s = \frac{1}{2} \lb \z^\mu \z^\nu - 2 (u^\r\xi_\r) u^{(\mu} \z^{\nu)} \rb \d_\scrB g_{\mu\nu}
  - \z^\mu \d_\scrB A_\mu - \z^\mu \N_\mu \d_\scrB \vf,
\end{equation}
\begin{equation}\nn
  \d_\scrB u^\mu = \frac{1}{2} u^\mu u^\r u^\nu \d_\scrB g_{\r\nu}, \qquad
  \d_\scrB u_\mu = \lb 2P_{\mu}^{\ (\r} u^{\nu)} - u_\mu u^\r u^\nu \rb \half \d_\scrB g_{\r\nu},
\end{equation}
\begin{equation}
  \d_\scrB \z^\mu =
  \lb u^\mu u^{(\r} \z^{\s)} - P^{\mu(\r} \xi^{\s)} \rb \d_\scrB g_{\r\s}
  +  P^{\mu\nu} \d_{\scrB} \xi_\nu, \quad
  \d_\scrB \z_\mu = (u^\nu \xi_\nu) u^{(\r} P^{\s)}_{\ \mu} \d_\scrB g_{\r\s}
  +  P_\mu^{\ \nu} \d_{\scrB} \xi_\nu.
\end{equation}
The first order parity-even free energy current $\cN^\mu$ in \cref{rel.even.N} has a term $2f_1
u^{[\mu} \xi^{\nu]} \frac{1}{T^2}\dow_\nu T$. We compute its divergence,
\begin{align}
  \N_\mu
  & \lb 2f_1 u^{[\mu} \z^{\nu]} \frac{1}{T^2}\dow_\nu T \rb
    =
    f_1 \z^{\nu} \frac{1}{2T}\dow_\nu T g^{\r\s} \d_\scrB g_{\r\s}
    + \d_\scrB \lb f_1 \z^{\nu} \frac{1}{T}\dow_\nu T \rb
    - \N_\mu \lb f_1 \z^{\mu} \frac{1}{T} \d_\scrB T \rb \nn\\
  &\qquad=
    f_1 \z^{\nu} \frac{1}{2T}\dow_\nu T P^{\r\s} \d_\scrB g_{\r\s}
    + f_1 \frac{1}{T}\dow_\nu T \d_\scrB\z^{\nu}
    + \z^{\nu} \frac{1}{T}\dow_\nu T \d_\scrB f_1 \nn\\
  &\qquad\qquad
    - f_1 \z^{\nu} \frac{1}{2T}\dow_\nu T u^\r u^\s \d_\scrB g_{\r\s}
    - f_1 \z^{\nu} \frac{1}{T^2}\dow_\nu T \d_\scrB T
    + f_1 \z^{\nu} \frac{1}{T}\dow_\nu \d_\scrB T
    - \N_\mu \lb f_1 \z^{\mu} \frac{1}{T} \d_\scrB T \rb \nn\\
  &\qquad=
    f_1 \z^{\nu} \frac{1}{2T}\dow_\nu T P^{\r\s} \d_\scrB g_{\r\s}
    + f_1 \frac{1}{T}\dow_\nu T \lB \lb u^\nu u^{(\r} \z^{\s)} - P^{\nu(\r} \xi^{\s)} \rb \d_\scrB g_{\r\s}
  +  P^{\nu\r} \d_{\scrB} \xi_\r \rB \nn\\
  &\qquad\qquad
    + \z^{\nu} \frac{1}{T}\dow_\nu T \lb \frac{\dow f_1}{\dow T} \d_\scrB T + \frac{\dow f_1}{\dow
    \nu} \d_\scrB \nu + \frac{\dow f_1}{\dow \hat\mu_s} \d_\scrB \hat\mu_s \rb \nn\\
  &\qquad\qquad
    - f_1 \z^{\nu} \frac{1}{2T}\dow_\nu T u^\r u^\s \d_\scrB g_{\r\s}
    - \N_\mu \lb f_1 \z^{\mu}\rb \frac{1}{T} \d_\scrB T \nn\\
  &\qquad=
    \bigg[
    u^\r u^\s \lb
    \a_{E,1} S_{e,1}
    - \frac{1}{T}\N_\mu \lb T f_1 \z^{\mu}\rb
    \rb
    + \lb \z^\r \z^\s - 2 (u^\mu\xi_\mu) u^{(\r} \z^{\s)} \rb S_{e,1} \a_{R_s,1} \nn\\
  &\qquad\qquad
    + \tilde P^{\r\s} f_1 S_{e,1}
    + 2 u^{(\r} \z^{\s)} f_1 S_5
    - f_1 2 \xi^{(\r} V^{\s)}_{e,1}
    \bigg] \half \d_\scrB g_{\r\s} \nn\\
  &\qquad\qquad
    + \bigg[ u^\r \a_{Q,1} S_{e,1} + f_1 V^\r_{e,1} - \z^\r \a_{R_s,1} S_{e,1} \bigg] \d_\scrB A_\r
    +  \bigg[ f_1 V^\r_{e,1} - \z^\r \a_{R_s,1} S_{e,1} \bigg] \dow_\r \d_\scrB \vf.
\end{align}
Performing a differentiation by parts,
\begin{align}
  \N_\mu
  & \lb 2f_1 u^{[\mu} \z^{\nu]} \frac{1}{T^2}\dow_\nu T + \cO(\dow^2) \rb \nn\\
  &\qquad=
    \bigg[
    u^\r u^\s \lb
    \a_{E,1} S_{e,1}
    - \frac{1}{T}\N_\mu \lb T f_1 \z^{\mu}\rb
    \rb
    + \lb \z^\r \z^\s - 2 (u^\mu\xi_\mu) u^{(\r} \z^{\s)} \rb S_{e,1} \a_{R_s,1} \nn\\
  &\qquad\qquad
    + \tilde P^{\r\s} f_1 S_{e,1}
    + 2 u^{(\r} \z^{\s)} f_1 S_5
    - f_1 2 \xi^{(\r} V^{\s)}_{e,1}
    \bigg] \half \d_\scrB g_{\r\s} \nn\\
  &\qquad
    + \bigg[ u^\r \a_{Q,1} S_{e,1} + f_1 V^\r_{e,1} - \z^\r \a_{R_s,1} S_{e,1} \bigg] \d_\scrB A_\r
    -  \N_\r \bigg[ f_1 V^\r_{e,1} - \z^\r \a_{R_s,1} S_{e,1} \bigg] \d_\scrB \vf.
\end{align}
From here we can read out the contributions to the constitutive relations
\cref{rel.even.N.consti}. Similarly divergence of the other term in \cref{rel.even.N} coupling to
$f_2$ can also be computed. Now, the first order parity-odd free energy current $\cN^\mu$ in
\cref{rel.odd.N} has a term $g_2 \b^\mu \tilde S_{e,2} + g_2 \tilde V^\mu_2$. We can compute its
divergence as,
\begin{align}
  \N_\mu
  &\lb g_2 \b^\mu \tilde S_{e,2} + g_2 \tilde V^\mu_2 \rb
    = \e^{\t\nu\r\s} \d_\scrB \lb g_2T \half \xi_\t u_\nu F_{\r\s} \rb
    - \N_\mu \lb \e^{\mu\t\nu\s} g_2T \xi_\t u_\nu \d_\scrB A_\s \rb \nn\\
  &\qquad=
    \frac{T}{2} \e^{\t\nu\r\s} \xi_\t u_\nu F_{\r\s} \d_\scrB g_2
    + \half g_2 \e^{\t\nu\r\s} \xi_\t u_\nu F_{\r\s} \d_\scrB T
    + \e^{\t\nu\r\s} g_2T \half \xi_\t F_{\r\s} \d_\scrB u_\nu \nn\\
  &\qquad\qquad
    + \frac{T}{2} g_2 \e^{\t\nu\r\s} u_\nu F_{\r\s} \d_\scrB \xi_\t
    + \e^{\t\nu\r\s} g_2T \xi_\t u_\nu \N_\r \d_\scrB A_\s
    - \N_\r \lb \e^{\t\nu\r\s} g_2T \xi_\t u_\nu \d_\scrB A_\s \rb \nn\\
  &\qquad=
    \frac{T}{2} \e^{\t\nu\r\s} \xi_\t u_\nu F_{\r\s}
    \lb \frac{\dow g_2}{\dow T} \d_\scrB T
    + \frac{\dow g_2}{\dow \nu} \d_\scrB \nu
    + \frac{\dow g_2}{\dow \hat\mu_s} \d_\scrB \hat\mu_s \rb
    + \e^{\t\nu\r\s} g_2T \half \xi_\t F_{\r\s} 2P_{\nu}^{\ (\r} u^{\s)} \half \d_\scrB g_{\r\s} \nn\\
  &\qquad\qquad
    + \lb g_2T \half \e^{\mu\nu\r\s} u_\nu F_{\r\s} 
    - \N_\r \lb \e^{\r\mu\t\nu} g_2T \xi_\t u_\nu \rb \rb \d_\scrB A_\mu
    + g_2T \half \e^{\t\nu\r\s} u_\nu F_{\r\s} \N_\t \d_\scrB \vf \nn\\
  &\qquad=
    \bigg[ u^\m u^\n \tilde\a_{E,2} \tilde S_{e,2}
    + 2g_2 u^{(\m} \tilde V^{\n)}_{e,4}
    + \lb \z^\m \z^\n - 2 (u^\r\xi_\r) u^{(\m} \z^{\n)} \rb \tilde\a_{R_s,i} \tilde S_{e,2}
    - \z^\m \z^\n \frac{g_2}{2\hat\mu_s} \tilde S_{e,2}
    \bigg] \half \d_\scrB g_{\m\n} \nn\\
  &\qquad\qquad
    + \bigg[ u^\mu \tilde\a_{Q,2} \tilde S_{e,2}
    + g_2 V_{e,2}^\mu
    - \z^\mu \tilde\a_{R_s,2} \tilde S_{e,2}
    - \N_\r \lb \e^{\r\mu\t\nu} g_2T \xi_\t u_\nu \rb
    \bigg] \d_\scrB A_\mu \nn\\
  &\qquad\qquad + \lB g_2 V_{e,2}^\mu
    - \z^\mu \tilde\a_{R_s,i} \tilde S_{e,2}
     \rB \N_\mu \d_\scrB \vf.
\end{align}
Performing a differentiation by parts,
\begin{align}
  \N_\mu
  &\lb g_2 u^\mu \tilde S_{e,2} + g_2 \tilde V_2 + \cO(\dow^2) \rb \nn\\
  &\qquad=
    \bigg[ u^\m u^\n \tilde\a_{E,2} \tilde S_{e,2}
    + 2g_2 u^{(\m} \tilde V^{\n)}_{e,4}
    + \lb \z^\m \z^\n - 2 (u^\r\xi_\r) u^{(\m} \z^{\n)} \rb \tilde\a_{R_s,i} \tilde S_{e,2}
    - \z^\m \z^\n \frac{g_2}{2\hat\mu_s} \tilde S_{e,2}
    \bigg] \half \d_\scrB g_{\m\n} \nn\\
  &\qquad\qquad
    + \bigg[ u^\mu \tilde\a_{Q,2} \tilde S_{e,2}
    + g_2 V_{e,2}^\mu
    - \z^\mu \tilde\a_{R_s,2} \tilde S_{e,2}
    - \N_\r \lb \e^{\r\mu\t\nu} g_2T \xi_\t u_\nu \rb
    \bigg] \d_\scrB A_\mu \nn\\
  &\qquad\qquad - \N_\mu \lB g_2 V_{e,2}^\mu
    - \z^\mu \tilde\a_{R_s,i} \tilde S_{e,2}
     \rB \d_\scrB \vf.
\end{align}
From here we can read out the contributions to the constitutive relations \cref{rel.odd.N.consti}.
Similarly divergence of the other term in \cref{rel.odd.N} coupling to $g_1$ can also be
computed. There is another term in the parity-odd free energy current $C_1 T^2 \o^\mu$; its
divergence is given as,
\begin{align}
  \N_\mu \lb C_1 T^2 \o^\mu \rb
  &=
    - 2 C_1 T \e^{\mu\nu\r\s} u_\mu \dow_\nu T \dow_\r u_\s
    + C_1 T^2 \e^{\mu\nu\r\s} \dow_\mu u_\nu \dow_\r u_\s \nn\\
  &=
    2 C_1 T^3 \o^{(\mu} u^{\nu)} \d_{\scrB} g_{\mu\nu}.
\end{align}
This can be matched with the constitutive relations \cref{rel.odd.N.consti}.

\paragraph*{Null Superfluid Free Energy Current:} 

We now move on to superfluids. The $\d_\scrB$ variation of hydrodynamic and superfluid
fields can be computed to be,
\begin{equation}\nn
  \d_\scrB T = T V^{(\sM} u^{\sN)}\d_\scrB g_{\sM\sN}, \qquad
  \d_\scrB \nu_n = \frac{1}{2T} u^{\sM} u^{\sN}\d_\scrB g_{\sM\sN}, \qquad
  \d_\scrB \nu = \frac{1}{T} u^\sM \d_\scrB A_\sM,
\end{equation}
\begin{equation}\nn
  \d_\scrB \mu_s = \frac{1}{2} \xi^\sM \xi^\sN \d_\scrB g_{\sM\sN} - \xi^\sM \d_\scrB A_\sM -
  \xi^\sM \N_\sM \d_\scrB\vf,
\end{equation}
\begin{equation}\nn
  \d_\scrB \hat\mu_s = \frac{1}{2} \lb \z^\sM \z^\sN + 2 \z^{(\sM} u^{\sN)} - 2 \z^{(\sM}
  V^{\sN)} (u^\sR \xi_\sR) \rb \d_\scrB g_{\sM\sN}
  - \z^\sM \d_\scrB A_\sM - \z^\sM \N_\sM \d_\scrB\vf,
\end{equation}
\begin{equation}\nn
  \d_\scrB u^\sM = \lb 2 u^\sM V^{(\sR} u^{\sS)} + V^\sM u^\sR u^\sS \rb \half \d_\scrB g_{\sR\sS}, \quad
  \d_\scrB u_\sM = \lb 2P_{\sM}^{\ (\sR} u^{\sS)}
  - V_\sM u^\sR u^\sS \rb \half \d_\scrB g_{\sR\sS},
\end{equation}
\begin{equation}\nn
  \d_\scrB \z^\sM = \lb
  - 2 \xi^{(\sR} P^{\sS)\sM}
  + 2 \z^{(\sR} V^{\sS)} u^\sM
  + 2 \z^{(\sR} u^{\sS)} V^\sM
  \rb \half \d_\scrB g_{\sR\sS}
  +  P^{\sM\sN} \d_\scrB \xi_\sN,
\end{equation}
\begin{equation}
  \d_\scrB \z_\sM = \lb
  2 (u^\sN \xi_\sN) P_{\sM}^{\ (\sR} V^{\sS)}
  - 2 P_{\sM}^{\ (\sR} u^{\sS)}
  \rb \half \d_\scrB g_{\sR\sS}
  +  P_\sM^{\ \sN} \d_\scrB \xi_\sN.
\end{equation}
The first order parity-even free energy current $\cN^\sM$ in \cref{null.even.N} has a term
$2 f_1 u^{[\sM} \z^{\sN]}\frac{1}{T^2} \dow_\sN T$. We compute its divergence,
\begin{align}
  \N_\sM
  &\lb 2 f_1 u^{[\sM} \z^{\sN]}\frac{1}{T^2} \dow_\sN T \rb
    = f_1 \z^{\sN}\frac{1}{2T} \dow_\sN T g^{\sR\sS} \d_\scrB g_{\sR\sS}
    + \d_\scrB \lb f_1 \frac{1}{T} \z^{\sN}\dow_\sN T \rb
    - \N_\sM \lb f_1 \z^{\sM}\frac{1}{T} \d_\scrB T \rb  \nn\\
  &\qquad= f_1 \z^{\sN}\frac{1}{2T} \dow_\sN T P^{\sR\sS} \d_\scrB g_{\sR\sS}
    + f_1 \frac{1}{T} \dow_\sN T \d_\scrB \z^{\sN}
    + \frac{1}{T} \z^{\sN}\dow_\sN T \d_\scrB f_1 \nn\\
  &\qquad\qquad
    - f_1 \z^{\sN}\frac{1}{T} \dow_\sN T V^{\sR} u^\sS \d_\scrB g_{\sR\sS}
    - f_1 \frac{1}{T^2} \z^{\sN} \dow_\sN T \d_\scrB T
    + f_1 \frac{1}{T} \z^{\sN}\dow_\sN \d_\scrB T
    - \N_\sM \lb f_1 \z^{\sM}\frac{1}{T} \d_\scrB T \rb \nn\\
  &\qquad= f_1 \z^{\sN}\frac{1}{2T} \dow_\sN T P^{\sR\sS} \d_\scrB g_{\sR\sS}
    + f_1 \frac{1}{T} \dow_\sM T \lB \lb
  - 2 \xi^{(\sR} P^{\sS)\sM}
  + 2 \z^{(\sR} V^{\sS)} u^\sM
  \rb \half \d_\scrB g_{\sR\sS}
  +  P^{\sM\sN} \d_\scrB \xi_\sN \rB \nn\\
  &\qquad\qquad
    + \frac{1}{T} \z^{\sN}\dow_\sN T \lb
    \frac{\dow f_1}{\dow T} \d_\scrB T
    + \frac{\dow f_1}{\dow \nu} \d_\scrB \nu
    + \frac{\dow f_1}{\dow \nu_n} \d_\scrB \nu_n
    + \frac{\dow f_1}{\dow \hat\mu_s} \d_\scrB \hat\mu_s
    \rb \nn\\
  &\qquad\qquad
    - f_1 \z^{\sN}\frac{1}{T} \dow_\sN T V^{\sR} u^\sS \d_\scrB g_{\sR\sS}
    - \N_\sM \lb f_1 \z^{\sM} \rb \frac{1}{T} \d_\scrB T \nn\\
  &\qquad=
    \bigg[
    2V^{(\sR} u^{\sS)} \lb \a_{E,1} S_{e,1} - \frac{1}{T}\N_\sM \lb T f_1
    \z^{\sM} \rb \rb
    + u^\sR u^\sS \a_{R_n,1} S_{e,1}
    + \tilde P^{\sR\sS} f_1 S_{e,1}
    - 2 f_1 \xi^{(\sR} V_{e,1}^{\sS)} \nn\\
  &\qquad\qquad
    + \lb \z^\sR \z^\sS + 2 \z^{(\sR} u^{\sS)} - 2 \z^{(\sR} V^{\sS)} (u^\sM \xi_\sM) \rb \a_{R_s,1} S_{e,1}
    + 2 \z^{(\sR} V^{\sS)} f_1 S_6
    \bigg] \half \d_\scrB g_{\sR\sS} \nn\\
  &\qquad\quad
    + \bigg[u^\sM \a_{Q,1} S_{e,1}
    - \z^\sM \a_{R_s,1} S_{e,1}
    + f_1 V^\sM_{e,1}
    \bigg] \d_\scrB A_\sM 
    + \bigg[ f_1 V^\sM_{e,1} - \z^\sM \a_{R_s,1} S_{e,1} \bigg] \N_\sM \d_\scrB \vf.
\end{align}
Performing a differentiation by parts,
\begin{align}
  \N_\sM
  &\lb 2 f_1 u^{[\sM} \z^{\sN]}\frac{1}{T^2} \dow_\sN T + \cO(\dow^2) \rb \nn\\
  &\qquad=
    \bigg[
    2V^{(\sR} u^{\sS)} \lb \a_{E,1} S_{e,1} - \frac{1}{T}\N_\sM \lb T f_1
    \z^{\sM} \rb \rb
    + u^\sR u^\sS \a_{R_n,1} S_{e,1}
    + \tilde P^{\sR\sS} f_1 S_{e,1}
    - 2 f_1 \xi^{(\sR} V_{e,1}^{\sS)} \nn\\
  &\qquad\qquad
    + \lb \z^\sR \z^\sS + 2 \z^{(\sR} u^{\sS)} - 2 \z^{(\sR} V^{\sS)} (u^\sM \xi_\sM) \rb \a_{R_s,1} S_{e,1}
    + 2 \z^{(\sR} V^{\sS)} f_1 S_6
    \bigg] \half \d_\scrB g_{\sR\sS} \nn\\
  &\qquad\quad
    + \bigg[u^\sM \a_{Q,1} S_{e,1}
    - \z^\sM \a_{R_s,1} S_{e,1}
    + f_1 V^\sM_{e,1}
    \bigg] \d_\scrB A_\sM 
    + \N_\sM \bigg[ \z^\sM \a_{R_s,1} S_{e,1} - f_1 V^\sM_{e,1} \bigg] \d_\scrB \vf.
\end{align}
From here we can read out the contributions to the constitutive relations
\cref{null.even.N.consti}. Similarly divergence of the other terms in \cref{null.even.N} coupling to
$f_2$, $f_3$ can also be computed. Now, the first order parity-odd free energy current $\cN^\sM$ in
\cref{null.odd.N} has a term $g_2 \b^{\sM} \tilde S_{e,2} + g_2 \tilde V^\sM_{3}$. We can compute its
divergence as,
\begin{align}
  \N_\sM
  &\lb g_2 \b^{\sM} \tilde S_{e,2} + g_2 \tilde V^\sM_{3} \rb
  = \half \e^{\sN\sR\sS\sT\sK} \d_\scrB \lb g_2 T \xi_\sN V_\sR u_\sS F_{\sT\sK} \rb
  - \N_\sT \lb \e^{\sN\sR\sS\sT\sK} g_2 T \xi_\sN V_\sR u_\sS \d_\scrB A_{\sK} \rb \nn\\
  &\qquad=
    \half \e^{\sN\sR\sS\sT\sK} T \xi_\sN V_\sR u_\sS F_{\sT\sK} \d_\scrB g_2
    + \half \e^{\sN\sR\sS\sT\sK}  g_2 T \xi_\sN V_\sR F_{\sT\sK} \d_\scrB u_\sS
    + \half \e^{\sN\sR\sS\sT\sK} g_2 \xi_\sN V_\sR u_\sS F_{\sT\sK}  \d_\scrB T \nn\\
  &\qquad\qquad
    + \half \e^{\sN\sR\sS\sT\sK}  g_2 T \xi_\sN u_\sS F_{\sT\sK} \d_\scrB V_\sR
    + \half \e^{\sN\sR\sS\sT\sK} g_2 T V_\sR u_\sS F_{\sT\sK} \d_\scrB \xi_\sN \nn\\
  &\qquad\qquad
    + \e^{\sN\sR\sS\sT\sK} g_2 T \xi_\sN V_\sR u_\sS \N_\sT \d_\scrB A_{\sK}
    - \N_\sT \lb \e^{\sN\sR\sS\sT\sK} g_2 T \xi_\sN V_\sR u_\sS \d_\scrB A_{\sK} \rb\nn\\
  &\qquad=
    \half \e^{\sN\sR\sS\sT\sK} T \xi_\sN V_\sR u_\sS F_{\sT\sK} \lb
    \frac{\dow g_2}{\dow T} \d_\scrB T
    + \frac{\dow g_2}{\dow \nu} \d_\scrB \nu
    + \frac{\dow g_2}{\dow \nu_n} \d_\scrB \nu_n
    + \frac{\d g_2}{\d \hat\mu_s}\d_\scrB \hat\mu_s
    \rb\nn\\
  &\qquad\qquad
    - u^{\sA} P_{\sM}^{\ \sB} g_2 T \half \e^{\sM\sN\sR\sT\sK} V_\sN u_\sR F_{\sT\sK}\d_\scrB g_{\sA\sB}
    + \half \e^{\sN\sR\sS\sT\sK}  g_2 T \xi_\sN u_\sS F_{\sT\sK} P_{\sR}^{
    \ \sB} V^\sA \d_\scrB g_{\sA\sB}  \nn\\
  &\qquad\qquad
    - \N_\sT \lb \e^{\sT\sM\sN\sR\sS} g_2 T \xi_\sN V_\sR u_\sS \rb \d_\scrB A_{\sM}
    + \half \e^{\sN\sR\sS\sT\sK} g_2 T V_\sR u_\sS F_{\sT\sK} \d_\scrB \xi_\sN \nn\\
  &\qquad=
    \bigg[
    2 \tilde\a_{E,2} V^{(\sM} u^{\sN)} \tilde S_{e,2}
    + \tilde\a_{R_n,2} u^{\sM} u^{\sN} \tilde S_{e,2}
    - 2g_2 u^{(\sM} \tilde V_{e,2}^{\sN)}
    - 2g_2 V^{(\sM} \tilde V_{e,4}^{\sN)}
    - \z^\sM \z^\sN \frac{g_2}{2\hat\mu_s} \tilde S_{e,2} \nn\\
  &\qquad\qquad\quad
    + \lb \z^\sM \z^\sN + 2 \z^{(\sM} u^{\sN)} - 2 \z^{(\sM}
    V^{\sN)} (u^\sR \xi_\sR) \rb \tilde\a_{R_s,2} \tilde S_{e,2}
    \bigg] \half \d_\scrB g_{\sM\sN}  \nn\\
  &\qquad\qquad
    + \bigg[ u^\sM \tilde\a_{Q,2} \tilde S_{e,2}
    + g_2 \tilde V_{e,2}
    - \z^\sM \tilde\a_{R_s,2} \tilde S_{e,2}
    - P^{\sM}_{\ \sK} \N_\sT \lb \e^{\sT\sK\sN\sR\sS} g_2 T \xi_\sN V_\sR u_\sS \rb \bigg] \d_\scrB
    A_{\sM} \nn\\
  &\qquad\qquad
    + \bigg[ g_2 \tilde V_{e,2}
    - \z^\sM \tilde\a_{R_s,2} \tilde S_{e,2}  \bigg] \N_\sM\d_\scrB\vf.
\end{align}
Performing a differentiation by parts,
\begin{align}
  \N_\sM
  &\lb g_2 \b^{\sM} \tilde S_{e,2} + g_2 \tilde V^\sM_{3} + \cO(\dow^2) \rb \nn\\
  &\qquad=
    \bigg[
    2 V^{(\sM} u^{\sN)} \tilde\a_{E,2} \tilde S_{e,2}
    + u^{\sM} u^{\sN} \tilde\a_{R_n,2} \tilde S_{e,2}
    - 2g_2 u^{(\sM} \tilde V_{e,2}^{\sN)}
    - 2g_2 V^{(\sM} \tilde V_{e,4}^{\sN)}
    - \z^\sM \z^\sN \frac{g_2}{2\hat\mu_s} \tilde S_{e,2} \nn\\
  &\qquad\qquad\quad
    + \lb \z^\sM \z^\sN + 2 \z^{(\sM} u^{\sN)} - 2 \z^{(\sM}
    V^{\sN)} (u^\sR \xi_\sR) \rb \tilde\a_{R_s,2} \tilde S_{e,2}
    \bigg] \half \d_\scrB g_{\sM\sN}  \nn\\
  &\qquad\qquad
    + \bigg[ u^\sM \tilde\a_{Q,2} \tilde S_{e,2}
    + g_2 \tilde V_{e,2}
    - \z^\sM \tilde\a_{R_s,2} \tilde S_{e,2}
    - P^{\sM}_{\ \sK} \N_\sT \lb \e^{\sT\sK\sN\sR\sS} g_2 T \xi_\sN V_\sR u_\sS \rb \bigg] \d_\scrB
    A_{\sM} \nn\\
  &\qquad\qquad
    + \N_\sM \bigg[ \z^\sM \tilde\a_{R_s,2} \tilde S_{e,2} - g_2 \tilde V_{e,2} \bigg] \d_\scrB\vf.
\end{align}
From here we can read out the contributions to the constitutive relations \cref{null.odd.N.consti}.
Similarly divergence of the other term in \cref{null.odd.N} coupling to $g_1$ can also be
computed. Divergence of the term coupling to $g_3$ is particularly simple,
\begin{align}
  \N_\sM \lb g_3 \tilde V^\sM_1 \rb
  &= \N_\sM \lb \frac{g_3}{T} \e^{\sM\sN\sR\sS\sT} V_\sN u_\sR \z_\sS \dow_\sT T \rb \nn\\
  &= - \N_\sM \lb g_3T \e^{\sM\sN\sR\sS\sT} V_\sN u_\sR \z_\sS \rb \dow_\sT \bfrac1T \nn\\
  &= V^{(\sM} P^{\sN)}_{\ \ \sP} \N_\sK \lb g_3T \e^{\sP\sK\sR\sS\sT} V_\sR u_\sS \z_\sT \rb \d_\scrB g_{\sM\sN}.
\end{align}
Finally the last term in parity-odd free energy current $C_1 T \o^\sM$ has divergence,
\begin{equation}
  \N_\sM \lb C_1 T \o^\sM \rb = C_1T^2 \o^{(\sM} V^{\sN)} \d_\scrB g_{\sM\sN}.
\end{equation}
This can be matched with the constitutive relations \cref{null.odd.N.consti}.

\bibliographystyle{utcaps}
\bibliography{aj-bib}

\end{document}